\documentclass[aps,pre,twocolumn,groupedaddress,longbibliography,showkeys,showpacs]{revtex4-1}

\usepackage{graphicx}
\usepackage{bm}
\usepackage{amssymb,amsfonts}
\usepackage{amsmath}
\usepackage{graphicx}
\usepackage[usenames,dvipsnames]{xcolor}
\usepackage{soul}
\usepackage{url}
\graphicspath{{Figures/}}
\newcommand{\one}{1 \hspace{-1.2mm}  {1}}
\newcommand{\an}[1]{{\color[rgb]{1,0,0}{#1}}} 
 
\newcommand{\hmb}[1]{{\color[rgb]{0,0,1}{#1}}}

\begin{document}

\title{Complex Networks in the Framework of Nonassociative Geometry}


\author{Alexander I. Nesterov}
   \email{nesterov@cencar.udg.mx}
\affiliation{Departamento de F{\'\i}sica, CUCEI, Universidad de Guadalajara,
Av. Revoluci\'on 1500, Guadalajara, CP 44420, Jalisco, M\'exico}

\author{Pablo H\'ector Mata Villafuerte}
   \email{themata@hotmail.com}
\affiliation{Departamento de F{\'\i}sica, CUCEI, Universidad de Guadalajara,
Av. Revoluci\'on 1500, Guadalajara, CP 44420, Jalisco, M\'exico}

\date{\today}

\begin{abstract}

In the framework of on nonassociative geometry, we introduce a new effective model that extends the statistical
treatment of complex networks with hidden geometry. The small-world property of the network 
is controlled by nonlocal curvature in our model. We use this approach to study the Internet as a 
complex network embedded in a hyperbolic space. The model yields a remarkable agreement 
with available empirical data and explains features of Internet connectance data that 
other models cannot. Our approach offers a new avenue for the study of a wide 
class of complex networks, such as air transport, social networks, biological networks, etc.

\end{abstract}


\maketitle

Due to its intrinsic interdisciplinary nature, Network Science can, and already has, contributed research in very diverse fields in both the natural sciences and the human world. Refinements in the techniques and methods of Network Science would therefore be of interest to a wide variety of researchers and, conceivably, policy makers and the general public.

Many real networks of large size, i.e. the Internet, the World Wide Web, airline networks, neural networks, citation networks, etc., are highly effective in exchanging information between distant nodes. This feature implies the existence of shortcuts between most pairs of nodes, \st{and} known as the small-world property \cite{WDSS,BSLV}.

Complex Networks (CNs) have benefitted from the adoption of statistical mechanics as a powerful framework to explain properties of real-world networks \cite{NewmanIntroBook,NMSW,MEJN1,ARB,PJNM}. The statistical physics approach has also been extended using geometric and topological ideas. Increasing attention to the geometrical and topological properties of CNs is focused on four main directions: characterization of the hyperbolicity of networks, emergence of network geometry, characterization of brain geometry, and network topology \cite{BG1}.  In particular, in 
\cite{KDPF1,KDPF2,Boguna2010} a duality between a highly heterogeneous degree distribution in a network and an underlying hyperbolic geometry was found and exploited for the realistic modeling of the Internet. 

\begin{figure}[tbh]
	\centering
	\includegraphics[width=0.6\linewidth]{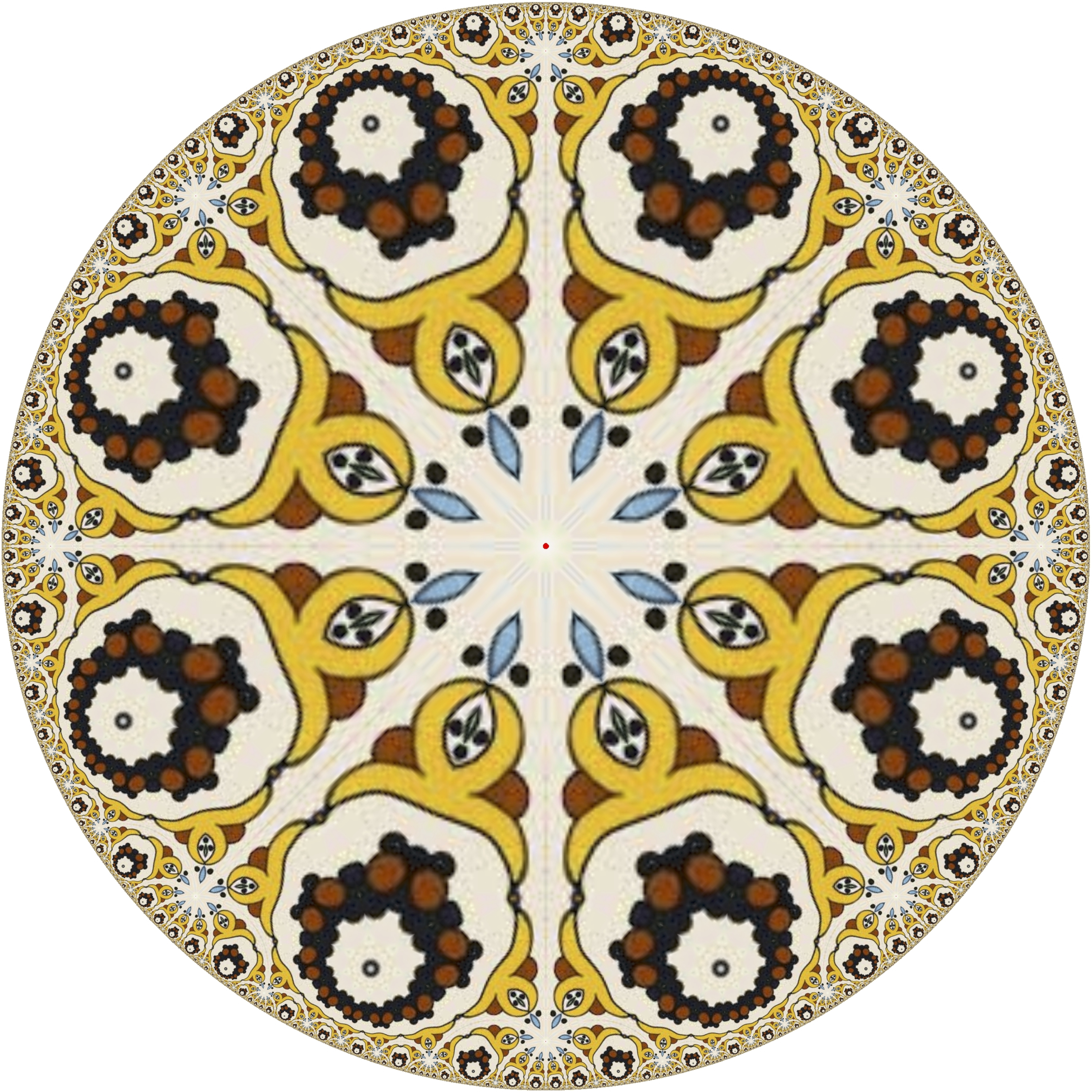} 
	\caption{Tiling of the Poincar\'e disk illustrating the exponential expansion of space. All patterns are of the same size in the hyperbolic space. The number of patterns exponentially increases with the distance from the origin, while their Euclidean size exponentially decreases. (Constructed with the \textit{Poincar\'e} tool \cite{poincare}.)}
	\label{fig1a}
\end{figure}

The exponential expansion of hyperbolic space illustrated in Fig.\ref{fig1a} allows one to map {an} exponentially growing network to a hyperbolic space. In this context, the emergence of scaling in CNs can be explained by the hidden hyperbolic geometry \cite{GDLM,SMKD1,PFKD,BABM,VKSS} (fundamental concepts concerning CNs, their statistical description and relation to hyperbolic geometry are treated in detail in 
\cite{NewmanIntroBook,NMSW,ARB,KDPF1,KDPF2,PJNM,GDLM,BABM,Boguna2010,BSLV,NOSI,BCRC,BMPSR,BOGACZ2006587,BG1,WANG}).

The successful embedding of a CN in a geometric space invites the possibility of further exploiting the geometric properties of such CNs, namely by the known methods of differential geometry. The insights and calculational benefits of statistical mechanics could thus be complemented with those from geometry to form a more complete model. However, it is not obvious how the methods of differential geometry would apply to networks, which are fundamentally discrete structures. The main challenge is to define the curvature of networks. This is a hot mathematical topic, and different approaches to resolve it can be found in the literature \cite{NOSI,BG1,OY1,SRKM,SESR,Keller2011,EE2012,EE2014}.
Nonassociative geometry \cite{Sab1,NS3,NS3a}, yielding an unified algebraic description of 
discrete spaces and smooth manifolds as well, opens a novel avenue for studying network 
geometry. The presence of curvature in a nonassociative space results in a non-trivial 
elementary holonomy, which is an equvialent of (nonlocal) curvature.

In this paper, we show how nonassociative geometry can be used to give a statistical 
description of CNs and reveal underlying geometry. We focus on the contribution from 
nonlocal curvature, described by elementary holonomy, to the statistical properties of CNs  and 
find that nonlocal curvature controls the formation of a small-world network.

As a particular example, we perform a detailed study of the Internet embedded in a hyperboloic 
space. 
Our model shows excellent agreement with the empirical Internet connectance data.
(All technical details concerning intermediate steps of our  paper are presented in the 
Supplemental Material (SM).)
\\

{\em Nonassociative geometry in brief.} --  The main  algebraic structures 
arising in nonassociative geometry are related to nonassociative 
algebra and the theory of quasigroups and loops (for details and review see Refs. 
\cite{Sab1,S5,S6,S7,NHM}).

Consider a loop $\langle Q,{\boldsymbol\cdot},e\rangle$, i.e. a set with a
binary operation (multiplication) $(a,b) \mapsto{} {a \boldsymbol\cdot b}$, and the condition 
that each of the equations $a{\boldsymbol\cdot} x=b,$ and $y{\boldsymbol\cdot} a=b$
has a unique solution:  $x=a\backslash b$, $y=b/a$. In addition, a two-sided identity holds: 
$a{\boldsymbol\cdot} e=
e{\boldsymbol\cdot} a=a$, where $e$ is a neutral element. A loop that
is also a differential manifold with an operation
$a{\boldsymbol\cdot} b$ that is a smooth map is called a {\it smooth loop}.

Nonassociativity of the operation is described by the identity $a{\mathbf\cdot} 
(b{\mathbf\cdot} c) =(a{\mathbf\cdot} b)l_{(a,b)}c$,
where $l_{(a,b)}$ is an {\em associator}. If $l_{(a,b)}= \one$, we obtain $a{\mathbf\cdot} (b{\mathbf\cdot} c) =(a{\mathbf\cdot} b)\cdot c$ and, thus, a loop $Q$ becomes a group.
The multiplication of elements $a, b \in Q $ can also be written as $ a{\boldsymbol\cdot} b = L_a b$,  where $L_a$ is a {\it  left translation}. In terms of left translations,  the  associator is given by $l_{(a,b)}=L^{-1}_{a{\boldsymbol\cdot} b}\circ L_a\circ L_b$. 
\begin{figure}[tbh]
\begin{center}
\includegraphics[width=5cm]{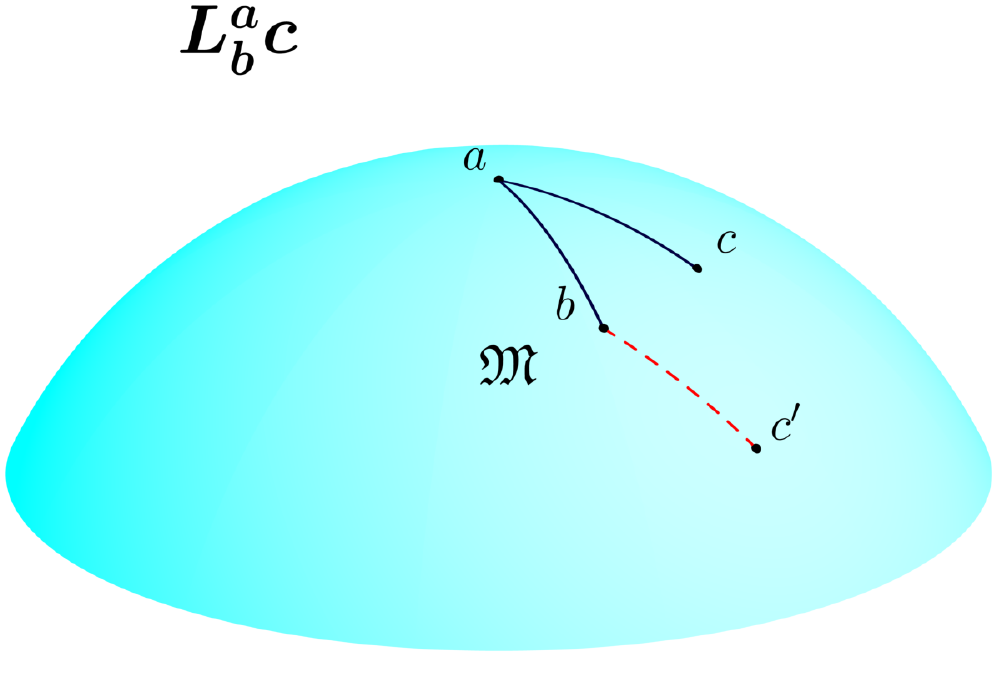}
\end{center}
\caption{ Parallel translation of the geodesic $(ac)$ along the geodesic $(ab)$. The result,  given by $c' =L^a_b c$, is presented by the geodesic $(bc')$ (red dashed curve).}
\label{fig1}
\end{figure}
\begin{figure}[tbh]
\begin{center}
\includegraphics[width=5.25cm]{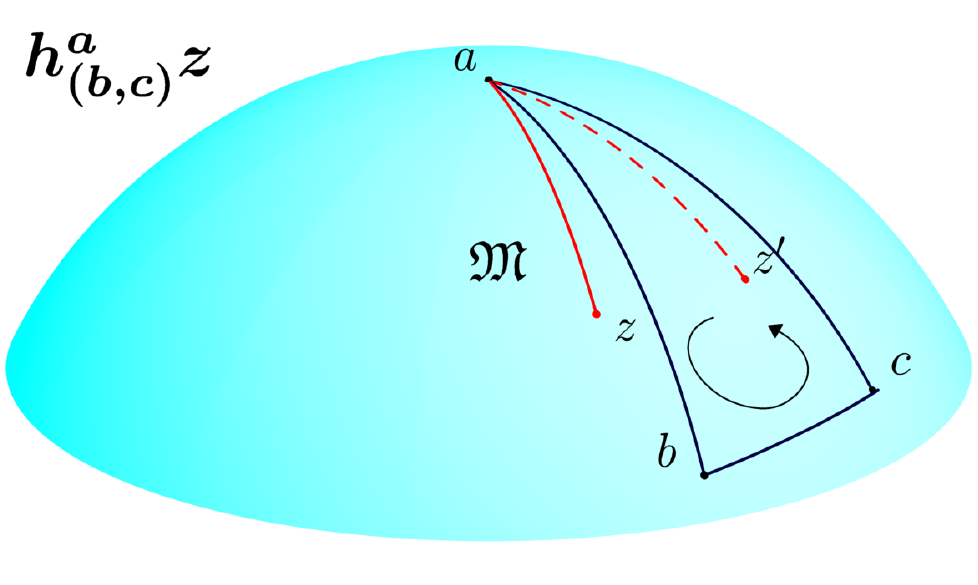}
\end{center}
\caption{Elementary holonomy $h^a_{(b,c)} $ describes the parallel translation of the geodesic $(az)$ along the geodesic triangle $(abc)$. The result is given by $z' = h^a_{(b,c)} z $ and presented by the geodesic $(az')$ (red dashed curve).}
\label{fig2}
\end{figure}

The foundations of nonassociative geometry are based on the fact
that in a neighborhood of an arbitrary point $a$ on a manifold $\mathfrak M$ with an
affine connection one can introduce the geodesic local loop, which
is uniquely defined by means of the parallel translation of
geodesics along geodesics (Fig. \ref{fig1}).

The curvature of a nonassociative space is 
described by {\em elementary holonomy},
$h^a_{(b,c)} = (L^a_c)^{-1}\circ L^b_c \circ L^a_b$,
where $L^a_b$ denotes a left translation with $a$ being a neutral element of the local loop. The elementary holonomy describes the parallel translation of the geodesic along the geodesic triangle (see Fig. \ref{fig2}). As one can see, it is some integral (nonlocal) curvature. If $h^a_{(b,c)} = \one $, we have a flat space.

As a particular example, we consider a nonassociative description of the two-dimensional hyperbolic space $\mathbb{H}^2$ presented by the Poincar\'e disk model. Let $D$ be the open unit disk: $D=\{\zeta\in\mathbb C:\left|\zeta\right|<1 \}$. We define the nonassociative binary operation $\ast$ as 
\begin{align}
L_\zeta\eta =\zeta\ast\eta=\frac{\zeta+\eta}{1+\bar\zeta\eta}, \quad\zeta,\eta\in D,
\label{L1}
\end{align}
where the bar denotes complex conjugation. The inverse operation is given by
\begin{align}
L^{-1}_\zeta\eta =\zeta^{-1}\ast\eta=\frac{\eta-\zeta}{1-\bar\zeta\eta}, \quad\zeta,\eta\in D.
\label{L1b}
\end{align}
Inside $D$, the set of complex numbers with the operation $\ast$ forms the two-sided loop QH(2) \cite{N1,N2}. 

The associator $l_{(\zeta,\eta)}$ on QH(2) is determined by
\begin{align}
l_{(\zeta,\eta)}\xi=\frac{1+\zeta\bar\eta}{1+\bar\zeta \eta }\xi.
\end{align}
 Since the hyperboloid is a symmetric space, the 
elementary holonomy is determined by the associator: $h_{(\zeta,\eta)} 
=l_{(\zeta,L^{-1}_\zeta\eta)}$\cite{S5}. The computation yields
\begin{equation}
h_{(\zeta,\eta)}\xi=\frac{1- \bar\zeta \eta }{1-\zeta\bar\eta}\xi.
\label{hol1}
\end{equation}

We define the left-invariant metric on ${H}^2$ as \cite{NHM}
\begin{align}
g(L_\eta\zeta,L_\eta \xi)=	g(\zeta,\xi) = \frac{4|\xi-\zeta|^2}{{(1 -|\zeta|^2)  (1 -|\xi|^2)}} .
\label{metric}
\end{align}

For a hyperbolic space ${\mathbb H}^2$ with curvature $K= -1/{\mathcal R}^2$ the previous formula should be modified to read
\begin{equation}
	g(\zeta,\xi) = \frac{4 {\mathcal R}^2|\xi-\zeta|^2}{{(1 -|\zeta|^2)  (1 -|\xi|^2)}} .
\label{eq:dma}
\end{equation}
Taking $\xi =  \zeta + d\zeta$, we find that
\begin{align}
	g(\zeta,\xi) \rightarrow ds^2 = \frac{4 {\mathcal R}^2d \zeta d\bar\zeta}{(1 -|\zeta|^2)^4} .
\label{eq:metric}
\end{align}

For each triplet of points, $\zeta_i,\zeta_j,\zeta_k \in D$, the elementary holonomy, 
$h^{i}_{jk}$, can be written as (see SM)
\begin{align}
h^{i}_{jk}=\frac{1-\bar\zeta_{ij}\zeta_{ik}}{1-\zeta_{ij}\bar \zeta_{ik}},
\label{hol3}
\end{align}
where
\begin{align}
\zeta_{ij}  = \frac{\bar\zeta_i\zeta_j (|\zeta_j| - |\zeta_i|)}{|\zeta_i||\zeta_j|(1-|\zeta_i||\zeta_j|)}.
\label{Z1}
\end{align}

Supposing that  $|\zeta_{ij}|,|\zeta_{jk}|,|\zeta_{ik}| \ll 1$,  we obtain
\begin{equation}
h^{i}_{jk}\approx 1-i\frac{\Delta(i,j,k) }{{\mathcal R}^2}.
\end{equation}
Here $\Delta(i,j,k) $ is the area of the geodesic triangle formed by the triplet of points $(i,j,k)$.

The phase gained by an arbitrary ``vector" $\zeta_{ip}$ during the parallel translation along the geodesic path $\gamma = \gamma_{ij}\cup \gamma_{jk} \cup  \gamma_{ki}$, where $\gamma_{ij}$ denotes the geodesic connecting the points $i$ and $j$, is given by
\begin{align}
\Delta \varphi = \frac{1}{i}\ln h^i_{jk} \approx -\frac{\Delta(i,j,k) }{{\mathcal R}^2}.
\end{align}
This is consistent with the formula for 
the parallel transportation of a vector $\mathbf V$ along a small contour $\mathcal C${ (}see SM):
\begin{align}
\Delta V^i = \frac{1}{2}R^i{}_{klm}V^k \Delta S^{lm}.
\end{align}
Here $R^i{}_{klm}$ is the curvature tensor and $\Delta S^{lm}$ is the area of the
segment restricted by $\mathcal C $. 

The loop QH(2) is isomorphic to the two-sheeted hyperboloid model (see SM for 
details). The isomorphism between the loop  QH(2) and the upper sheet $H^+$ of the 
hyperboloid is established by $\zeta=e^{\rm i\varphi}\tanh(\theta/2)$, where 
$(\theta,\varphi)$ are inner coordinates on $H^+$.  In the new variables, 
\eqref{eq:metric} yields the conventional metric on the hyperbolic 
space: $ds^2 =  {\mathcal R}^2( d\theta^2 + \sinh^2 \theta \,d\varphi^2)$.

To each pair of points $\zeta_i,\zeta_j\in D $ one can assign the hyperbolic distance, $d_{ij}$, as follows \cite{BCRC}:
\begin{align} 
\cosh(\kappa d_{ij}) = \cosh \theta_i \cosh \theta_j - \sinh \theta_i \sinh \theta_j \cos \varphi_{ij} ,
\label{Deq}
\end{align}
where  $\kappa = \sqrt{-K}= 1/{\mathcal R}$ and $\varphi_{ij}= \varphi_{j} - \varphi_{i}$. The straightforward calculation shows that
\begin{align}
\sinh \frac{ d_{ij}}{2\mathcal R} = \frac{ \ell_{ij}}{2\mathcal R}
\end{align}
where $\ell_{ij}= \sqrt{g(\zeta_i, \zeta_j)}$, and  for $d  \ll 
 \mathcal R$ we obtain $d \approx \ell$.\\

{\em Complex networks in the framework of nonassociative geometry.} -- A network 
is a set of $N$ nodes (or vertices) connected by $L$ links (or edges). One can describe 
the network by an adjacency matrix, $a_{ij}$, where each existing or nonexisting link 
between pairs of nodes ($ij$) is indicated by a 1 or 0 in the $i,j$ entry. Individual 
nodes possess local properties such as node degree (or connectivity) $k_i = \sum_j 
a_{ij}$, and clustering coefficient $c_i = \sum_{jk} a_{ij}a_{jk}a_{ki}/k_i(k_i -1)$ 
\cite{WDSS,BSLV,ARB}. The network as a whole can be described quantitatively 
by its degree distribution $P(k)$ and connectance. The connectance is characterized by the 
connection probability $p_{ij}$, i.e. the probability that a pair nodes $(ij)$ is connected.  

The most general statistical description of an undirected network in equilibrium, with a 
fixed 
number of vertices $N$ and a varying number of links, is given by the grand canonical ensemble 
\cite{PJNM,CDLM,CDAS}. For a particular graph $G$, the probability of obtaining this graph, 
$P(G)$, can be written as 
\begin{align}
P(G) = \frac{e^{-\beta H(G)}}{Z} ,
\end{align}
where $H(G)$ is the graph Hamiltonian, $Z$ denotes the partition function, and  $\beta =1/T$ stands for inverse ``temperature" of the network.

In what follows we restrict ourselves to consideration of a two-star model,  one of the simplest 
and fundamental CN models. We assume that the CN is embedded in a hyperbolic space of 
constant curvature. The Hamiltonian describing the network generalizes the weighted two-star 
Hamiltonian introduced in \cite{PJNM} and takes the form
\begin{align}
	H= \frac{4J}{N-1}\sum_{ijk} h^{i}_{jk} a_{ij}a_{ik} - 2B \sum_{ij} \alpha_{ij}a_{ij},
	\label{eq:H}
\end{align}
where $\alpha_{ij}$ is the weight of the edge $\langle ij \rangle$, $J$ and $B$ are coupling 
constants, and $h^{i}_{jk} $ denotes the elementary holonomy associated with the nodes 
$(i,j,k)$. In our approach the weights  are  determined  by  the elementray holonomy and 
connectivity of the nodes.

The first term in \eqref{eq:H} describes inhomogeneity in the distribution of links, resulting in 
natural clustering of nodes into cliques. Thus, one can expect that the holonomy (non-local 
curvature) is responsible for formation of the communities inside the CN \cite{GN}. To clarify 
this issue, let us rewrite \eqref{eq:H} as $H=\sum_{ij} E_{ij} a_{ij}$, where the energy of the link 
$\langle ij \rangle$ is
\begin{align}
 E_{ij}= \frac{4J}{N-1}\sum_{k} h^{i}_{jk}a_{ik} - 2B  \alpha_{ij}.
	\label{eq:H1}
\end{align}
The first term in this expression describes the contribution to the energy to the link $\langle ij 
\rangle$ from the remaining nodes $(k \neq i,j)$ connected with the node $i$ by the shortest 
path. This leads to a mesoscopic inhomogeneity in the distribution of links, in a such way
that nodes inside of the same group  have very high degree, but between groups the connection 
is low.

The variables $a_{ij}$ can be thought of as Ising pseudo-spins, $\sigma_{ij} $, representing the edges connecting $(ij)$ pairs of nodes in a network. We can thus map the network to the Ising model by setting $\sigma_{ij}= 2 a_{ij} -1$, such that 
\begin{align}
	\sigma_{ij} = \left \{  
	\begin{array}{rl}
		1 &{\rm  if }\, i \,{\rm  is \, connected \, to\, }  j\\
	   -1& \rm otherwise
	\end{array}
	  \right .  
\end{align}
 
Inserting $\sigma_{ij} $ into Eq. (\ref{eq:H}), after some algebra we obtain
\begin{align}
	H= \frac{J}{N-1}\sum_{ijk} h^{i}_{jk}\sigma_{ij}\sigma_{ik} 
	- \sum_{ij} B_{ij}\sigma_{ij},
	\label{eq:H1}
\end{align}
where
\begin{align}
B_{ij}=  B\alpha_{ij}-\frac{2J}{N-1}\sum_{ k} h^{i}_{(jk)},
\end{align}
and we have used the notation $h^{i}_{(jk)} = \frac{1}{2} \big(h^{i}_{jk}+h^{i}_{kj}\big )$.

Within the mean field (MF) approximation, the Hamiltonian  (\ref{eq:H1}) is
replaced by
\begin{align}
 {\mathcal H}= &\frac{J}{N-1}\sum_{i,j,k}  h^{i}_{jk} \langle \sigma_{ij}\rangle\langle \sigma_{ik}\rangle - \sum_{ij}\sigma_{ij}  {h}^{(e)}_{ij} ,
\label{H2}
\end{align}
where $\langle\dots \rangle$ denotes an expectation value, and the effective field, ${h}^{(e)}_{ij}$, is given by
\begin{align}
{h}^{(e)}_{ij}=  B_{ij} -\frac{2J}{N-1}\sum_{ k} h^{i}_{(jk)} \langle \sigma_{ik}\rangle .
\label{H2a}
\end{align}
The total Hamiltonian of the system can be rewritten as ${\mathcal H}= \sum_{ij}{\mathcal H}_{ij} $, where ${\mathcal H}_{ij} ={\mathcal H}^0_{ij}- \sigma_{ij}  {h}^{(e)}_{ij}$ is the
 Hamiltonian for a single pseudo-spin located on the edge $(ij)$, and
\begin{align}
{\mathcal H}^0_{ij}= \frac{2J}{N-1}\sum_{k} h^{i}_{(jk)} \langle \sigma_{ij}\rangle\langle \sigma_{ik}\rangle.
 \label{H3b}
\end{align}

Since the pseudo-spins in the MF approximation are decoupled, the partition function factorizes into a product of independent terms: $Z = \prod Z_{ij}$. We obtain
\begin{align}
	Z_{ij}=2\cosh(\beta h^e_{ij})e^{-\beta{\mathcal H}^0_{ij }} .
\end{align}

The computation of the expectation value for the pseudospin,  $\langle \sigma_{ij} \rangle = \partial Z_{ij}/\partial (\beta  h^e_{ij})$, yields
\begin{align}
\langle \sigma_{ij} \rangle =\tanh (\beta {h}^{(e)}_{ij}).
	\label{L1a}
\end{align}
Inserting $\langle \sigma_{ij} \rangle$ into Eq. (\ref{H2a}), we obtain a self-consistent system 
of transcendental equations to determine the effective field,
\begin{align}
{h}^{(e)}_{ij}= & B_{ij} -\frac{2J}{(N-1)}\sum_{ k} h^{i}_{(jk)}\tanh \big (\beta {h}^{(e)}_{ik} \big )  .
 \label{H5}
\end{align}

We are now in position to calculate the connectance of the network described by the connection probability, $ 
p_{ij} \equiv\langle a_{ij}\rangle = (1/2)(1+\langle \sigma_{ij}\rangle)$. Employing Eq. 
(\ref{L1a}), we obtain	
\begin{align}
	p_{ij} = \frac{1}{2}\big( 1+\tanh (\beta{h}^{(e)}_{ij})\big) = \frac{1}{1 + e^{-2\beta  {h}^{(e)}_{ij}} }.
	\label{eq_pij}
\end{align}

{\em The Internet as a complex hyperbolic network.} -- We turn now to the study of the 
Internet as a particular case of a scale-free CN embedded in a hyperbolic space $\mathbb H^2$, 
as considered in \cite{KDPF1,KDPF2,Boguna2010}. A scale-free network is characterized by a
power-law degree distribution, $P(k) \sim (\gamma -1)k^{-\gamma}$, where $k$ is the node 
degree. 

The Internet nodes are mapped to a hyperbolic space of curvature $K<0$ by assigning 
to each a random angular coordinate $\varphi$, and a radial coordinate $r=\theta/\kappa 
$ ($\kappa= \sqrt{-K}$)  according to the radial node density
\begin{align}
\rho(r) = \frac{\alpha e^{\alpha (r - R/2)}}{2\sinh (\alpha R/2) },
\quad  0\leq r \leq R, 
\label{eq:radialDensity}
\end{align}
where $\alpha =\kappa (\gamma -1)/2$.

The size of the network is given by 
 \begin{align}
 	R= \frac{2}{\kappa} \ln \bigg (\frac{N}{\bar k} \Big (\frac{\gamma -1}{\gamma -2} 
 	\Big)\bigg ){,}
 	\label{R}
 \end{align}
where $\bar k$ is the average degree in the whole network and
\begin{align}
   	\kappa =  1 -  \frac{\ln \Big (\frac{2}{\pi } \Big (\frac{\gamma 
 	-1}{\gamma -2} \Big)\Big)}{\ln \Big (\frac{2N}{\pi \bar k} \Big (\frac{\gamma -1}{\gamma 
 	-2} \Big)^{2} \Big)}.
   \end{align}

To adapt our model to empirical Internet data we consider $d$ as the independent variable 
in our calculations, thus allowing direct comparison to the results in \cite{KDPF1}.
We specify our model writing $B \alpha_{ij}= (\kappa /4)(R-d_{ij}) $, where $d_{ij}$ is the hyperbolic distance between nodes $i$ and $j$. This yields the connection probability \eqref{eq_pij} in the form of the Fermi-Dirac distribution,
\begin{align}
	p_{ij} = \frac{1}{ e^{\beta (\varepsilon_{ij}- \mu)} +1 },
	\label{FD}
\end{align}
where $\mu = \kappa R/2$ is the chemical potential, and 
\begin{align}
	\varepsilon_{ij} =\frac{\kappa d_{ij}}{2} + \frac{8J}{(N-1)}\sum_{ k} h^{i}_{(jk)}\langle a_{ik} \rangle .
\end{align}
The second term in this expression includes the contribution to the energy $\varepsilon_{ij} $ of the 
link $(ij)$ from all nodes in the network and, thus, leads to the formation of ``small world"  
communities \cite{WDSS}.

Taking the distance $d$ between nodes as the independent variable, we find that the connection probability can be written as 
\begin{align}
	p =  \frac{1}{ e^{\beta (\varepsilon - \mu)} + 1},
	\label{eq_pd} 
\end{align}
where
\begin{align}
	\varepsilon = \frac{\kappa d}{2}+ \sum^2_{a=0}\frac{4J\delta_a \Big (1+\displaystyle\tanh \Big(\beta \Big (\frac{\kappa d}{4}-\frac{\mu}{2} \Big )\Big)\Big ) }{\displaystyle\cosh^2 
	\left( \frac{\kappa( d - r_a)}{2} \right)} , 
	\label{heq3}
\end{align}
and $ \sum^2_{a=0} \delta_a=1$ (for technical details see SM). When the coupling constant $J =0$, our model 
simplifies to the model presented in \cite{KDPF1} and describes the homogeneous scale-free 
network with link energy $\varepsilon = \kappa d/2$.

In the framework of our model, the Internet temperature is defined from the following equation 
(for details see SM):
\begin{widetext}
\begin{align}
\bar k =  &\frac{ N\sigma (\gamma -1)}{2\sinh(\beta_c(\gamma -1) \mu)}\Big(e^{\beta_c (\gamma -1)\mu} \Phi\big (- e^{\beta \mu} ,1, \sigma  (\gamma -1)\big )-e^{-\beta_c (\gamma -1)\mu}  \Phi\big ( -e^{-\beta \mu} ,1,  \sigma  (\gamma -1)\big )\Big ),
\label{MSG}
	\end{align}
\end{widetext}
where $\sigma  = \beta_c/\beta$, $\bar k$ is the average node degree of the CN, and 
$\Phi(z,a,b)$ denotes the Lerch transcendent \cite{AEWM}. The chemical potential is given by
\begin{align}
	 \mu = T_c \ln \bigg (\frac{N}{\bar k} \Big (\frac{\gamma -1}{\gamma -2} \Big)\bigg ).
\end{align}
 \begin{center}
 \begin{tabular}{|l|c|c|}
  \hline 
  \rule[-1ex]{0pt}{2.5ex} \textbf{Empirical Internet data}& \textbf{BGP }& \textbf{CAIDA} \\ 
  \hline 
  \rule[-1ex]{0pt}{2.5ex} Number of nodes $(N)$ & 17,446 &  23,752 \\ 
  \hline 
  \rule[-1ex]{0pt}{2.5ex} Number of links $(L)$ & 40,805 & 58,416  \\ 
  \hline
  \rule[-1ex]{0pt}{2.5ex} Average node degree $(\bar k)$ & 4.68& 4.92 \\ 
  \hline 
  \rule[-1ex]{0pt}{2.5ex} Exponent of  degree distribution $(\gamma)$&2.16 & 2.1 \\ 
  \hline \hline
  \rule[-1ex]{0pt}{2.5ex} \textbf{ Model parameters} &\textbf{BGP} &\textbf{ CAIDA} \\ 
  \hline
  \rule[-1ex]{0pt}{2.5ex} Curvature of the hyperbolic space $(K)$&-0.76 & -0.72 \\ 
  \hline
  \rule[-1ex]{0pt}{2.5ex} Coupling constant $(J)$& 0.26 & 2.44 \\ 
  \hline
  \rule[-1ex]{0pt}{2.5ex} Size of the network $(R)$&23.47 & 25.65 \\ 
  \hline
  \rule[-1ex]{0pt}{2.5ex} Temperature of the Internet $(T )$&1.037 & 1.067 \\ 
  \hline 
  \rule[-1ex]{0pt}{2.5ex}  Critical temperature $(T_c )$ &1 & 1 \\ 
  \hline 
  \end{tabular} 
  \end{center}
  {Table 1: Empirical Internet data and model parameters. BGP and CAIDA data are 
  extracted from Refs. \cite{MKF,Boguna2010}.}\\

\begin{figure}[tbh]
	\centering
	\includegraphics[width=0.9\linewidth]{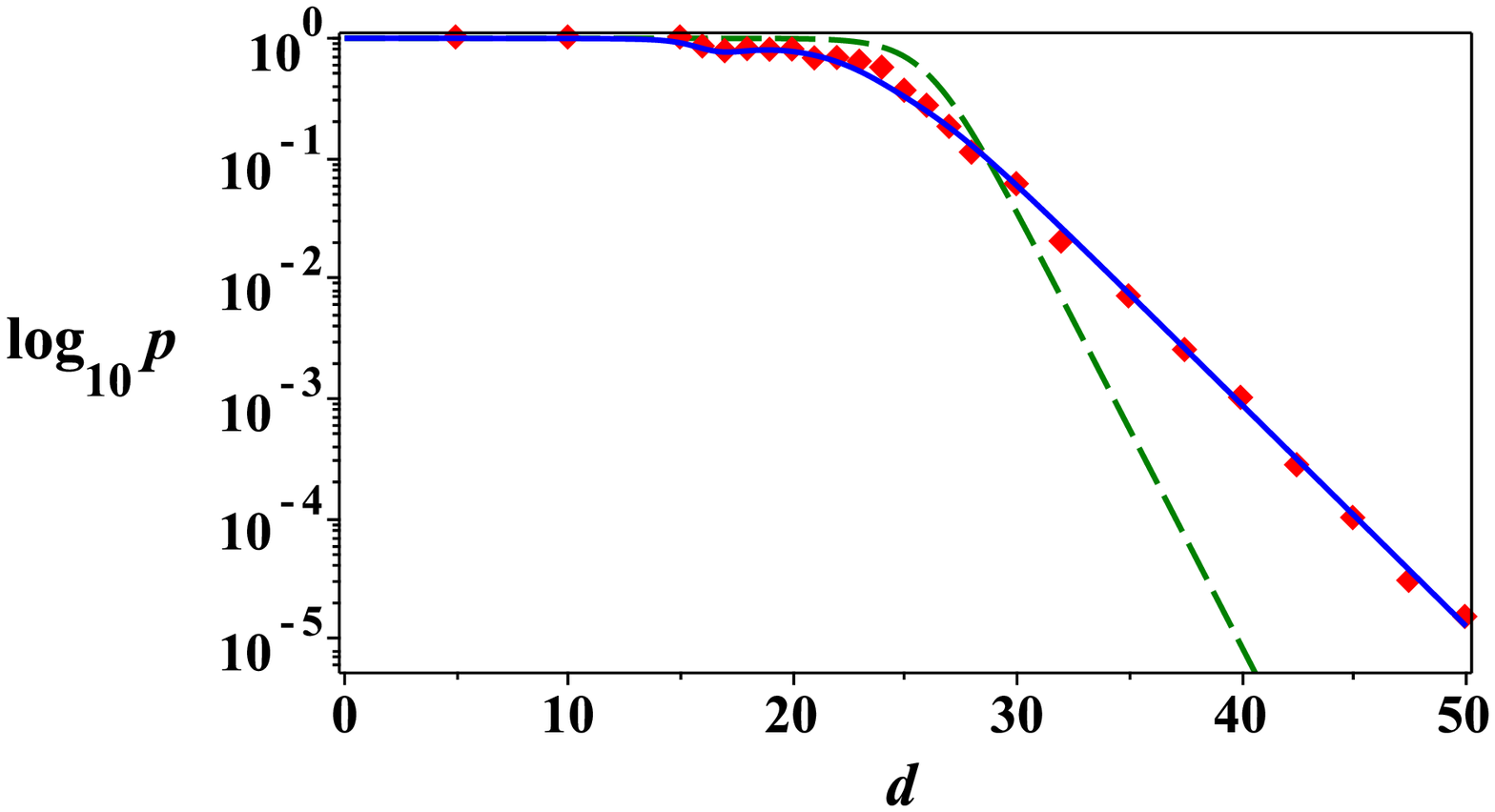} \\
		\includegraphics[width=0.9\linewidth]{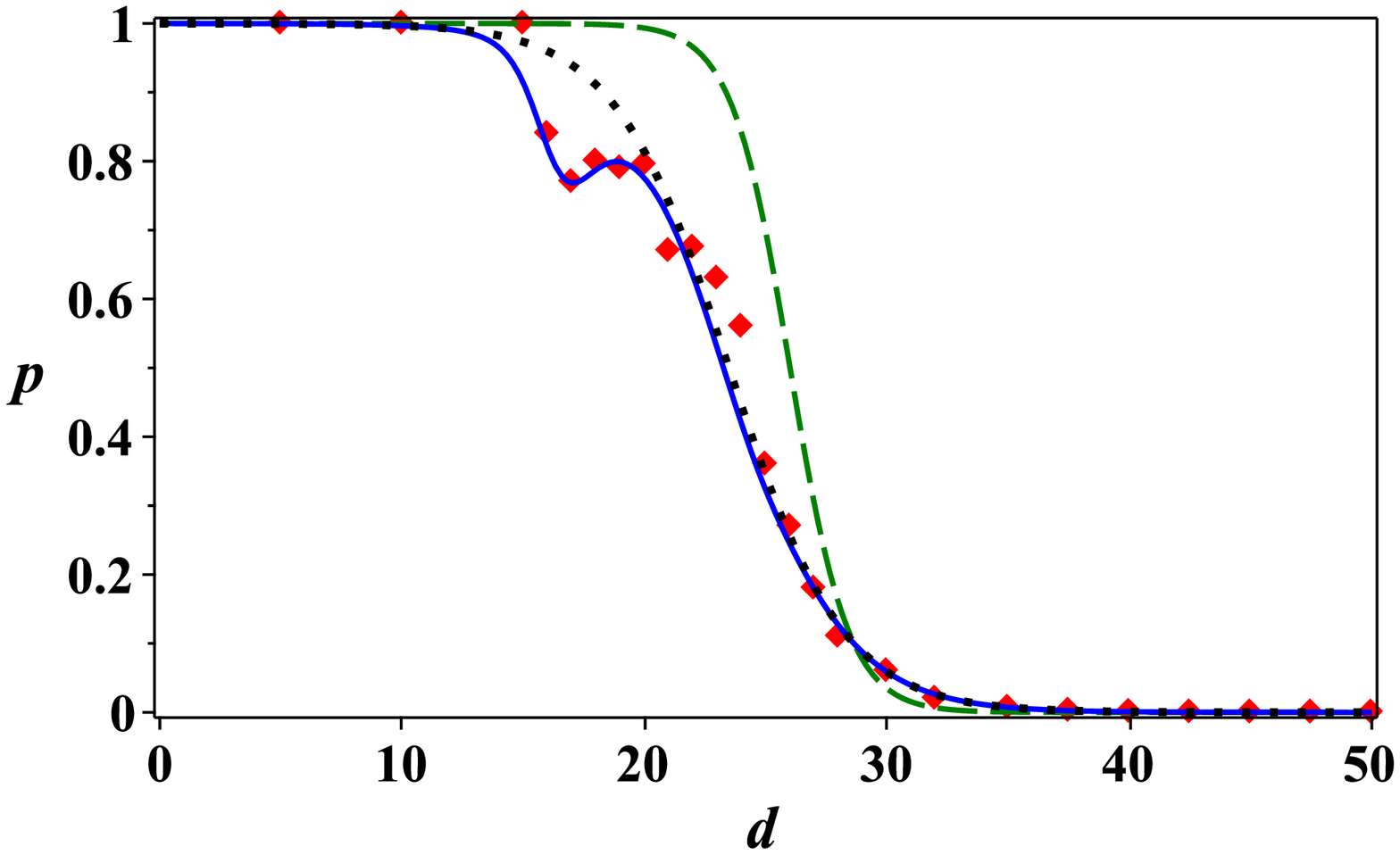} 
	\caption{ (Color online) Top: Connection probability in a logarithmic scale for the 
	BGP data (red diamons) compared to the fitted model from expression (\ref{eq_pd}) 
	(blue). The connection probability for the homogeneous model ($J=0$) is depicted by the black 
	dotted curve. Parameters: $J=0.26$ (blue solid), $J=0$ (black dotted), $T = 1.037$, 
	$K=-0.76$, $R=23.47$, $\delta_0 = 0$, 
	$\delta_1 = 0.87$, $\delta_2 = 0.13$, $r_1=16.36$, $r_2=25$. The results 
	obtained in \cite{KDPF1}  are presented by green dashed curves. The values of $T$, 
	$K$ and $R$ are taken {as} $T=0.6$, $K=-0.83$ and $R=26$. The details of the fit 
	can be better appreciated on a linear scale (bottom).}
	\label{fig3}
\end{figure}

As shown in the SM, at the point $T=T_c $ the  system experiences a phase transition.  Near the critical point, the chemical potential behaves as $\mu \sim -\ln(T- T_c)$, and its 
derivative as $d\mu/dT \sim -1/(T-T_c)$. This agrees with conclusions made in \cite{KDPF2} 
on the behavior of the Internet size near the critical temperature. 

Below the critical temperature the graph is completely disconnected, $\bar k =0$.  In the limit of $T \rightarrow \infty$, we obtain 
\begin{align}
	\mu \rightarrow\beta_c^{-1} \ln  \Big (\frac{2(\gamma -1)}{\gamma -2} \bigg )\quad {\rm and } 
	\quad \frac{\bar k}{N}  \rightarrow \frac{1}{2} .
\end{align}

\begin{figure}[tbh!]
	\centering
	\includegraphics[width=0.9\linewidth]{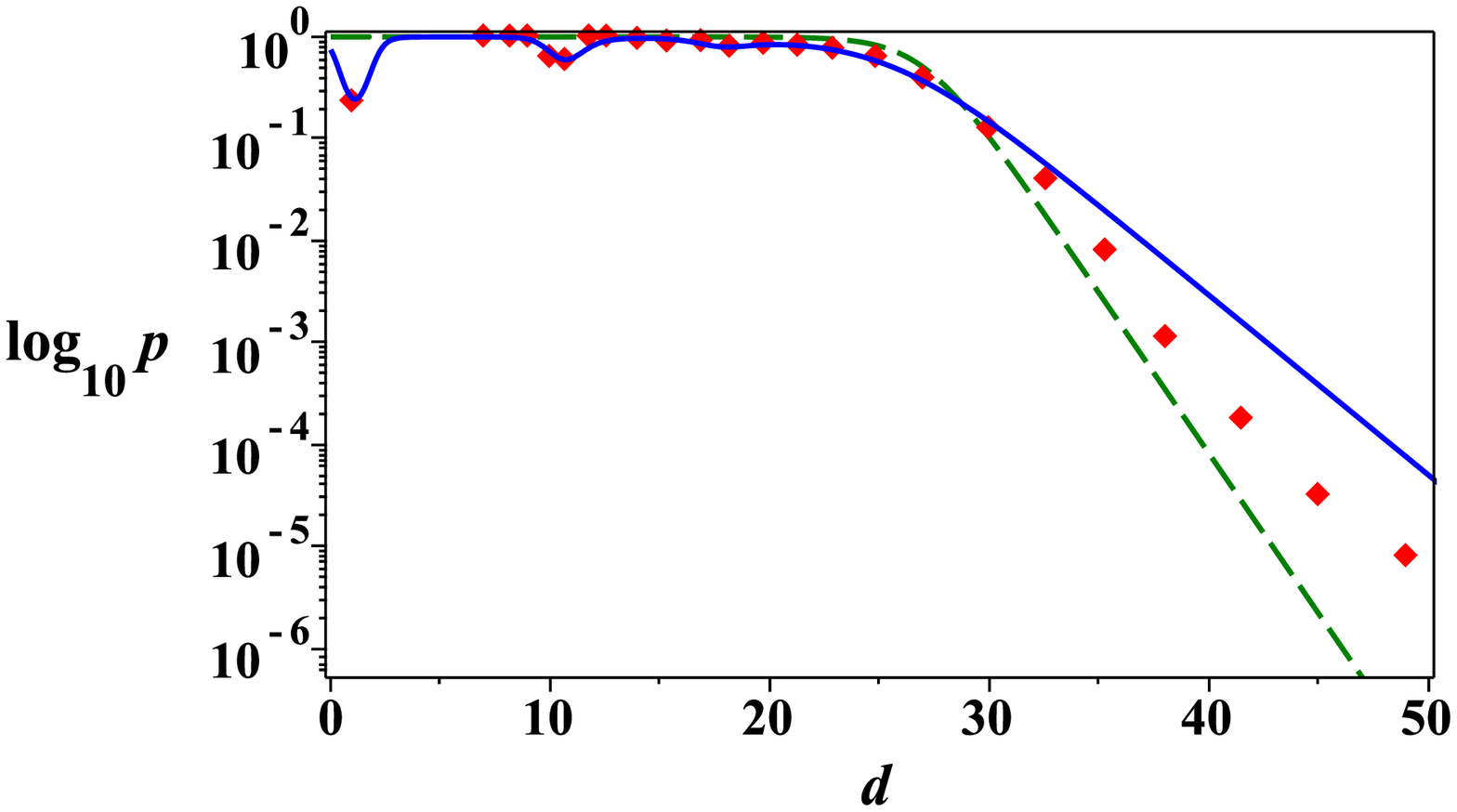}\\
	\includegraphics[width=0.9\linewidth]{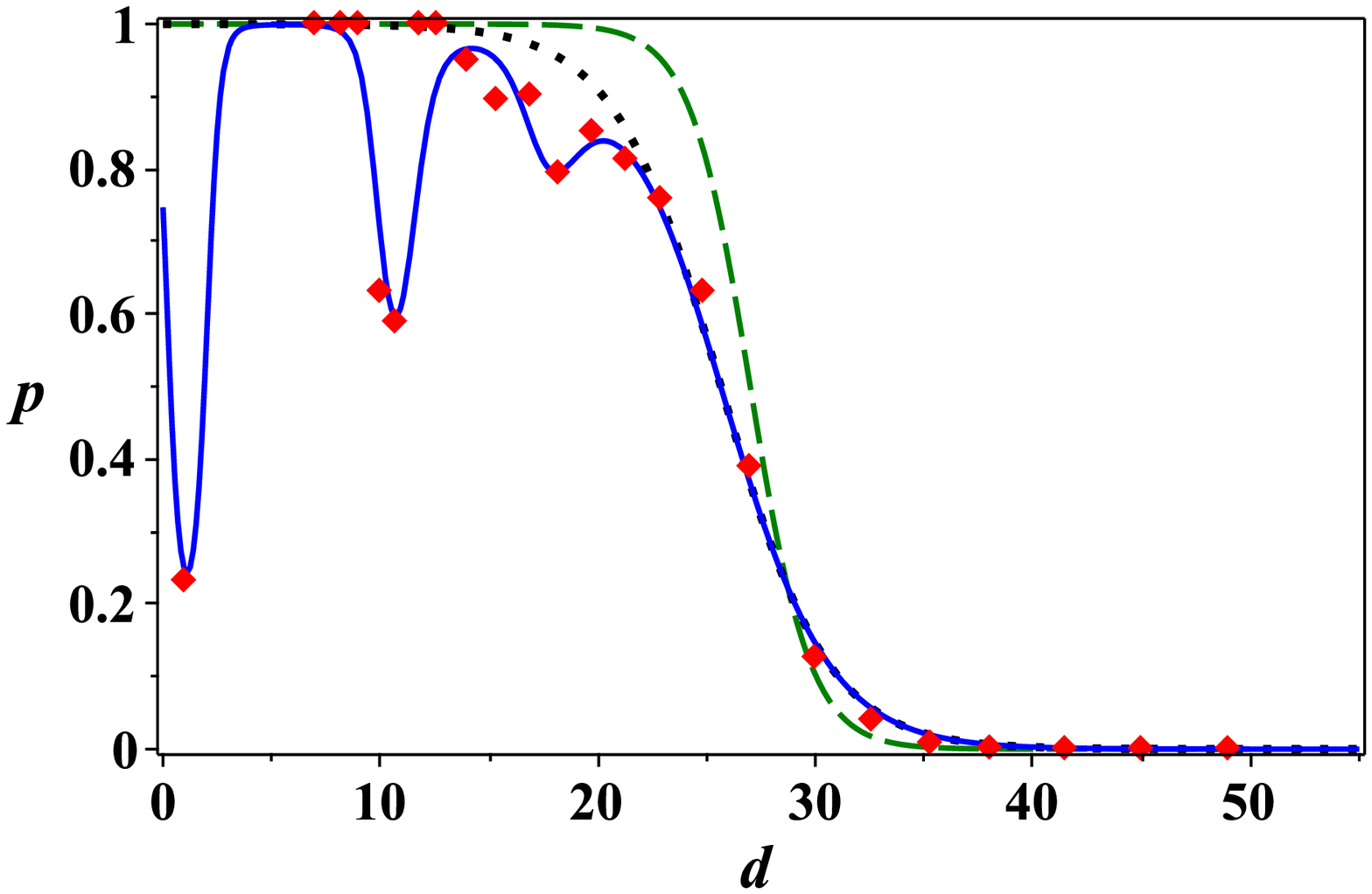} 
	\caption{(Color online) Connection probability for the 
	Internet Archipelago data from \cite{Boguna2010} (red diamonds) compared to the 
	holonomy-inclusive model for expression (\ref{eq_pd}) (blue). {The} connection probability for 
	the homogeneous model ($J=0$) is depicted by the black dotted curve. Parameters: $J=2.44$ 
	(blue solid),  $J=0$ (black dotted), $T = 1.067$, $K=-0.72$, 
	$R=25.65$, $\delta_0 = 0.595$, $\delta_1 = 0.305$, $\delta_2 = 0.1$, 
	$r_0=1$, $r_1=10.5$ and $r_2=17.5$. Numerical results obtained in 
	\cite{Boguna2010} are presented by green-dashed curves, with $R=27$, $T=0.69$ and 
	$K = -1$. }
	\label{fig4}
\end{figure}

We use Border Gateway Protocol (BGP) data and the Internet Archipelago data 
collected by the Cooperative Association for Internet Data Analysis (CAIDA), extracted from 
Refs. \cite{MKF,Boguna2010}, to estimate the size and temperature of the Internet embedded 
in the hyperbolic space, and the curvature of the space as well. Table 1 summarizes the empirical Internet data together with values of key model parameters.

Figs. \ref{fig3} and \ref{fig4} show the results of 
our numerical simulations and compare them with BGP data, CAIDA, and predictions by the model presented in \cite{KDPF1,Boguna2010}. We adapted the empirical connectance data 
for the BGP and CAIDA views of the Internet directly from \cite{KDPF1,Boguna2010,MKF}, and 
plotted them (red diamonds) along with the graph obtained from Eq.\eqref{eq_pd} (blue curves)  
and numerical results presented in \cite{KDPF1,Boguna2010} (green dashed curves). (For details 
see Methods in SM.)
 
Our findings show that a homogeneous model ($J=0$) yields (in general) a good 
agreement with available empirical data (black dotted curve), but can not explain noticeable anomalies in the 
connection probability that break the scale-free behavior of the Internet. As one can see, the predictions of our complete (heterogeneous) model (blue curves) are in excellent agreement with the empirical data (red diamonds). The local minima in the connection probability around $d\approx 16.5$ in the BGP case (Fig.\ref{fig3}), and $d\approx 1, 10.5,17.5 $ in the CAIDA case (Fig.\ref{fig4}), are not artifacts in the empirical data but rather effect of small-world communities described by  holonomy (nonlocal curvature).

{\em Small world properties.} -- The small-world notion refers to the fact that for the 
most real 
networks the typical length, $\ell$,  defined as number of steps required to pass along the 
shortest path connecting two randomly chosen pair of nodes, could be relatively small $\ell 
\propto\ln N$ \cite{WDSS}.

We found that the homogeneous model $(J=0)$ reproduces all the small-world properties with 
remarkable accuracy. The  contribution of the holonomy to the small-world geodesic distance is 
described by the corrections $\propto \ln\ln N$. Thus, one can say that the non-local curvature 
(described by the elementary holonomy) is responsible for formation of the ultra-small world 
effect \cite{CRHS,CFL}. However, the corrections are tiny and this point requires more thorough 
study to support our conjecture (For the technical details see SM.).

{\em Communities formation.} -- The most real networks, including the Internet, exhibit inhomogeneity in the link distribution leading to the natural clustering of the network into groups or communities. Within the same community vertex-vertex connections are dense, but between groups connections are less dense  \cite{GN}. 

We found that for BGP and  CAIDA experimental data, the holonomy exclusive model yields high 
level of the connection between nodes,  $\bar k \approx N/2$ (see SM for details).  However, in 
the complete model we have $\bar k  \ll N$. This means that there are many vertices with low 
degree and a small number with high degree. Our findings show that the holonomy is 
responsible for formation of the community structure of the Internet. \\

{\em Concluding remarks.}-- Inspired by theoretical studies of networked systems that employ 
the methods of statistical physics and geometry, we introduced a general, flexible, and viable 
model for CNs with hidden geometry. Our approach incorporates the effects of nonlocal 
curvature and extends the statistical treatment of CNs. While we have considered CNs with 
hidden hyperbolic geometry,  our model can be applied to 
the study of CNs with hidden geometry of space with arbitrary curvature as well. 

We studied the Internet as a CN embedded in a hyperbolic space to explain features of Internet 
connectance data not only unexplained in previous studies, but completely unmentioned. We 
found an impressive agreement with available empirical data. To our best knowledge, this is the 
first model that explains all features of the Internet connectance data.

We show that non-local curvature is responsible for the formation of communities and 
ultra-small world network effects inside of the Internet. 
However, the corrections are tiny and our conjecture  on the ultra-small world network 
formation requires more thorough study. This point will be addressed in future work.
 
\begin{acknowledgements}
{The authors acknowledge the support by the CONACYT.}
\end{acknowledgements}

\begin{widetext}
\section*{Supplemental Material}

\appendix


\section{Poincar\'e disk model}

The Poincar\' e disk model is related to the two-dimensional hyperbolic model as follows. 
Consider the two-sheeted hyperboloid, $H^2$, defined in Cartesian coordinates $(x,y,z)$ by the 
equation: $x^2 + y^2 -z^2 =-\mathcal R^2$, the curvature of the 
hyperboloid being  $K= -1/{\mathcal R}^2$. We introduce the inner coordinates on the upper 
sheet of $H^2$ as follows:
\begin{align}
	x=& \mathcal R\sinh \theta \cos\varphi, \\
	y= &\mathcal R\sinh \theta \sin\varphi, \\
	z= &\mathcal R\cosh \theta,
\end{align}
where the radial coordinate is $0 \leq  \theta < \infty$ and $0 \leq \varphi < 2\pi$.
Taking a point \hmb{$P$} on the upper sheet of the hyperboloid, we project it to the plane 
$z=0$ by using the conversion formulas
\begin{align}
	\Re \zeta = &\frac{x}{{\mathcal R}+z} = \tanh\Big (\frac{\theta}{2} \Big )\cos \varphi, \\
	\Im \zeta = &\frac{y}{{\mathcal R}+z} = \tanh\Big (\frac{\theta}{2} \Big )\sin \varphi, \quad 
	\zeta \in D.
\end{align}
\begin{figure}
	\centering
	\includegraphics[width=0.3\linewidth]{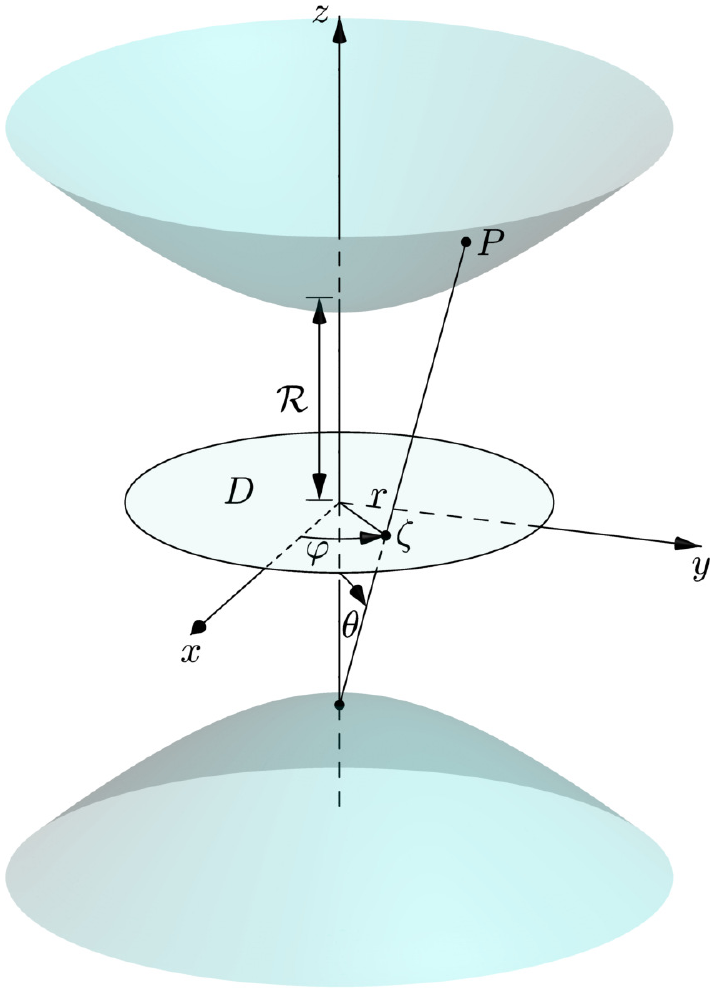}
	\caption{A diagram of how points on the upper sheet of a two-sheeted hyperboloid are 
	mapped to the unit disc.}
	\label{fig:hyperbola}
\end{figure}

The nonassociative binary operation  
\begin{align}
L_\zeta\eta =\frac{\zeta+\eta}{1+\bar\zeta\eta}, \quad L^{-1}_\zeta\eta 
=\frac{\eta-\zeta}{1-\bar\zeta\eta}, \quad\zeta,\eta\in D,
\label{L1}
\end{align}
where the bar denotes complex conjugation, is defined for the neutral element located at the 
origin of coordinates. If the neutral element is chosen at the point $\zeta_0$,  the operation 
\eqref{L1} is modified as follows:
\begin{align}
L^{\zeta_0}_\zeta\eta =\frac{\tilde \zeta+\tilde \eta}{1+\bar{\tilde \zeta}\tilde \eta}, \quad 
\big ( {L^{\zeta_0}_\zeta}\big )^{-1}\eta =\frac{\tilde \eta-\tilde \zeta}{1+\bar{\tilde 
\zeta}\tilde \eta},
\label{L2}
\end{align}
where
\begin{align}
\tilde\zeta =&\tanh\Big( \frac {\theta -\theta_0}{2}\Big) e^{i(\varphi - \varphi_0)}= \frac{\zeta 
\bar\zeta_0(|\zeta| - |\zeta_0|)}{|\zeta_0\zeta|(1-|\zeta_0\zeta|)}, \\
\tilde\eta =&\tanh\Big( \frac {\theta' -\theta_0}{2}\Big) e^{i(\varphi' - \varphi_0)}
= \frac{\eta \bar\eta_0(|\eta| - |\eta_0|)}{|\eta_0\eta|(1-|\eta_0\eta|)}.
\end{align}
The computation of the associator and elementary holonomy yields
\begin{align}
l^{\zeta_0}_{(\zeta,\eta)}\xi=\frac{1+\tilde \zeta\bar{\tilde\eta}}{1+\tilde 
\eta\bar{\tilde\zeta}}\tilde\xi, \quad
h^{\zeta_0}_{(\zeta,\eta)}\xi=\frac{1- \bar{\tilde \zeta}{\tilde\eta}}{1- \bar{\tilde 
\eta}{\tilde\zeta}}\tilde\xi.
\label{hol2}
\end{align}

In general, for any three vertices $i$, $j$, and $k$, the elementary holonomy with respect to $i$ 
can be written as
\begin{align}
h^{i}_{jk}=\frac{1-\bar\zeta_{ij}\zeta_{ik}}{1-\zeta_{ij}\bar \zeta_{ik}},
\label{hol3}
\end{align}
where
\begin{align}
\zeta_{ij} =\tanh\Big( \frac {\theta_{ij}}{2}\Big) e^{i\varphi_{ij}} = \frac{\zeta_j 
\bar\zeta_i(|\zeta_j| - |\zeta_i|)}{|\zeta_i||\zeta_j|(1-|\zeta_i||\zeta_j|)},
\label{Z1}
\end{align}
and we set $\theta_{ij}=\theta_j-\theta_i$, $\varphi_{ij}=\varphi_j-\varphi_i$. 

For each triplet of nodes, $(i,j,k)$, the elementary holonomy, $h^{i}_{jk}$, can  be 
used as a measure of nonlocal curvature around $i$. Indeed,  for  an arbitrary ``vector", 
$\zeta_{ip}$, we obtain
\begin{align}
\zeta'_{ip} =	 h^i_{jk}\zeta_{ip}.
\label{zeta}
\end{align}
The phase gained by  $\zeta_{ip}$ is found to be
\begin{align}
\Delta \varphi_{ip} = \frac{1}{i}\ln h^i_{jk}.
\end{align}

\hmb{If we} assume that $|\zeta_{ij}|,|\zeta_{jk}|,|\zeta_{ik}| \ll 1$, then we obtain
\begin{equation}
h^{i}_{jk}\approx 1- i\frac{\Delta(i,j,k) }{{\mathcal R}^2} \Longrightarrow \Delta \varphi_{ip} 
\approx -\frac{\Delta(i,j,k) }{{\mathcal R}^2}.
\label{SMh1}
\end{equation}
Here $\Delta(i,j,k)=(1/2)\mathcal R^2|\theta_{ij}||\theta_{ik}||\varphi_{jk}| $ is the area of 
the geodesic triangle formed by the triplet of points $(i,j,k)$. \hmb{Employing} Eqs. \eqref{zeta} 
and \eqref{SMh1}, we obtain
\begin{align}
\Delta\zeta_{ip} =	 (h^i_{jk} -1)\zeta_{ip} =-i\frac{\Delta(i,j,k) }{{\mathcal R}^2} \zeta_{ip}.
\label{zeta1}
\end{align}
This is consistent with the formula for the parallel transport of the vector $\mathbf V$ along a 
small contour $\gamma$:
\begin{align}
\Delta V^i = \frac{1}{2}R^i{}_{jlm}V^j \Delta S^{lm},
\end{align}
where $R^i{}_{klm}$ is the curvature tensor and $\Delta S^{lm}$ is the area of the
segment enclosed by $\gamma$. Indeed, for a space of  constant curvature $K$, we have 
$R^i{}_{jkl} = K(\delta^i_l g_{jk} -  \delta^i_k g_{jl})$. For
$K= -1/\mathcal R^2$ we obtain
\begin{align}
\Delta V^i = -\frac{1}{\mathcal R^2}V_j\Delta S^{ij}.
\end{align}

\section{Temperature of complex networks}

The most general statistical description of an undirected network in equilibrium, with a
fixed number of vertices $N$ and a varying number of links is given by the grand 
canonical ensemble \cite{PJNM,CDLM,CDAS}. For a graph model with energy given 
by $E =\sum_{i<j} \varepsilon_{i j} a_{i j}$, the connection probability between nodes $i$ and 
$j$ (i.e., the probability that the link exists), has the usual form of the Fermi-Dirac 
distribution:
\begin{align}
	p_{i j}=\frac{1}{e^{\beta \left(\varepsilon_{i j}-\mu\right) }+1},
\end{align}
where $\varepsilon_{ij}$ is the energy of the link $\langle i,j \rangle$, and $\mu$ is the 
chemical potential. The chemical potential controls the link density and the 
connection probability, while the temperature, $T=\beta^{-1}$,  controls clustering in 
the network. Below we will focus {on} a particular case, related to scale-free networks, {where} 
the link 
energy has a simple form: $\varepsilon_{ij} = \varepsilon_{i}  + \varepsilon_{j} $ and 
$0 \leq \varepsilon_{i,j} \leq \mu$.

Let us assign to each node a 
``hidden variable''  $x_i = e^{\beta_c (\mu - \varepsilon_i)}$. Then one can write the 
connection probability as
\begin{equation}
p_{i j}=\frac{1}{e^{\beta \left(\varepsilon_{i }+ \varepsilon_j-\mu\right) 
}+1}=\frac{(z x_{i} x_{j})^{1/\sigma }}{1+(z x_{i} x_{j})^{1/\sigma }},
\end{equation}
where $\sigma = \beta_c/\beta$ and $z= e^{-\beta \mu}$.  Suppose that $x_i$ is distributed 
as $\rho(x_i)\sim (\gamma-1) 
x_i^{-\gamma}$, where $1\leq x_i \leq x_0$ and $\gamma >1$ \cite{CDAS}. This yields the 
distribution of $\varepsilon_i = \mu - T_c \ln x_i$ according to $\varrho(\varepsilon_i )\sim 
\beta_c(\gamma -1)e^{\beta_c(\gamma- 1)(\varepsilon_i - \mu)}$. 

We denote by $p(\varepsilon)$ the  connection probability between two nodes 
with the fixed energy $\varepsilon = \varepsilon_i + \varepsilon_j$,
\begin{align}
p(\varepsilon)=\frac{1}{L_{\varepsilon}}\sum_{\langle i, j \rangle \in  
\Lambda_\varepsilon}  p_{ij} ,
\end{align}
where $ \Lambda_\varepsilon$ denotes the set of all pairs $\langle i, j \rangle $ 
having the energy $\varepsilon=\varepsilon_{ij}$, and $L_\varepsilon$ is the total 
number of links (edges) belonging to  $ \Lambda_\varepsilon$. Replacing the sum by {an} 
integral, we obtain
\begin{align}
p(\varepsilon)=\frac{\varrho(\varepsilon)}{e^{\beta (\varepsilon-\mu) }+1},
\end{align}
where
\begin{align}
	\varrho(\varepsilon) = \frac{1}{L_{\varepsilon}} \iint 
	\varrho(\varepsilon_i)\varrho(\varepsilon_j) \delta(\varepsilon - \varepsilon_i 
	-\varepsilon_j)d\varepsilon_i d\varepsilon_j = C e^{\beta_c(\gamma- 1)\varepsilon}.
\end{align}
The constant $C$ is defined by the normalization condition: 
$\int\limits_0^{2\mu}\varrho(\varepsilon) d\varepsilon =1$. The computation yields
\begin{align}
	\varrho(\varepsilon) = \frac{\beta_c(\gamma -1)e^{\beta_c (\gamma 
	-1)(\varepsilon - \mu)}}{2\sinh(\beta_c(\gamma -1) \mu)}.
\end{align}

To find the average node degree {for the whole network, $\bar k$,} we use the relationship $L  
= \bar k N/2$, where $L$ is the number of total existing links,
\begin{align}
	L= \frac{N(N-1)}{2} \int_0^{2\mu} \frac{\varrho (\varepsilon) d 
	\varepsilon}{e^{\beta (\varepsilon-\mu) }+1}.
	\end{align}
Assuming $N \gg 1$, we obtain
\begin{align}
\bar k = N\int_0^{2\mu} \frac{\varrho (\varepsilon) d \varepsilon}{e^{\beta 
(\varepsilon-\mu) }+1}.
\label{SMSGa}
	\end{align}
Performing the integration, we find
\begin{align}
\bar k =  &\frac{ N\sigma (\gamma -1)}{2\sinh(\beta_c(\gamma -1) \mu)}\Big(e^{\beta_c 
(\gamma -1)\mu} \Phi\big (- e^{\beta \mu} ,1, \sigma  (\gamma -1)\big )-e^{-\beta_c 
(\gamma -1)\mu}  \Phi\big ( -e^{-\beta \mu} ,1,  \sigma  (\gamma -1)\big )\Big ),
\label{SMSG}
	\end{align}
where $\sigma  = \beta_c/\beta$ and $\Phi(z,a,b)$ is the Lerch transcendent \cite{AEWM}. 
	
Let us consider a particular model with the chemical potential defined as 
$\mu = T_c\ln(\nu N/\bar k )$, where $\nu$ is an unknown, temperature-independent 
parameter. Substituting $\bar k =N\nu e^{-\beta_c \mu}$ in Eq.\eqref{SMSG} and assuming 
$\beta\mu \gg 1$, we obtain  
\begin{align}
\nu  \approx  \frac{\gamma -1}{\gamma -2}e^{-(\beta 
 - \beta_c)\mu} -e^{-2\beta (\gamma -1)\mu + \beta_c\mu} .
	\label{SGa}
\end{align}
{We further} assume that $\mu(T) \rightarrow \infty$ and $(\beta 
 - \beta_c)\mu \rightarrow 0$ while $T\rightarrow T^{+}_c $. Then from Eq. \eqref{SGa} it 
 follows {that} $\nu =(\gamma -1)/(\gamma -2)$. 
 
At the point $T=T_c$ the system experiences a phase transition. Below the 
critical temperature, $T< T_c$, the graph is completely disconnected, $\bar k =0$. Near the 
critical point, for $T \gtrsim T_c$, the chemical potential behaves as $\mu \sim -\ln(T- T_c)$ 
and $d\mu/dT \sim -1/(T-T_c)$. In the limit of $T \rightarrow \infty$, we obtain $\bar k   
\rightarrow {N}/{2}$ and 
\begin{align}
 \mu \rightarrow T_c\ln  \Big (\frac{2(\gamma -1)}{\gamma -2} \bigg ).
\end{align}
 
In summary, the average node degree is given by $\bar k =N\nu e^{-\beta_c \mu}$, where
\begin{align}
 	\nu =\frac{\gamma -1}{\gamma -2}.
 	\label{Nu}
\end{align}  

{\em Comment.} Since the temperature, $T$, is {an} undetermined parameter, one can take 
the  value of the critical temperature to be $T_c =1$ without loss of generality. If we 
have empirical information about the energy of the nodes and chemical potential, then we can 
define the temperature for a given network employing  Eq.\eqref{SMSGa} (or Eq. \eqref{SMSG}).

\section{ Approximation of the connection probability }

In what {follows} we consider a network with a large number of nodes, $N \gg 1$. First, we 
would like to calculate the connection probability between {any} two nodes $i$ and $j$ taking
elementary holonomy into account. We can already state the form of $p^e_{ij}$ as 
\begin{align}
p^e_{ij} = \frac{1}{1 + e^{-2\beta h^{e}_{ij} }},
\label{eq:pijFerminoic}
\end{align}
where $\beta$ is an inverse ``temperature''. The effective field $h^{e}_{ij}$ satisfies 
equation (24) of the main paper, which can be rewritten as follows:
\begin{align}
h^e_{ij}= h^0_{ij}  -\frac{2J}{ N-1}\sum_{ k} h^{i}_{(jk)}(1+\tanh (\beta h^e_{ik}) ) , 
\label{eq:hijFormal}
\end{align}
where $ h^0_{ij} = B \alpha_{ij} $ and $h^{i}_{(jk)}=( {1}/{2} )\big(h^{i}_{jk}+h^{i}_{kj} \big )$. 
Computation of 
$h^{i}_{(jk)}$ yields
	\begin{align}
h^{i}_{(jk)} =1- \frac{2 |\zeta_{ji}|^2 |\zeta_{ki}|^2\sin^2\varphi_{jk} }{1 -2 |\zeta_{ji}| 
|\zeta_{ki}| \cos\varphi_{jk} +|\zeta_{ji}|^2 |\zeta_{ki}|^2},
\end{align}
where $\varphi_{jk} = \varphi_{k} - \varphi_{j}$. Using the identity
	\begin{align}
		\frac{1}{2} \big( 1 + \cosh x \cosh y\big) = \cosh^2 \frac{x}{2} \cosh^2 \frac{y}{2} +   
		\sinh^2 \frac{x}{2} \sinh^2 \frac{y}{2} ,
		\label{SA1}
	\end{align}
one can show that
\begin{align}
h^{i}_{(jk)} =1- \frac{4 \sinh^2\frac{\theta_{ij}}{2} 
\sinh^2\frac{\theta_{ik}}{2}\sin^2\varphi_{jk} 
}{1 + \cosh \theta_{ij} \cosh \theta_{ik} - \sinh \theta_{ij} \sinh \theta_{ik} \cos\varphi_{jk} }.
\end{align}

Our important assumption, essential for further estimations, is that nodes are 
densely and uniformly distributed in their angular coordinates. This implies that the effective 
field depends only on the ``radial'' coordinates: $h^e_{ij} = h^e(\theta_i,\theta_j)$. Then we can 
replace the sum 
over $\varphi_k$ in \eqref{SA1} by an integral in the angular coordinate to {get, after} some 
algebra,
\begin{align}
h^e_{ij}= h^0_{ij} - \frac{2J}{N -1}\sum_{ \theta_k} N_k \Big(1- \tanh^2\Big( \frac {\theta_{ij} 
}{2}\Big)  \tanh^2\Big( \frac {\theta_{ik} }{2}\Big)  \Big ) (1+\tanh (\beta h^e_{ik}) ),
\label{eq:AijInterm}
\end{align}
where $\theta_{ij} =\theta_j -\theta_i$, and  $N_k$ is the number of nodes located at the 
distance $\theta_k $ from the origin of coordinates.  Further, it is convenient to rewrite 
\eqref{eq:AijInterm} as
\begin{align}
h^e_{ij}= h^0_{ij} - \frac {1}{\cosh^2\frac {\theta_{ij}}{2} }\frac{2J}{N -1}\sum_{ \theta_k} N_k 
(1+\tanh (\beta h^e_{ik}) ) +\frac{2J}{N -1} \sum_{ \theta_k} N_k \Bigg(1- \frac{\tanh^2\Big( 
\frac {\theta_{ij} }{2}\Big)}{  \cosh^2\Big( \frac {\theta_{ik} }{2}\Big)  }\Bigg ) (1+\tanh (\beta 
h^e_{ik}) ).
\label{SA2}
\end{align}

To proceed further we use the \textit{anzatz} $h^e_{ij}= 
h_{ij}+\Delta h_{ij}$, where
\begin{align}
	h_{ij}=  h^0_{ij}- \frac {2J}{\cosh^2\frac {\theta_{ij}}{2} }(1+\tanh \beta h^0_{ij}),
	\label{SMeqH}
\end{align}
and $\Delta h_{ij}$ is a perturbation of the effective field. Employing Eq.\eqref{eq:hijFormal}, in 
the linear approximation we obtain
\begin{align}
\Delta h_{ij} \approx\frac {2J}{\cosh^2\frac {\theta_{ij}}{2} }(1+\tanh \beta h^0_{ij}) 
-\frac{2J}{N-1}\sum_{ k}  h^{i}_{(jk)} (1+\tanh \beta h_{ik}).
\label{SMeq3a}
\end{align}
Writing the connection probability as $p^e_{ij}  = p_{ij}  + \Delta p_{ij} $, where
\begin{align}
p_{ij} = \frac{1}{2 }\big (1 + \tanh(\beta h_{ij} )\big) = \frac{1}{1 + e^{-2\beta h_{ij} }},
\label{SMeq1}
\end{align}
we find that
\begin{align}
\Delta p_{ij} = \frac{\beta  \Delta h_{ij}}{2\cosh^2(\beta h_{ij} )}.
\label{SMeq2}
\end{align}

 When $|\Delta p_{ij} |/ p_{ij} \ll 1$ one can neglect the perturbation 
$\Delta p_{ij}$ and use Eq.\eqref{SMeq1} instead of the exact expression given by 
Eq.\eqref{eq:pijFerminoic}. The computation of {this} quotient yields
\begin{align}
Z_{ij} =\left|\frac{\Delta p_{ij}}{p_{ij}}\right|= \left|\frac{\beta \Delta h_{ij}}{(1+\tanh (\beta 
h_{ij})) \cosh^2 (\beta h_{ij})}\right|
=\left|\beta \Delta h_{ij} (1- \tanh (\beta h_{ij}))\right|.
\end{align}
When $Z_{ij}\ll 1$ over a large range of variables $\theta_{ij}$ and $h^0_{ij}$, we can neglect the 
contributions from $\Delta h_{ij}$ and use Eq. \eqref{SMeqH} for calculation of the effective 
field $h_{ij}$ in $p_{ij}$.

\section{The Internet embedded in  hyperbolic space}

To adapt our model to empirical Internet data, such as BGP and CAIDA, we 
consider $d$ as the independent variable in our calculations, thus allowing direct 
comparison to the results in \cite{KDPF1} (the authors there use $x$ for distance 
rather than $d$). We average the connection probability over all pairs of {nodes with 
fixed distance $d$ between them}, writing
\begin{align}
\bar p = \frac{1}{L_d}\sum_{\langle i, j \rangle \in  \Lambda_d}  p^e_{ij} =\frac{1}{2 
L_d}\sum_{\langle i, j \rangle \in  \Lambda_d} \big( 1+\tanh (\beta{h}^e_{ij})\big),
	\label{eq_p}
\end{align}
where $\Lambda_d$ denotes the set of all {$\langle i, j \rangle $ pairs with distance $d$ 
between them}, and $L_d$ is the total number of links (edges) in $\Lambda_d$. 

We write the effective field as $h^e_{ij}  = h  + \Delta h_{ij} $, where 
\begin{align}
h = \frac{1}{L_d}\sum_{\langle i, j \rangle \in  \Lambda_d}  h_{ij},
\label{SMh}
\end{align}
and $h_{ij}$ is given by \eqref{SMeqH}. Substituting  $h^e_{ij}$ in Eq. \eqref{eq_p}, we find that 
$\bar p = p + \Delta p$, where
\begin{align}
p = \frac{1}{2}\big( 1+\tanh (\beta h)\big) 
\label{SMp1}
\end{align}
and 
\begin{align}
\Delta p = \frac{\beta   \Delta h}{2\cosh^2(\beta h )}, \quad \Delta h = 
\frac{1}{L_d}\sum_{\langle i, j \rangle \in  \Lambda_d}  \Delta h_{ij}.
\label{SMeq2g}
\end{align}
The {approximate} formula for the connection probability \eqref{SMp1} will be valid when 
$|\Delta p |/ p \ll 1$. This validity is discussed below in Sec. A. Further, we assume that 
$h^0_{ij}$ is a homogeneous field, $h^0_{ij}= h_0$ and we set $h_0 = (\kappa/4) (R-d)$. Using 
Eq. \eqref{SMeqH}, we find
\begin{align} 
 h_{ij}= h_0- 2J\big (1+\tanh (\beta  h_0 )\big )\frac {1}{\cosh^2\frac {\theta_{ij}}{2} }.
 \label{SMeq5}
\end{align}
Now substituting $h_{ij}$ {into} Eq. \eqref{SMh}, we obtain
\begin{align} 
 h =  h_0- 2J\big (1+\tanh (\beta  h_0 )\big )\Bigg \langle\frac {1}{\cosh^2\frac {\Delta 
 \theta}{2} }\Bigg \rangle,
 \label{SMeq5g}
\end{align}
where
\begin{align}
 \Bigg \langle\frac {1}{\cosh^2\frac {\Delta\theta}{2} }\Bigg \rangle 
 =\frac{1}{L_d}\sum_{\langle i, j \rangle \in  \Lambda_d} \frac {1}{\cosh^2\frac {\theta_{ij}}{2} } 
 .
 \label{SMeq8}
\end{align}

What we now need is a practical way of evaluating this sum. We begin with the hyperbolic 
distance $d$ between a pair of nodes $\langle i, j\rangle $, as defined by Eq. (13) from the 
main 
paper:
\begin{align} 
	\cosh\theta = \cosh \theta_i \cosh \theta_j - \sinh \theta_i \sinh \theta_j \cos 
	\varphi_{ij} ,
	\label{SMeq4}
\end{align}
where $\theta =\kappa d$. To find the dependence of the effective field $h$ on this distance 
we use Eq.\eqref{SMeq4}, treating it as a constraint, $f(\theta_i,\theta_j,\theta,\varphi_{ij}) 
=0$, which allows us to eliminate either one of the variables $\theta_i$ or $\theta_j$ from 
consideration.  
\begin{figure}[tbh]
\begin{center}
\includegraphics[width=0.35\linewidth]{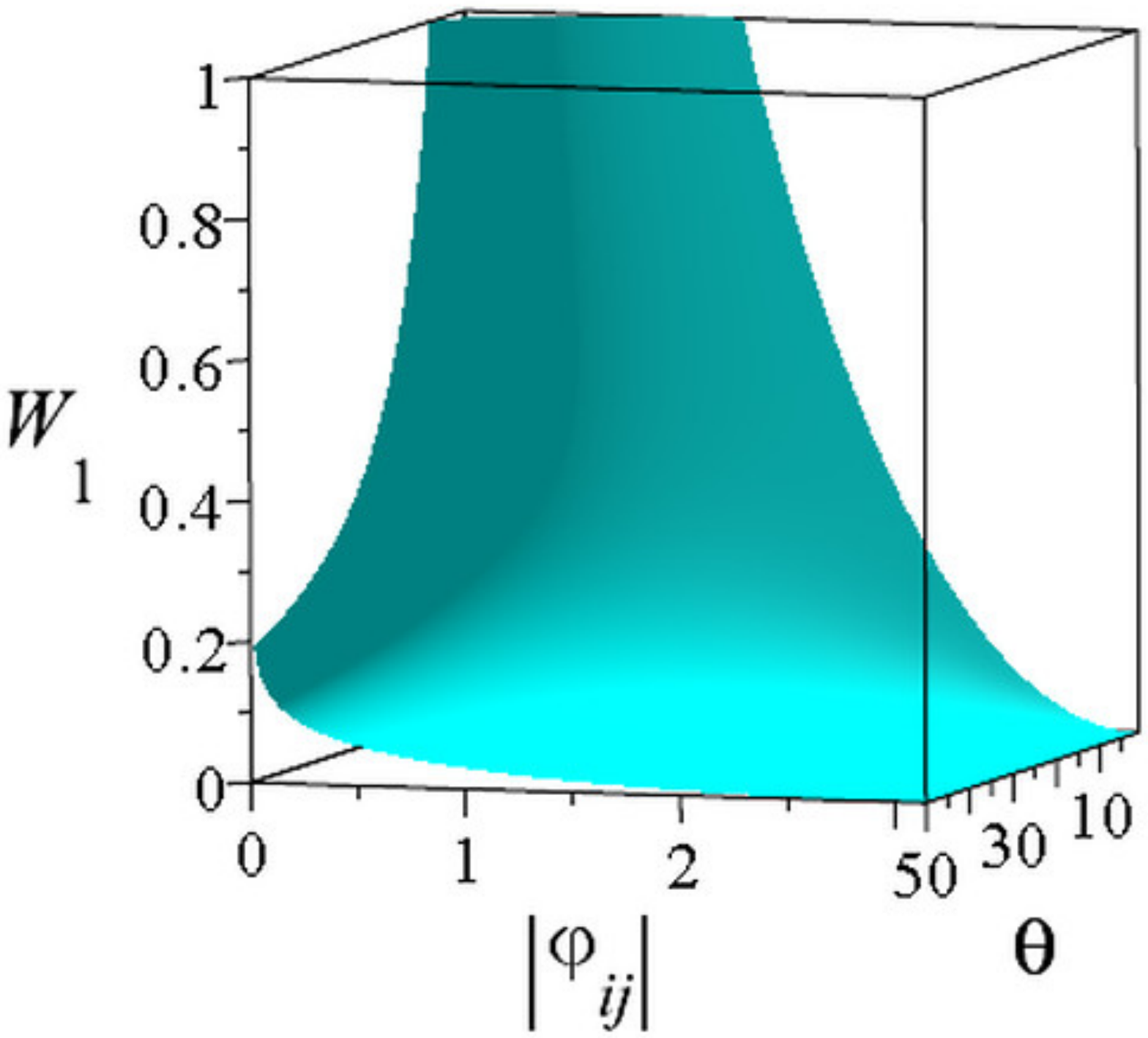} 
\includegraphics[width=0.425\linewidth]{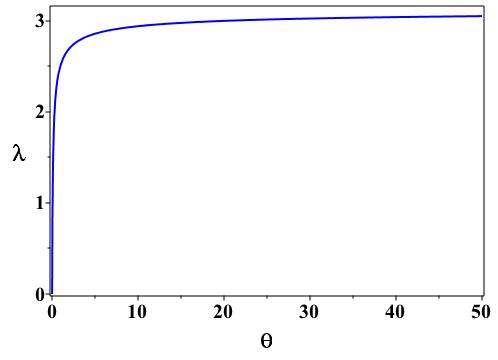} 
	\end{center}
	\caption{Left panel: graph of the function  $W_1 (\varphi_{ij}, \theta)$, $d \approx r_i + r_j$. 
	{$W_1$ is small for large radii and tends to zero as the angle $\varphi_{ij}$ between nodes 
	increases.} Right panel: graph of the function $\lambda(\theta)$, $\varepsilon = 0.1$, $ \pi 
	-\lambda \leq|\varphi_{ij}| \leq \pi +\lambda$. {The angle between nodes where the 
	approximation is valid increases with distance $\theta$, up to nodes that are completely 
	opposite.} }
	\label{SMfig3a}
\end{figure}

One can recast (\ref{SMeq4}) in an equivalent form, writing
\begin{align} 
\cosh\theta = \cosh (\theta_i+ \theta_j) -2 \sinh \theta_i \sinh \theta_j \cos^2\frac{ 
\varphi_{ij}}{2},
\label{SMeq4c}
\end{align}
where $\theta = \kappa d$. When $\theta_i, \theta_j \gg 1$, we obtain
\begin{align} 
d\approx  r_i + r_j + \frac{2}{\kappa}
\ln\Big(\Big |\sin\frac{ \varphi_{ij}}{2}\Big |\Big) .
\label{SMeq4a}
\end{align}
Thus, one can approximate the distance as $d\approx  r_i + r_j$ {(i.e., the sum of the radial 
coordinates) if, for some threshold value $\varepsilon$,} $|W_1(\theta, \varphi_{ij})|\leq 
\varepsilon \ll 1$, where
\begin{align} 
W_1= \frac{2}{\theta}
\ln\Big(\Big |\sin \frac{\varphi_{ij}}{2}\Big |\Big) .
\label{SMW}
\end{align}
The graph of the function $|W_1(\theta, \varphi_{ij})|$ is depicted in Fig. \ref{SMfig3a} (left 
panel). For a given $\varepsilon$, the approximation $d\approx  r_i + r_j$ is valid for $ \pi 
-\lambda \leq|\varphi_{ij}| \leq \pi +\lambda$, where $\lambda=  2\arcsin 
e^{-\varepsilon/2\theta}$ (see Fig. \ref{SMfig3a}, right panel). {These figures show how the 
approximation works best at large distances between nodes, and is also valid for nearly any 
angle between them.}

Now, we rewrite Eq. \eqref{SMeq4} as
\begin{align}
\cosh\theta = \cosh \theta_{ij} +2 \sinh \theta_i \sinh \theta_j \sin^2\frac{ \varphi_{ij}}{2}.
\label{SMeq4b}
\end{align}
First, consider the case when $\theta_i \ll 1$ and  $\theta_j \gg 1$. Then the distance 
between nodes can be approximated as $d \approx r_j$,  if $|W_2(\theta_i, \theta_j)|  
\ll 1$, where
\begin{align} 
W_2= \frac{2\sinh \theta_i\sinh \theta_j}{\cosh \theta_{ij}}\sin^2\frac{ \varphi_{ij}}{2}.
\label{SMWa}
\end{align}
In the opposite case, when $\theta_i \gg 1$ and  $\theta_j \ll 1$, we obtain $d 
\approx r_i$. For a given $\varepsilon$, this approximation is valid for  $|\varphi_{ij}| < \nu$, 
where
\begin{align} 
\nu=  2\arcsin\sqrt{  \frac {\varepsilon\cosh \theta}{2\sinh {(\theta +\theta_j)}\sinh (\theta_j)} 
}
\label{SMW1}
\end{align}
(see Fig. \ref{SMfig3}, right panel). Next, for $\theta_i,\theta_j \lesssim 1$, we approximate the 
distance 
as 
$\theta = |\theta_i -\theta_j| + \varepsilon_{ij}$, where $\varepsilon_{ij}$ is given by
\begin{align}
\varepsilon_{ij} =  \cosh^{-1}\Big(\cosh \theta_{ij} +2 \sinh \theta_i \sinh \theta_j 
\sin^2\frac{ \varphi_{ij}}{2}\Big) -  |\theta_i -\theta_j| .
\end{align}
In Fig. \ref{SMfig2} the function $\varepsilon_{ij}(\theta_i ,\theta_j)$ is depicted for 
$|\varphi_{ij}|= \pi$ (cyan surface) and $|\varphi_{ij}|= \pi/8$ (red surface).

\begin{figure}[tbh]
\includegraphics[width=0.35\linewidth]{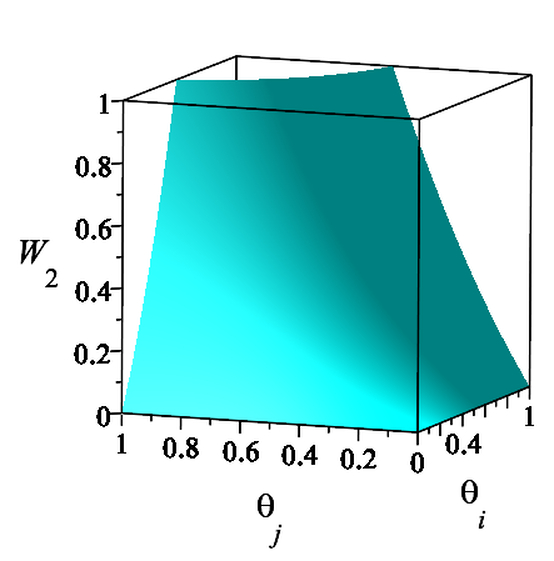} 
\includegraphics[width=0.46\linewidth]{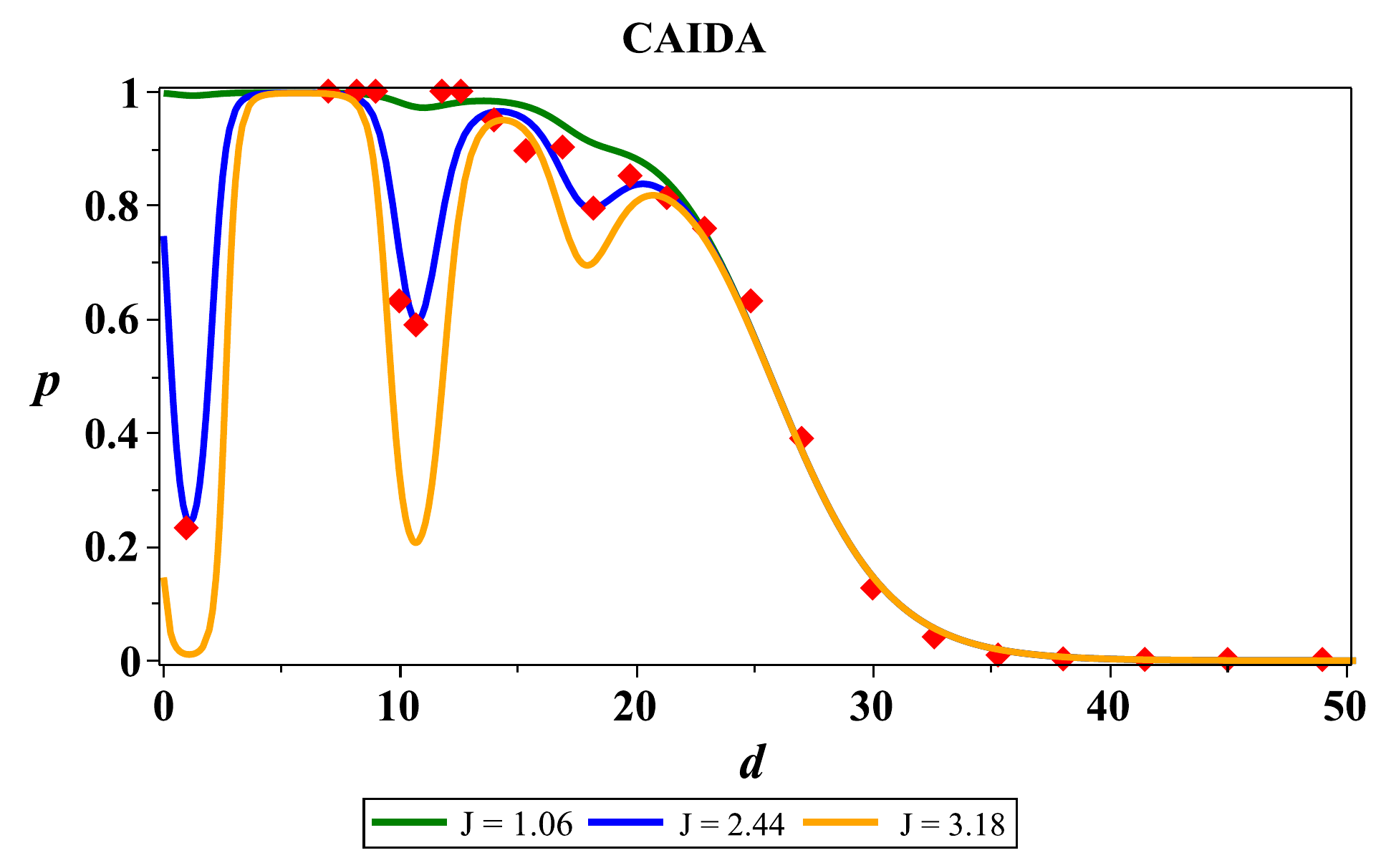} 
	\caption{Left panel: graph of the function $W_2 (\theta_i, \theta_j)$, $d\approx |r_j - r_i|$, 
	$|\varphi_{ij}|< \pi$. Right panel: graph of the function $\nu(d)$, $d\approx |r_j - r_i|$, 
	$|\varphi_{ij}| \leq\nu$. From top to bottom:  $\theta_j = 0.2,1,5$ ($\varepsilon = 0.1$). }
	\label{SMfig3}
\end{figure}

\begin{figure}[tbh]
\includegraphics[width=0.5\linewidth]{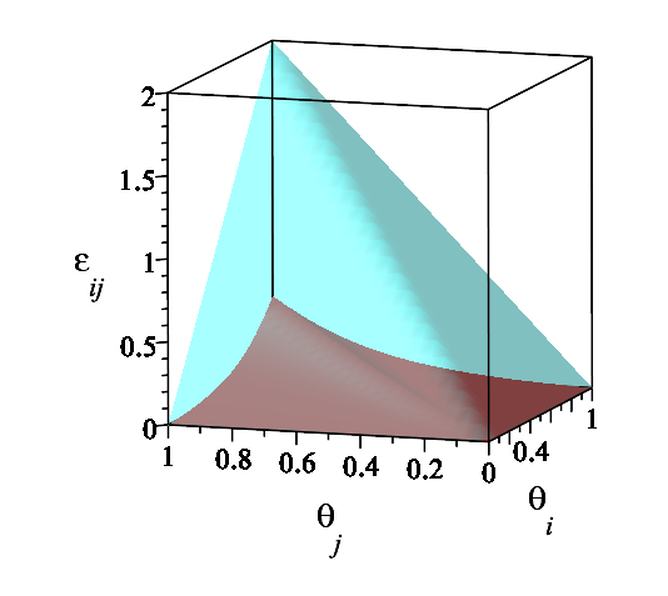} 
	\caption{Graph of the function $\varepsilon_{ij} (\theta_i, \theta_j)$, 
	$\theta\approx 	|\theta_j - \theta_i| + \varepsilon_{ij}$. Upper outer surface (cyan): 
	$|\varphi_{ij}|= \pi$. Lower inner surface (red): $|\varphi_{ij}|= \pi/8$ }
	\label{SMfig2}
\end{figure}

Finally, according to the above analysis, the sum \eqref{SMeq8} can be approximated as the 
sum of three general cases which we can calculate approximately:
\begin{align}
\Bigg \langle\frac {1}{\cosh^2\frac {\Delta\theta}{2} }\Bigg \rangle \approx 
\frac{1}{L_d} \sum^{N_0}_{\langle i,j \rangle } \frac {1}{\cosh^2\frac {\theta- \varepsilon_{ij}}{2} 
} +  \frac{1}{L_d} \sum^{N_1}_{\langle i,j \rangle }\frac {1}{\cosh^2\frac {\theta-\theta_{j}}{2} 
}   +  \frac{1}{L_d} \sum^{N_2}_{\langle i,j \rangle }\frac {2}{\cosh^2\frac {\theta - 
2\theta_{j}}{2} }  ,
\end{align}
where $N_0$ is the number of pairs $\langle i,j \rangle $ in  the 
interval $\theta_i, \theta_j \lesssim 1$, $N_{1,2}$ is the number of pairs inside the interval 
$\theta_i \ll 1, \theta_j \gg 1$ (or $\theta_i \gg 1, \theta_j \ll 1$ ) and 
$\theta_i, \theta_j \gg 1$, respectively. Applying the mean value theorem, we obtain 
\begin{align}
\Bigg \langle\frac {1}{\cosh^2\frac {\Delta\theta}{2} }\Bigg \rangle \approx 
\sum^2_{a=0}\frac{\delta_a}{\cosh^2 \left( \frac{\theta - \theta_a}{2} \right)},
\label{SMeq6}
\end{align}
where $\delta_a = N_a/L$ ($a=0,1,2$), and as one can see 
$\delta_0+\delta_1+\delta_2=1$. The unknown parameters, $\theta_a$, determine 
the reference points in the application of the mean value theorem. 
Taking into account that $\theta = \kappa d$ and setting $\theta_a = \kappa 
r_a$, we obtain
 \begin{align}
 \Bigg \langle\frac {1}{\cosh^2\frac {\Delta\theta}{2} }\Bigg \rangle \approx 
 \sum^2_{a=0}\frac{\delta_a}{\cosh^2 \left( \frac{\kappa( d - r_a)}{2} \right)} .
 \end{align}
Thus, we find that the connection probability is given by 
\begin{align}
p = \frac{1}{2}\big( 1+\tanh (\beta h)\big) 
\label{SMp}
\end{align}
where
\begin{align}
	h =h_0 - 2J(1+\tanh \big(\beta h_0)\big )  \sum^2_{a=0}\frac{\delta_a}{\cosh^2 
	\left( \frac{\kappa( d - r_a)}{2} \right)} ,
	\label{heq3}
\end{align}
and $h_0 = (\kappa/4) (R-d)$. The fitting parameters $\delta_a$ and $r_a$ should be 
fixed by comparing with available empirical data.

The connection probability  \eqref{SMp} can be rewritten in the form of the 
Fermi-Dirac distribution:
\begin{align}
p =  \frac{1}{ e^{\beta (\varepsilon - \mu)} + 1},
\label{eq_pd} 
\end{align}
 where $\mu = \kappa R/2$ is the chemical potential, 
and
\begin{align}
		\varepsilon = \frac{\kappa d}{2}+ \sum^2_{a=0}\frac{4J\delta_a \Big 
		(1+\displaystyle\tanh 
		\Big(\beta \Big (\frac{\kappa d}{4}-\frac{\mu}{2} \Big )\Big)\Big ) }{\displaystyle\cosh^2 
	\left( \frac{\kappa( d - r_a)}{2} \right)} .
	\label{heq3}
\end{align}

\subsection{Validity of approximation}

The approximation \eqref{SMp} is valid when $|\Delta p |/ p \ll 1$. The computation yields
\begin{align}
Y =\frac{|\Delta p|}{p}= \beta\Delta h (1- \tanh (\beta h)) ,
\end{align}
where 
\begin{align} 
 h =  h_0- 2J\big (1+\tanh (\beta  h_0 )\big )\Bigg \langle\frac {1}{\cosh^2\frac {\Delta 
 \theta}{2} }\Bigg \rangle,
 \label{SMeq5g}
 \end{align}
 and
\begin{align} 
\Delta h = \frac{1}{L_d} \sum_{\langle i,j \rangle } \Delta h_{ij}
= \frac{1}{L_d} \sum_{\langle i,j \rangle }\Big( \frac {2J}{\cosh^2\Big( \frac 
{\theta_{ij}}{2}\Big ) }(1+\tanh \beta h_0) -\frac{2J}{N-1}\sum_{ k}  h^{i}_{(jk)} (1+\tanh 
\beta h_{ik})  \Big).
\label{ASMeq1}
\end{align}
Here $ \Delta h_{ij}$ is taken from Eq. \eqref{SMeq3a}. Next, using the relation
\begin{align}
\frac{2J}{N-1}\sum_{ k}  h^{i}_{(jk)} = \frac{2J}{N -1}\sum_{ \theta_k} N_k \Big(1- \tanh^2\Big( 
\frac {\theta_{ij} }{2}\Big)  
\tanh^2\Big( \frac {\theta_{ik} }{2}\Big)  \Big ),
\label{eq:AijInterm3}
\end{align}
one can recast Eq. \eqref{ASMeq1} as
\begin{align} 
\Delta h =& \frac{1}{L} \sum_{\langle i,j \rangle } \Delta h_{ij}
= \frac{2J}{L_d} \sum_{\langle i,j \rangle } \bigg (\frac {1}{\cosh^2\Big( \frac 
{\theta_{ij}}{2}\Big ) }\Big(1+\tanh (\beta h_0) - F(\theta_i) \Big ) -  Q(\theta_i)\bigg) ,
\label{ASMeq2}
\end{align}
where 
\begin{align} 
F(\theta_i) =& {\frac{1}{N -1}\sum_{ \theta_k} N_k  
\tanh^2\Big( \frac {\theta_{ik} }{2}\Big)\big (1+\tanh 
(\beta h_{ik}))},\nonumber \\ 
Q(\theta_i) =&\frac{1}{N-1}\sum_{ \theta_k}  \frac 
{N_k}{\cosh^2\Big( \frac 
{\theta_{ik}}{2}\Big ) } \big (1+\tanh 
(\beta h_{ik})\big ).
\label{ASMeq2}
\end{align}
Replacing the sum by an integral, we we have
\begin{align} 
F(\theta_i) =& \int^{\theta_0}_0 d\theta \rho(\theta)
\tanh^2\Big( \frac {\theta_{i}-\theta }{2}\Big)(1+\tanh (\beta h(d,\theta_i- 
\theta))\big ) , \\
Q(\theta_i) =&\int^{\theta_0}_0  \frac {d\theta \rho(\theta)\big 
(1+\tanh (\beta h(d,\theta_i- \theta))\big )}{\cosh^2\Big( \frac 
{\theta_i- \theta}{2}\Big ) } ,
\label{ASMeq4}
\end{align}
where
\begin{align}
\rho(\theta) = \frac{\alpha e^{\alpha (\theta - \theta_0/2)}}{2\sinh 
(\alpha\theta_0/2) }\quad {\rm and} \quad
 h(d,\theta_i- \theta)=  h(d)- \frac {2J}{\cosh^2\big(\frac {\theta_i- \theta}{2}\big) 
 }\big 
 (1+\tanh (\beta  h(d) )\big 
).
\label{ASMh}
\end{align}
\begin{figure}[tbh]
\includegraphics[width=0.6\linewidth]{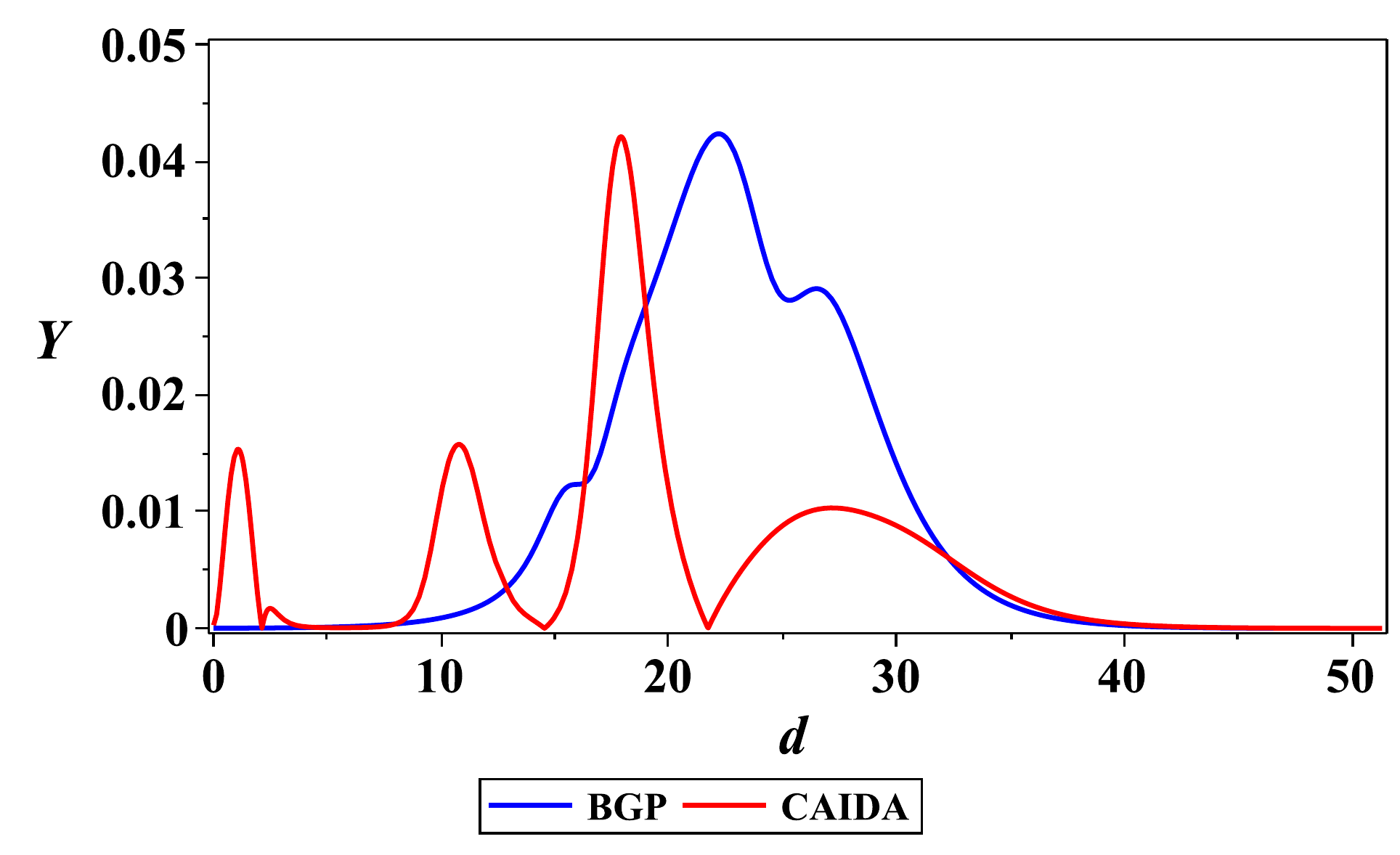}
	\caption{Graph of the function $Y(d)$ {representing the relative size of $\Delta p$ and $p$ as 
	an indicator of approximation error}. The blue curve corresponds to Fig. 4 and the 
	red curve corresponds to Fig. 5 in the main text.
}
	\label{SMfig6}
\end{figure}

Once again applying the mean value theorem, we obtain 
\begin{align}
Y (d)= 2\beta J\big (1- \tanh (\beta h)\big )\sum^2_{a=0}{\delta_a}
	\bigg (\frac {1}{\cosh^2\Big( \frac 
	{d-\theta_{a}}{2}\Big ) }\Big(1+\tanh (\beta h_0) - F(\theta_a) \Big ) -  
	Q(\theta_a)\bigg)  .
\end{align}
 
{Fig. \ref{SMfig6} depicts the function $Y(d)$.} The choice of parameters 
corresponds to Figs. 4 and 5 of the main text. As one can see, 
the approximation is valid for a wide range of distances, and one can {safely} use 
Eq.\eqref{SMp} for computation of the connection probability. 

\section{Methods}

{Parameters in our model} are: temperature $T$, size of the network $R$, curvature $K$, 
coupling constant $J$; fitting parameters $r_a$ and $\delta_a$ ($a=0,1,2$). Taking into account 
that $\delta_0 +\delta_1 + \delta_2 =1$, we obtain nine independent parameters. {These} 
parameters should {then} be adjusted to make the model fit the available Internet data {and, as 
we will see in the next section, can be reduced to six parameters if we have empirical values for 
$\bar{k}$, $\gamma$, and $N$.}

{In our work, we} use the empirical connectance data for the BGP and CAIDA views of the 
Internet{, and map} these data {onto} hyperbolic geometry, as
described in \cite{MKF,KDPF1,Boguna2010} (see Fig. \ref{fig3}).  
\begin{figure}[tbh]
	\centering
	\includegraphics[width=0.52\linewidth]{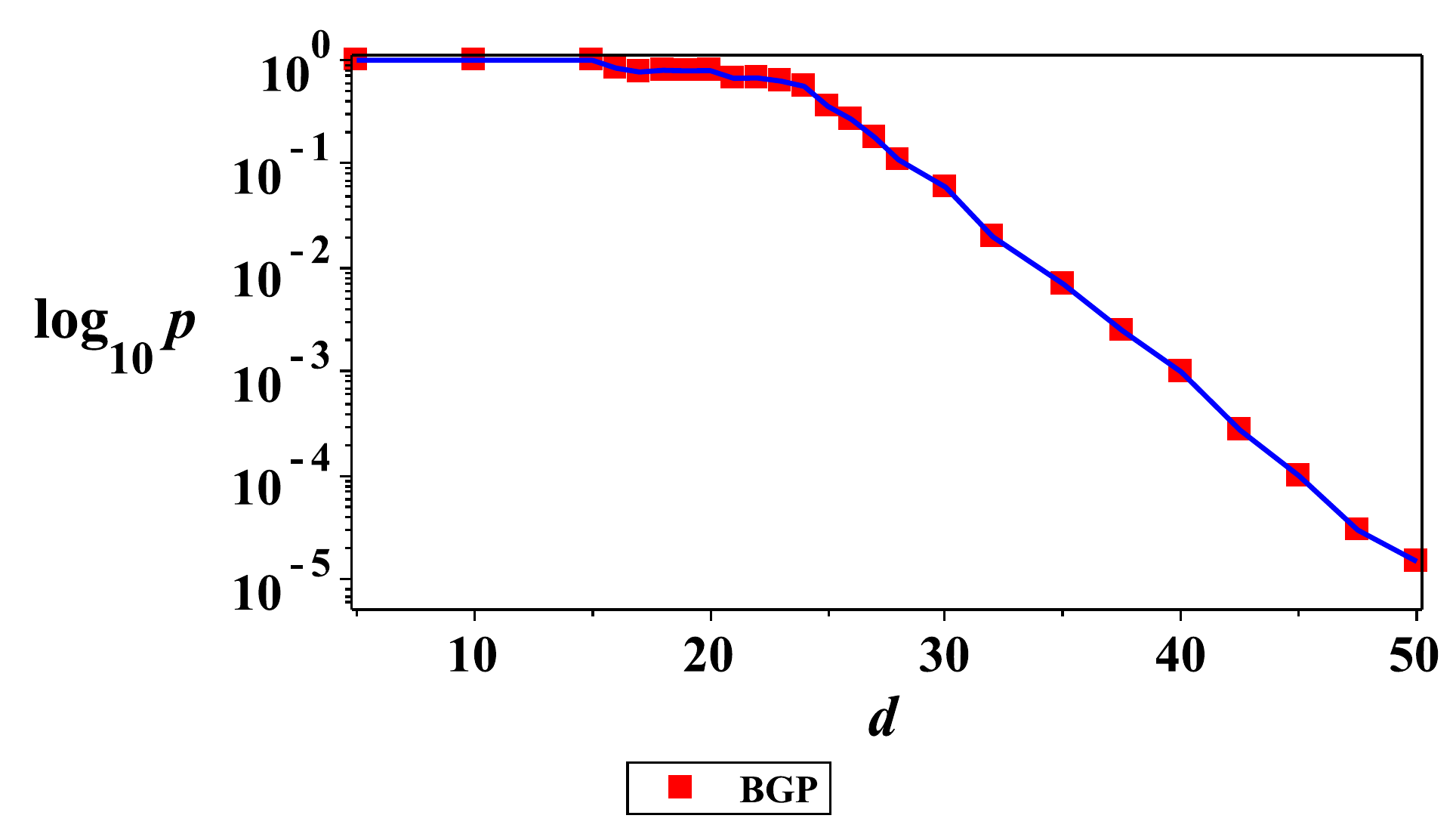} 
	\includegraphics[width=0.425\linewidth]{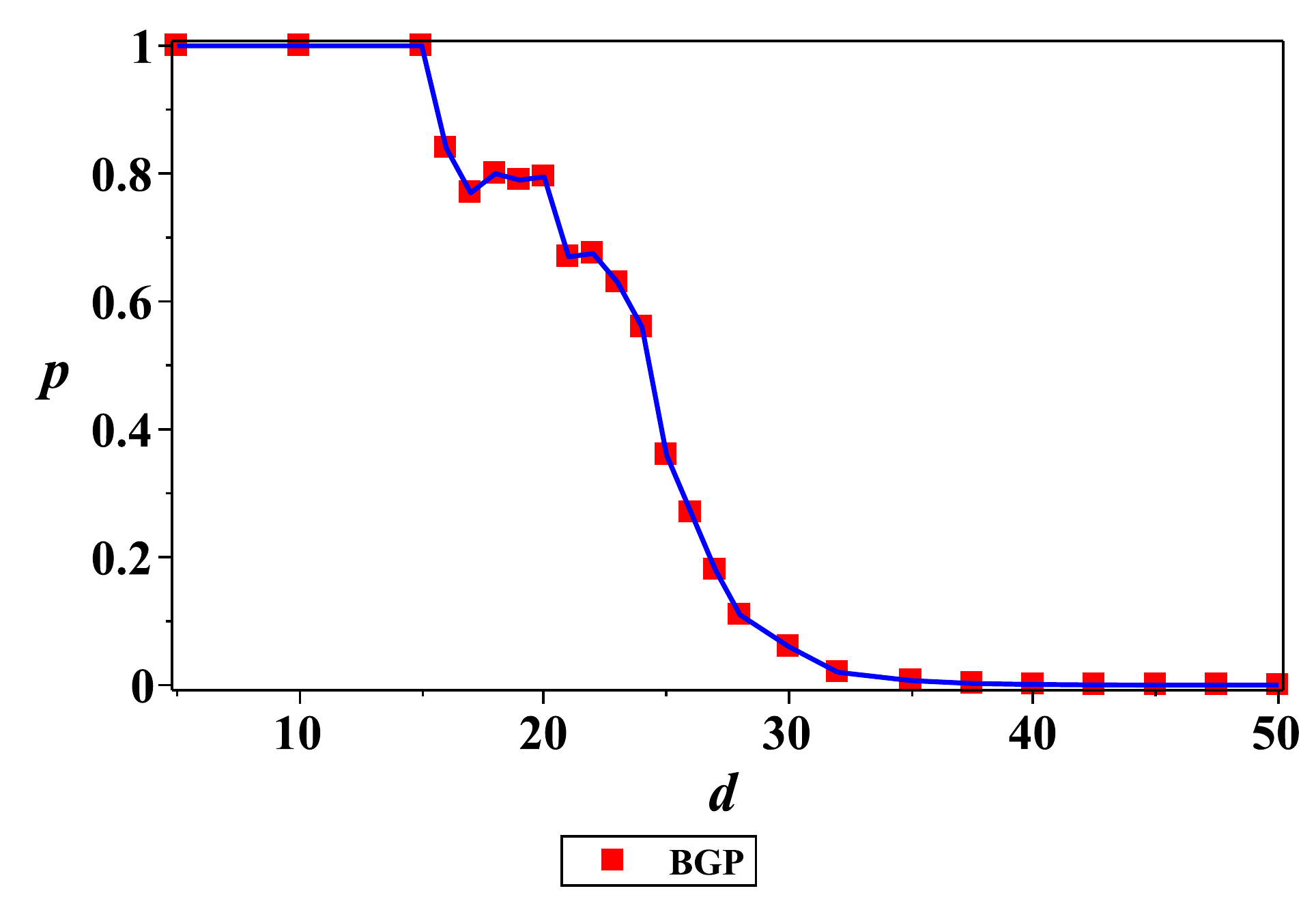} 
		\includegraphics[width=0.51\linewidth]{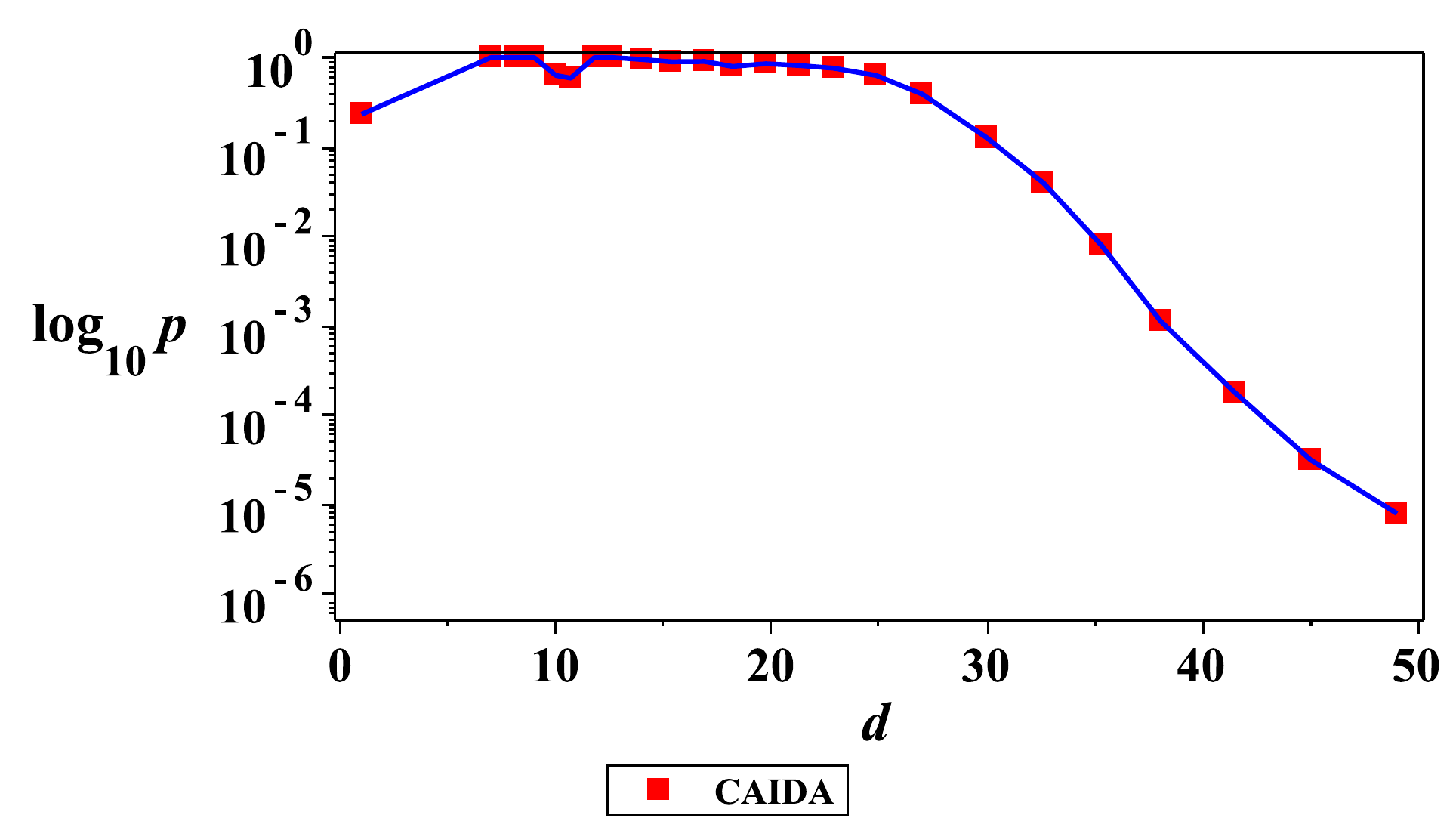}
			\includegraphics[width=0.425\linewidth]{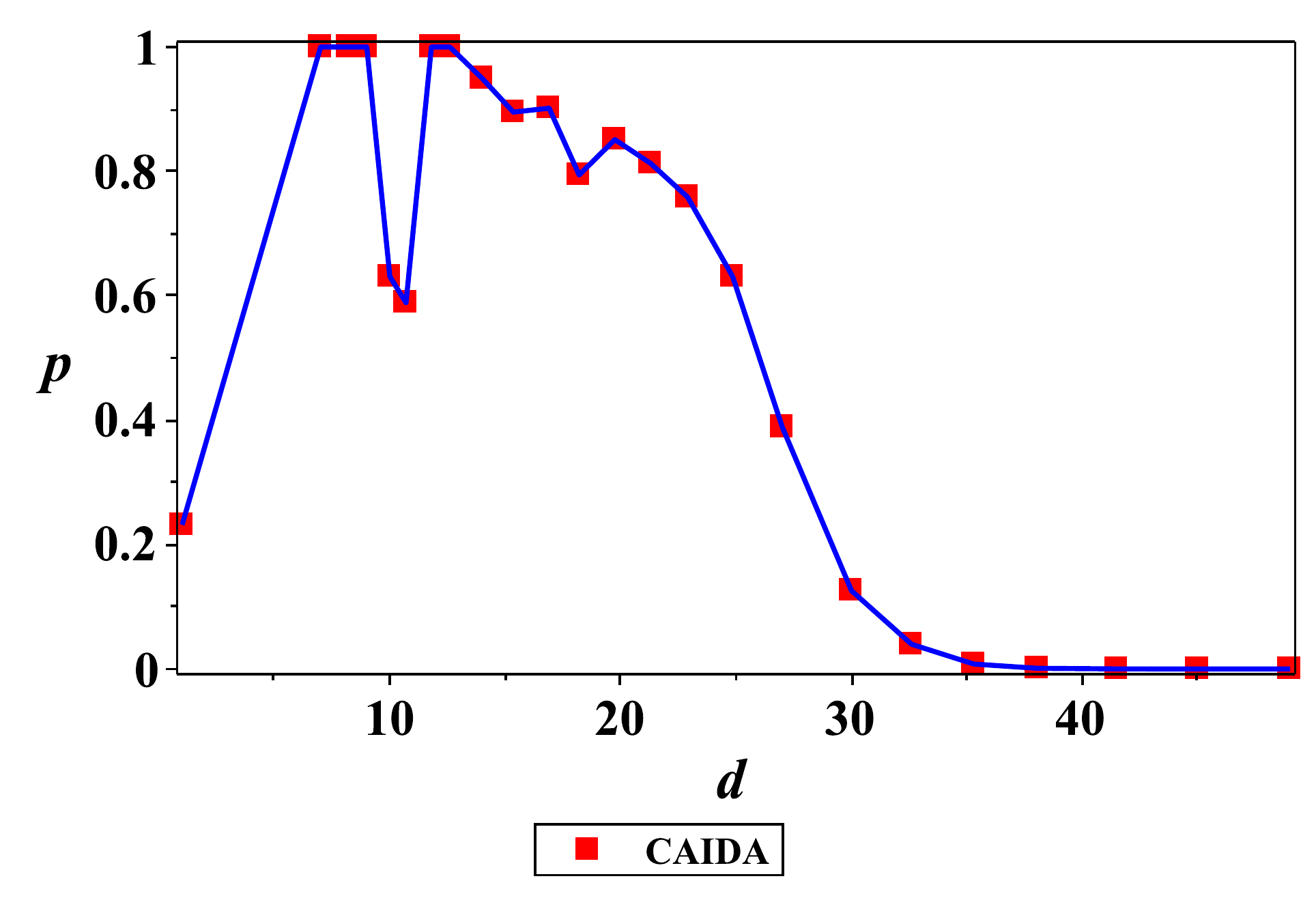}  
	\caption{Top: Empirical connection
probability for the Internet BGP based on data obtained in Ref. \cite{KDPF1}. {Bottom:} Empirical 
connection
probability for the Internet Archipelago data (CAIDA) extracted from Ref. \cite{Boguna2010}. }
	\label{fig3}
\end{figure}

 \subsection*{Size  and temperature of the Internet embedded in hyperbolic space}
 
The Internet nodes are mapped to a hyperbolic space of  curvature $K<0$ by assigning to each 
node a random angular coordinate $\varphi $, and a radial coordinate $r=\theta/\kappa $ 
according to the radial node density:
\begin{align}
\rho(r) = \frac{\alpha e^{\alpha (r - R/2)}}{2\sinh (\alpha R/2) },
\quad  0\leq r \leq R, 
\label{eq:radialDensity}
\end{align}
where $\alpha =\kappa (\gamma -1)/2$ and  $\kappa= \sqrt{-K}$. As was shown above, the 
connection probability is given by the 
Fermi-Dirac distribution:
\begin{align}
p =  \frac{1}{ e^{\beta (\varepsilon - \mu)} + 1},
\label{eq_FD} 
\end{align}
 where $\mu = \kappa R/2$ and
\begin{align}
		\varepsilon = \frac{\kappa d}{2}+ \sum^2_{a=0}\frac{4J\delta_a \Big 
		(1+\displaystyle\tanh 
		\Big(\beta \Big (\frac{\kappa d}{4}-\frac{\mu}{2} \Big )\Big)\Big ) }{\displaystyle\cosh^2 
	\left( \frac{\kappa( d - r_a)}{2} \right)} .
	\label{Energy}
\end{align}
When the coupling constant $J=0$, the model describes the homogeneous  scale-free network 
with the link energy $\varepsilon = {\kappa d}/{2}$.

The key formulae for embedding the Internet in the hyperbolic space are: {the} power-law 
degree distribution in the network, $P(k) \sim (\gamma -1)k^{-\gamma}$, and {the} expression 
for the average degree in the whole network, $\bar{k}= N \nu e^{-  \kappa R/2}$. In our model 
the control parameter is $\nu =(\gamma -1)/(\gamma -2)$, and the size of the network is 
given by
  \begin{align}
  	R= \frac{2}{\kappa} \ln \bigg (\frac{N}{\bar k} \Big (\frac{\gamma -1}{\gamma -2} 
  	\Big)\bigg ).
  	\label{R}
  \end{align}
To  determine $\kappa$ we use the expression obtained in \cite{KDPF2,CDAS} for the Internet 
embedded in the (universal) hyperbolic space with curvature $K=-1$,
\begin{align}
 	R= 2\ln \bigg (\frac{2N}{\pi \bar k} \Big (\frac{\gamma 
 	-1}{\gamma -2} \Big)^{2} \bigg ).
 	\label{SR}
 \end{align} 
The curvature, $K = -\kappa^2$, of the target hyperbolic space is determined now by 
comparing \eqref{R} with  \eqref{SR}. We obtain 
   \begin{align}
   	\kappa = \frac{\ln \Big (\frac{N}{\bar k} \Big (\frac{\gamma -1}{\gamma -2} \Big)\Big)}{\ln 
   	\Big (\frac{2N}{\pi \bar k} \Big (\frac{\gamma 
 	-1}{\gamma -2} \Big)^{2} \Big)} = 1 -  \frac{\ln \Big (\frac{2}{\pi } \Big (\frac{\gamma 
 	-1}{\gamma -2} \Big)\Big)}{\ln \Big (\frac{2N}{\pi \bar k} \Big (\frac{\gamma -1}{\gamma 
 	-2} \Big)^{2} \Big)}.
   \end{align}

\subsection*{Internet temperature}
   
{\em Homogeneous model.} --   To estimate the Internet temperature, first we consider a 
purely homogeneous network ($J=0$). Since the Internet is a scale-free sparse network, one 
can use Eq.\eqref{SMSG} for the average node degree to determine the temperature of the 
Internet
\begin{align}
\bar k =  &\frac{N(\gamma -1)}{2\sinh(\beta(\gamma -1) \mu)}\Big(e^{\beta 
(\gamma -1)\mu} \Phi\big ( -e^{\beta \mu} ,1, \gamma -1\big )-e^{-\beta (\gamma 
-1)\mu}  \Phi\big ( -e^{-\beta \mu} ,1, \gamma -1\big )\Big ),
\label{SG}
\end{align}
where the chemical potential is given by
\begin{align}
\mu =  \ln \bigg (\frac{N}{\bar k} \Big (\frac{\gamma -1}{\gamma -2} 
\Big)\bigg ).
\end{align}
Substituting $\mu$ in \eqref{SG}, one can rewrite it {as} $F(N,\bar k,\gamma,T) =0$.
For given $N,\bar k$ and $\gamma$, solution of this equation yields $T$.  \\

{\em Heterogeneous model.} --   In our complete model the coupling constant $J \neq 0$. 
Then, employing Eq. \eqref{SMSGa}, we obtain 
\begin{align}
\bar k = N\int_0^{2 R} \frac{\varrho (r) d r}{e^{\beta (\varepsilon - \mu)}+1},
\end{align}
where $\varrho (r )$ and $\varepsilon (r )$ are given by Eqs. \eqref{eq:radialDensity} and  
\eqref{Energy}, respectively.
We use this expression to calculate the temperature of the Internet for the heterogeneous model.

In Table 1 we present the results of our computation of the Internet temperature for BGP and 
CAIDA data. The empirical data are extracted from  Refs. \cite{KDPF1,Boguna2010}. As one can 
see, {temperatures are very close for both models (homogeneous and heterogeneous)}.
\vspace{0.25cm}
   \begin{center}
   \begin{tabular}{|c|c|c|c|c|c|c|}
   \hline
         \rule[-1ex]{0pt}{2.5ex}\textbf{Homogeneous model $(J=0)$}& Size $(N)$  & Number of 
         links $(L)$ &  
               Average node degree $ (\bar k )$ &$ \gamma$ & Temperature (T) \\ 
               \hline
          \rule[-1ex]{0pt}{2.5ex} BGP & 17,446 &40,805  &4.68 & 2.16 & 1.03602\\ 
            \hline 
          \rule[-1ex]{0pt}{2.5ex} CAIDA &  23,752 & 58,416 &  4.92 & 2.1 & 1.06718 \\ 
            \hline \hline 
                     \rule[-1ex]{0pt}{2.5ex}\textbf{Complete model $(J\neq0)$}& Size $(N)$  & 
                     Number of 
                     links $(L)$ &  
                           Average node degree $ (\bar k )$ &$ \gamma$ & Temperature (T) \\ 
                           \hline
                      \rule[-1ex]{0pt}{2.5ex} BGP & 17,446 &40,805  &4.68 & 2.16 & 1.0365\\ 
                        \hline 
                      \rule[-1ex]{0pt}{2.5ex} CAIDA &  23,752 & 58,416 &  4.92 & 2.1 & 1.06725 \\ 
                        \hline 
    \end{tabular} 
    \end{center}
  \vspace{0.25 cm}
  {Table 1: The characteristics of the Internet: number of nodes ($N$),  number of links $(L)$, 
average degree $ (\bar k )$,  exponent of  degree distribution $(\gamma)$ and temperature 
($T$). Empirical data are extracted from Refs. 
\cite{KDPF1,Boguna2010}}
\vspace{0.25 cm}

 {\em Impact of parameters. } -- {Fig. \ref{fig4} presents} our numerical results ({black} curve) 
 and {compares} them with the theoretical predictions of the model presented in 
 \cite{KDPF1,Boguna2010} (green-dashed 
curve) and  empirical Internet data obtained for BGP and CAIDA (red diamonds). Our findings 
show that a homogeneous model yields (in general) a good agreement with available empirical 
data, but can not explain evident anomalies in the connection probability that break the 
scale-free behavior of the Internet. 

We found that parameters ${\beta, R, \kappa}$ control the homogeneous, sigmoidal shape of 
the curve only, whereas the ${r_i}$s and $\bm{\delta}$s regulate the  location and depth of the 
local minima. Finally, the coupling constant $J$ controls the  height of the local maxima.
\begin{figure}[tbh!]
		     	\centering
		     	\includegraphics[width=0.45\linewidth]{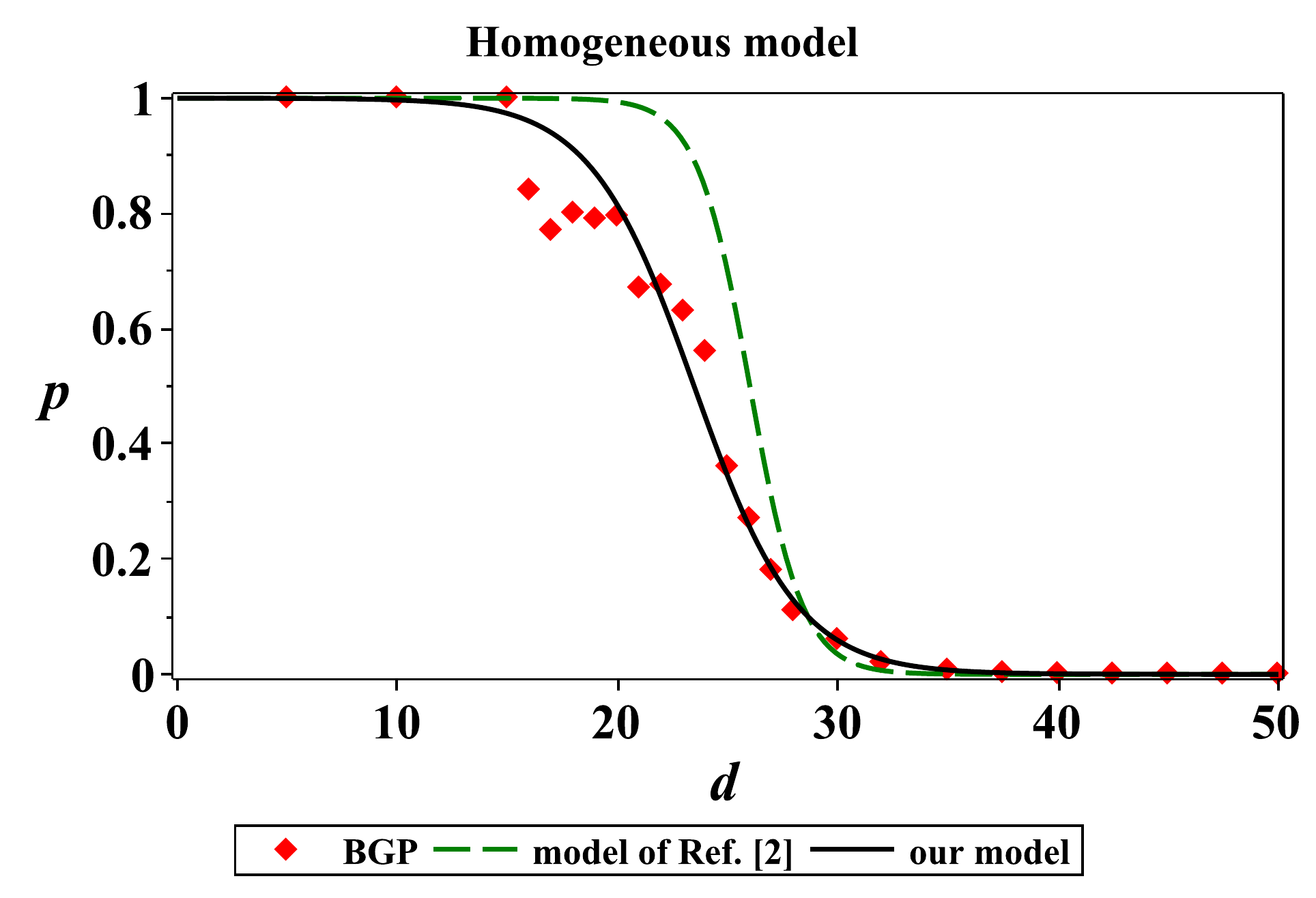}
		     	\includegraphics[width=0.45\linewidth]{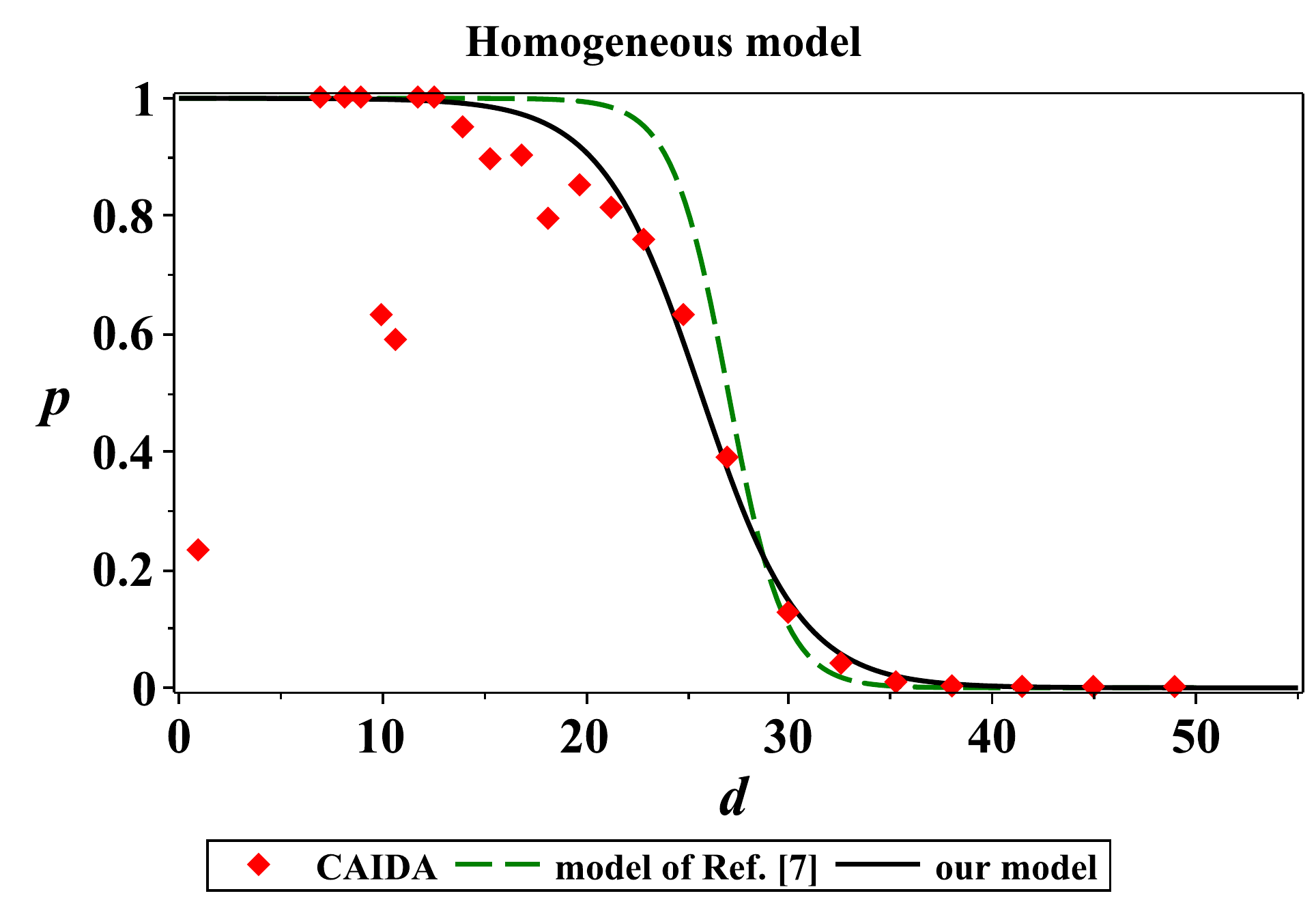} 
		     	\caption{Connection probability for the Internet empirical data (red diamonds) 
		     	compared to the homogeneous model ($J=0$, black curve) and numerical results 
		     	(green-dashed curve) obtained in \cite{KDPF1,Boguna2010}. Left: BGP. Right: CAIDA. }
		     	\label{fig4}
		     \end{figure}
		     	\begin{figure}[tbh!]
		     	\centering
		     		(a)
		     	\includegraphics[width=0.425\linewidth]{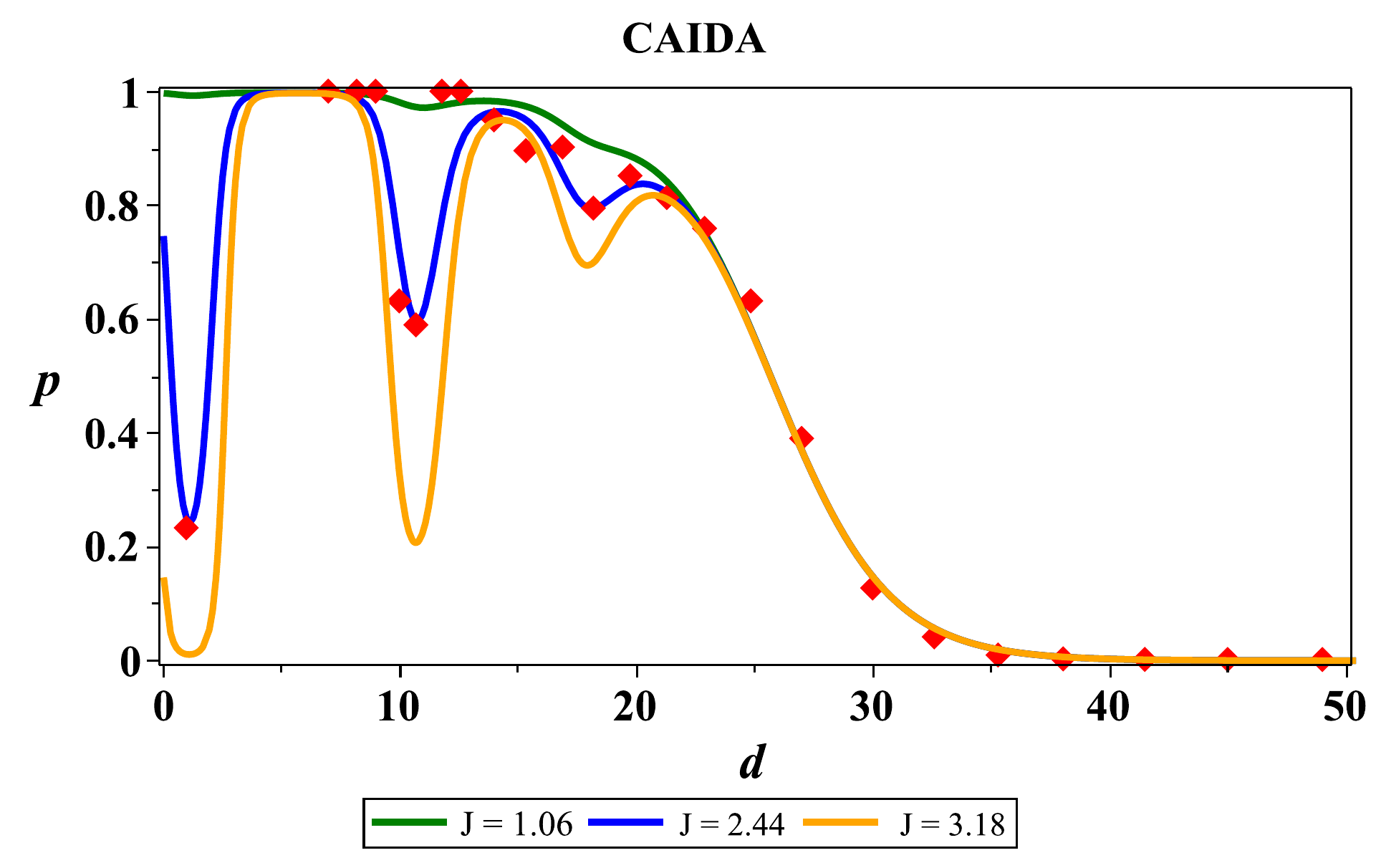} 
		     (b)
		     	\includegraphics[width=0.4\linewidth]{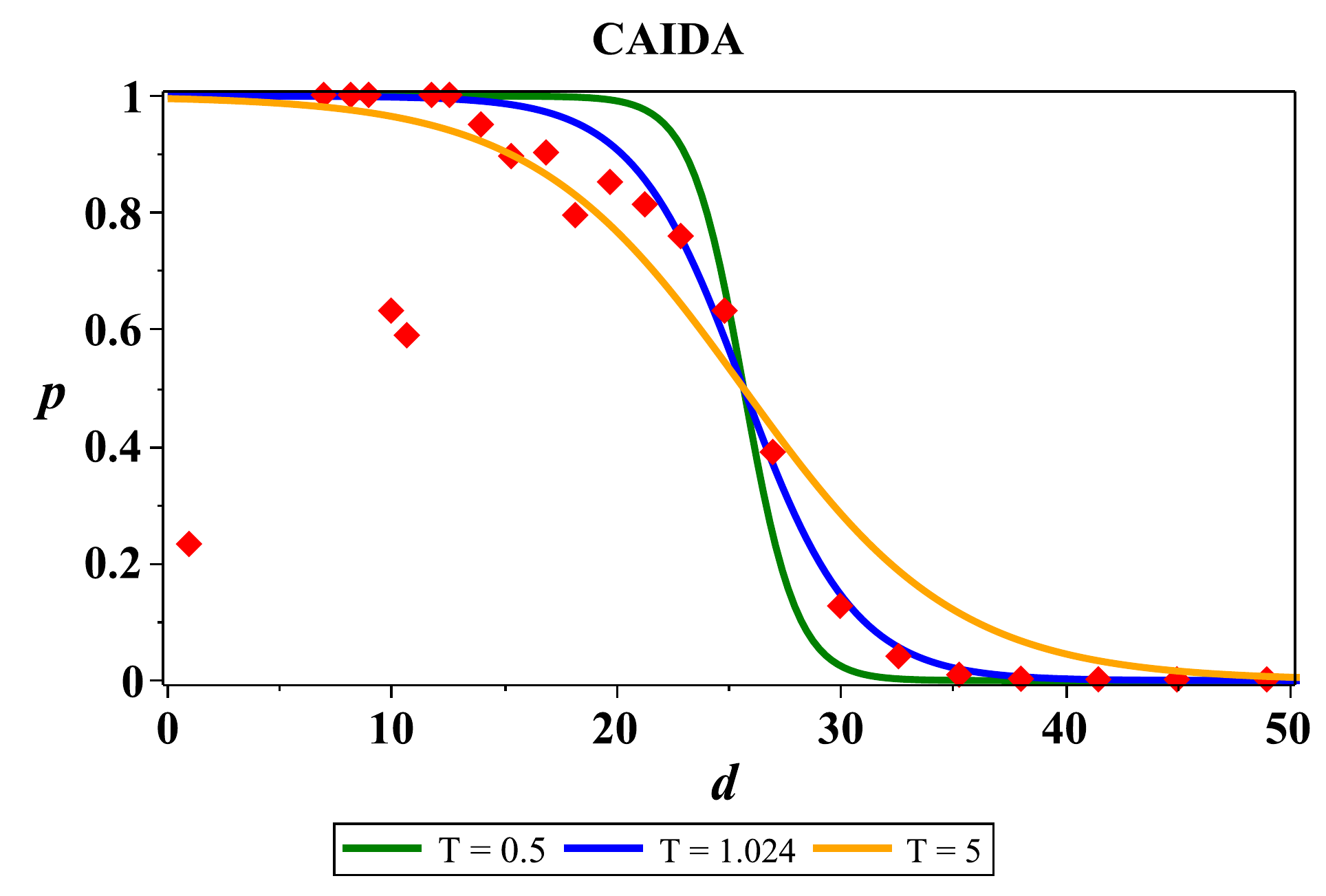} \\
		     	(c)
		     	\includegraphics[width=0.4\linewidth]{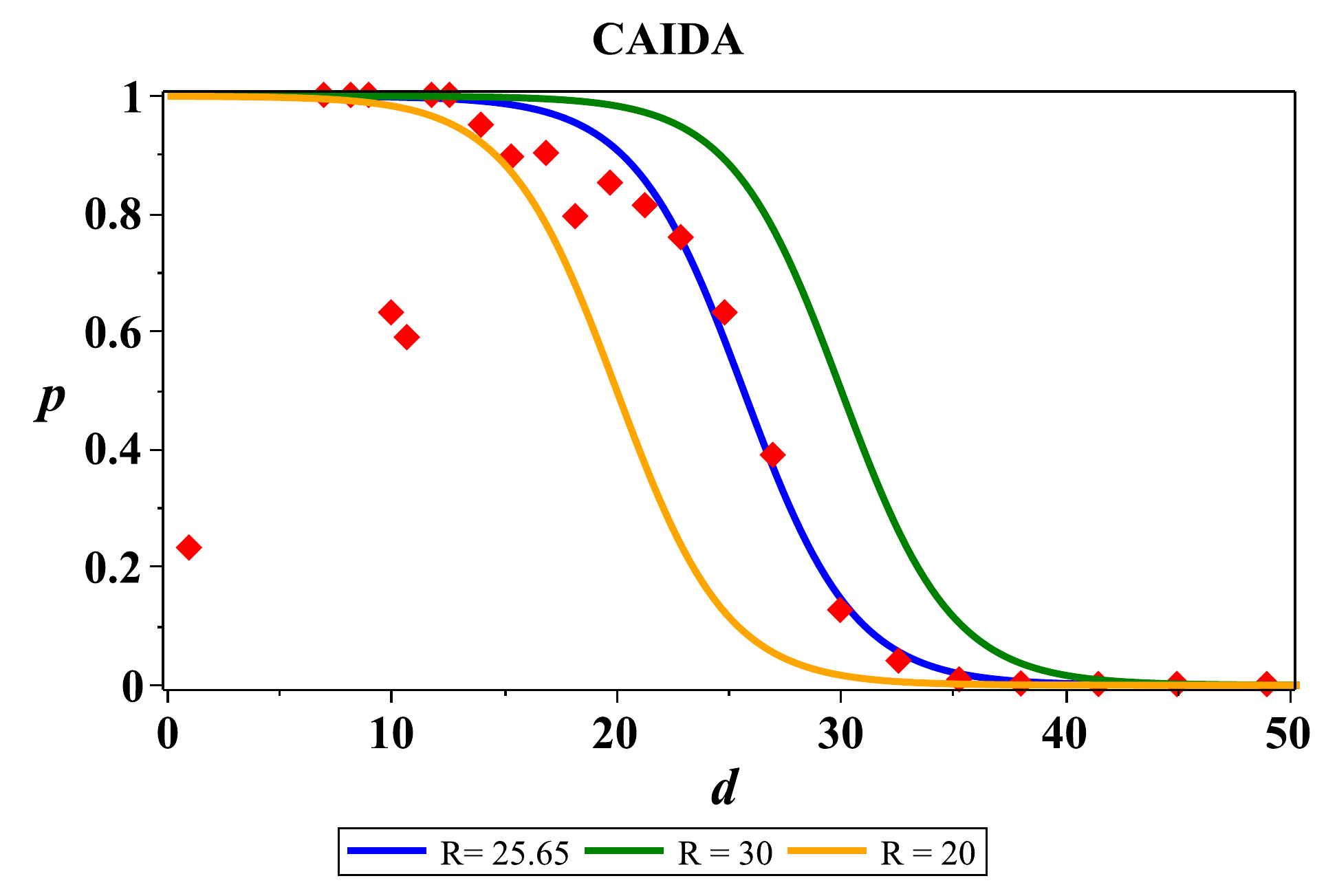}
		     	\caption{Comparison of the effect of parameters J (a), T (b), and R (c) on the fit to 
		     	CAIDA 
		     	data. }
		     	\label{fig4a}
		     \end{figure}
 In Fig. \ref{fig4a} we depict the connection 
		 probability for the homogeneous model with vanishing contribution {from} the holonomy 
		 ($J=0$) and compare {the} homogeneous model {to} the heterogeneous one for different 
		 values of paramaters $J,T$ and $R$. 

\subsection*{Community structure and small-world properties of the Internet}

{\em Community structure of the Internet.} --  Many real networks exhibit inhomogeneity in the 
link {distribution,} leading to the natural clustering of the network into groups or communities. 
Within the same community {vertex-to-vertex} connections are dense, but {connections are less 
dense between groups} \cite{GN}. 

To study {the} contribution of  holonomy to the formation of small communities, {first} we 
compare the connection probability for the {exclusively holonomic} model with the 
connection 
probability for homogeneous and heterogeneous models and {empirical} data (see Fig. \ref{fig5}).
Additionally, we calculated the average node degree for {the exclusively holonomic} model. We 
found 
that for BGP and CAIDA experimental data, the model yields {a} high level of the connection 
between nodes,  $\bar k \approx N/2$. This is close to the value of $\bar k $ in the limit of  
high {temperatures}. However, in the complete model we have $\bar k  \ll N$. This 
means that 
there are many vertices with low {degrees} and a small number with high {degrees}. Thus, our 
results 
show that holonomy (non-local curvature) is responsible for {the} formation of the 
community structure of the Internet: 

\vspace{0.25 cm}
\begin{figure}[tbh!]
	\centering
		(a)
	\includegraphics[width=0.465\linewidth]{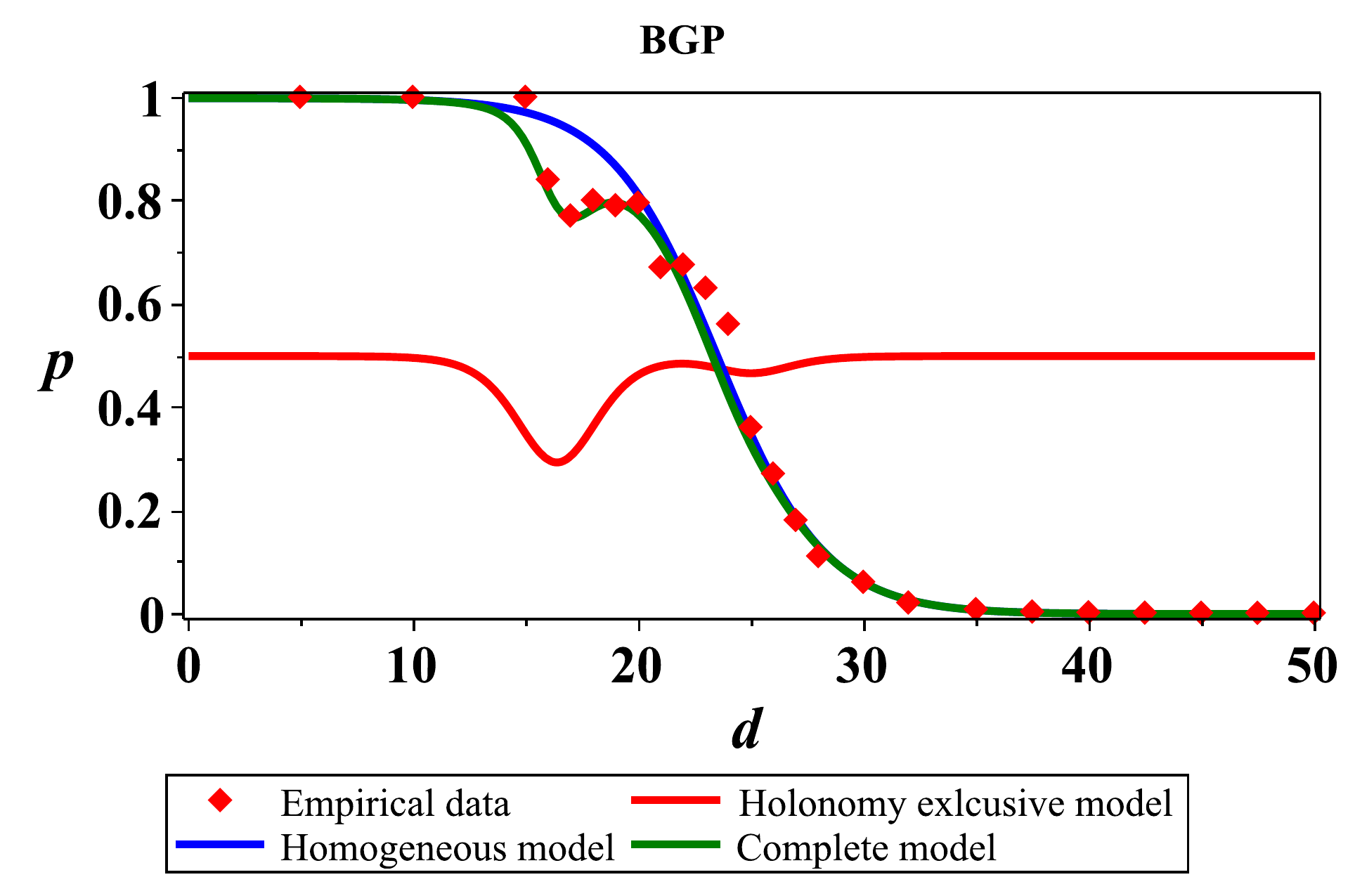} 
(b)
	\includegraphics[width=0.465\linewidth]{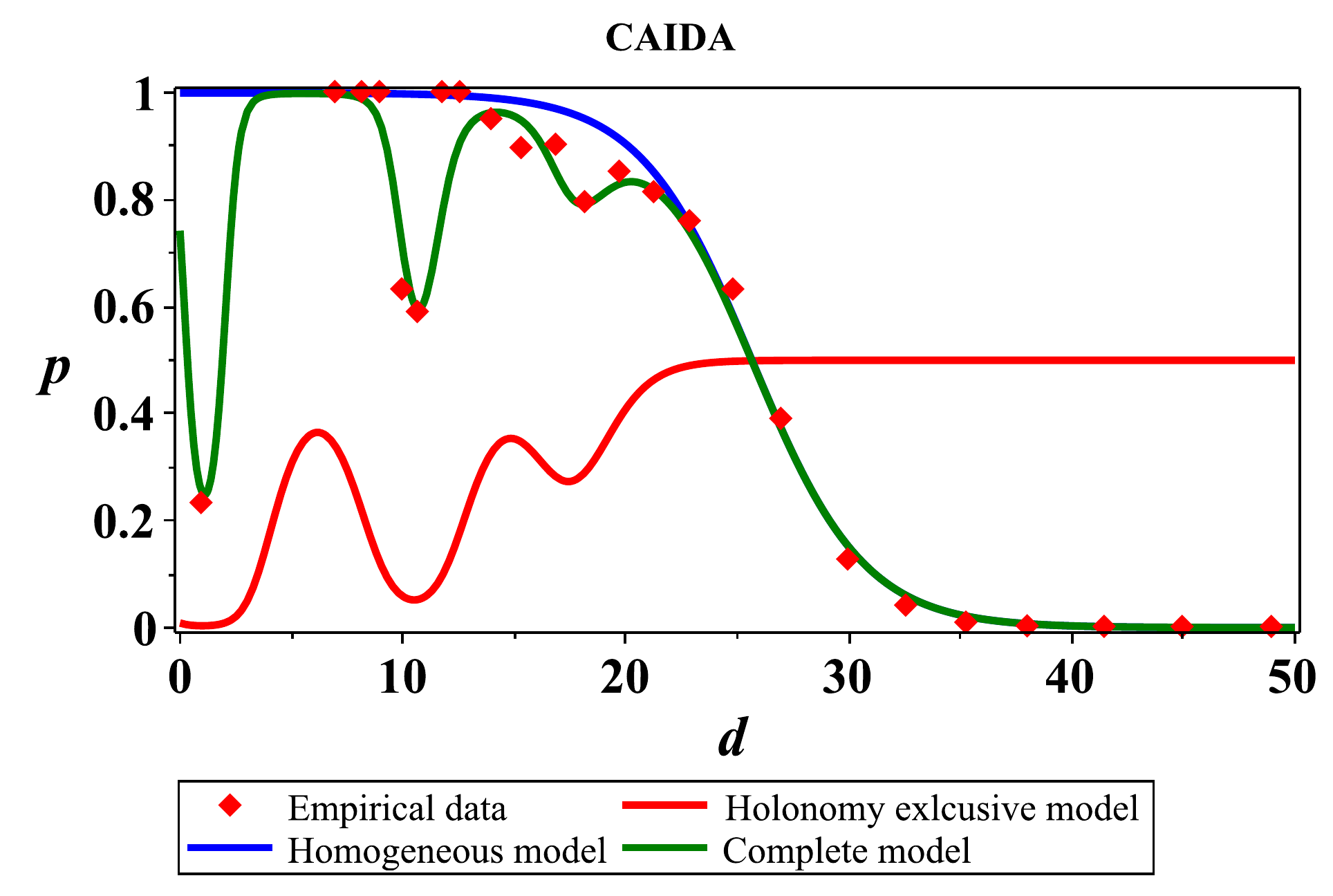} 
	\caption{Connection probability. (a) BGP: $\bar k\approx N/2 =8723$ (purely  holonomic
	model), $\bar k = 4.68$ (complete model);  (b) CAIDA: $\bar k \approx N/2= 11875$ 
	({purely holonomic} model), $\bar k = 4.92$ (complete model).}
	\label{fig5}
\end{figure}

{\em Small-world properties of the Internet. } -- {The small-world property of a network refers 
to a }
relatively short distance between randomly chosen {pairs} of nodes. In {a} small-world network 
{the typical distance between nodes, $\ell$,} ({ required to connect them by passing through 
other nodes}) increases  proportionally to the logarithm of the number 
of nodes, $\ell \propto\ln N$, {as long as} the clustering coefficient is not small 
\cite{WDSS}. For a {scale-free} network with power-law degree distribution, $P(k) \sim (\gamma 
-1)k^{-\gamma}$ and for $2 < \gamma < 3$, this dependence 
{is}
modified as follows: the shortest path 
between two randomly chosen nodes grows {as}
\begin{align}
\ell\propto \frac{2}{|\ln(\gamma -2)|}\ln \ln N.
\end{align}
{The presence of this behavior is known as the ultra small-world property of the scale-free 
network} \cite{CRHS,CFL}.

To study small-world properties of the Internet embedded in hyperbolic space, we 
calculated the mean geodesic distance between two nodes for homogeneous ($J=0$) and 
heterogeneous models. The results are presented in Fig. \ref{fig4b}. We found that {$l \propto 
\ln N$ for the homogeneous model}. The contribution of the holonomy to the small-world 
{effect} is described by the corrections $\propto \ln\ln N$. Thus, one can say that the 
non-local curvature (described by the elementary holonomy) is responsible for formation of the 
{ultra-small-world networks within the} Internet. Since the corrections are tiny, {support for our 
conjecture requires more thorough study}.
\begin{figure}[tbh!]
	\centering
		(a)
	\includegraphics[width=0.425\linewidth]{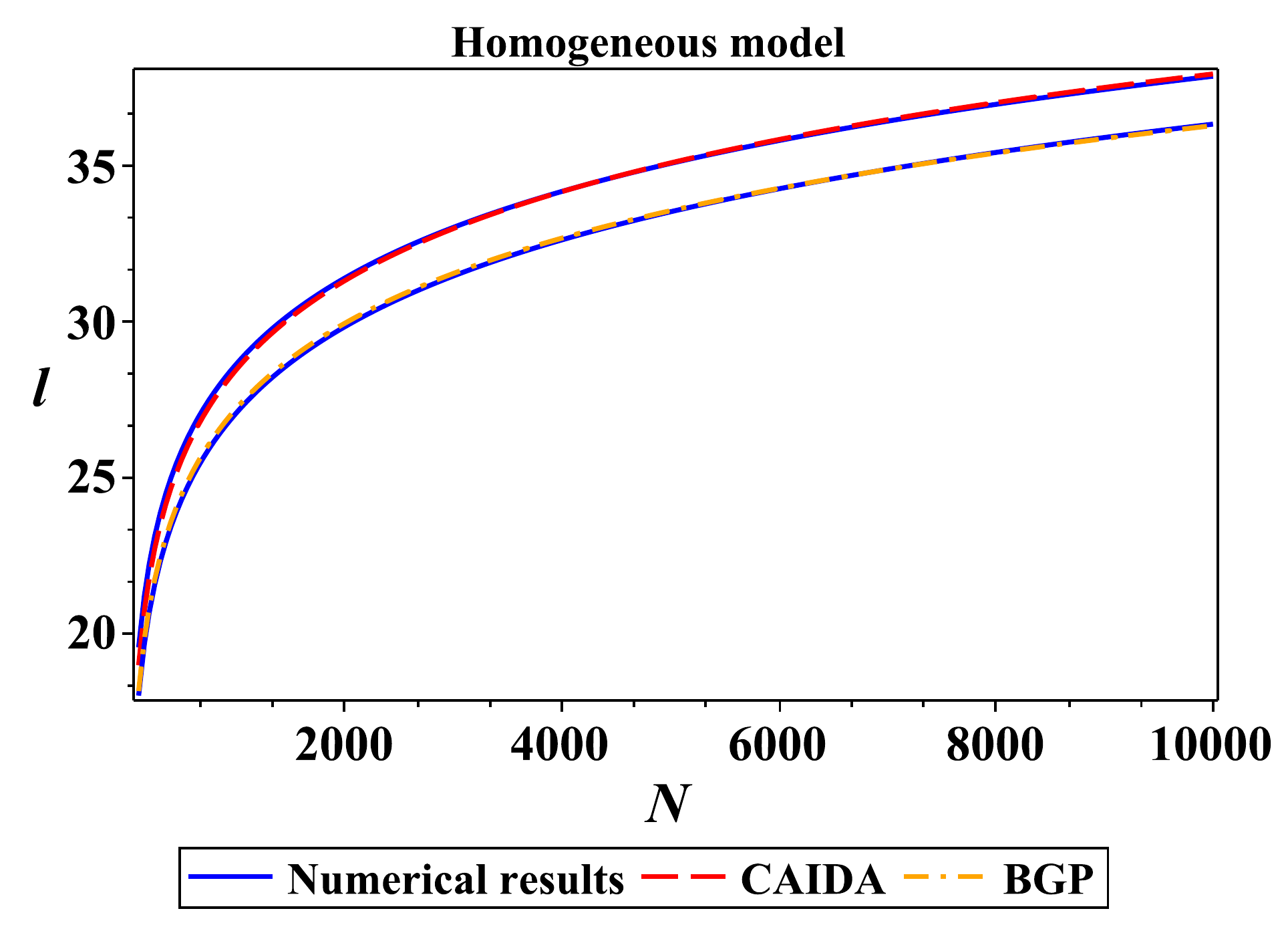} 
(b)
	\includegraphics[width=0.425\linewidth]{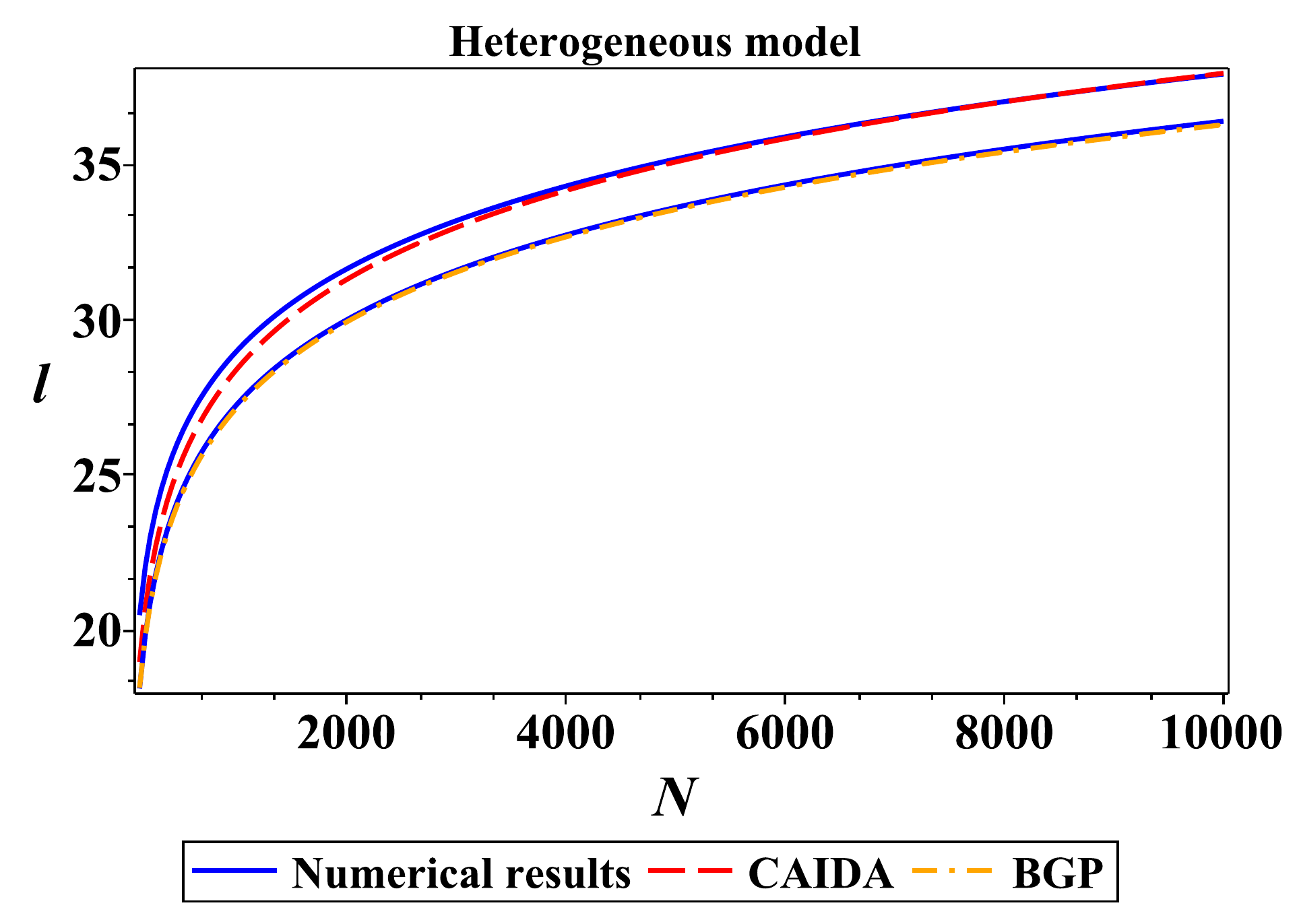} \\
	(c)
		\includegraphics[width=0.425\linewidth]{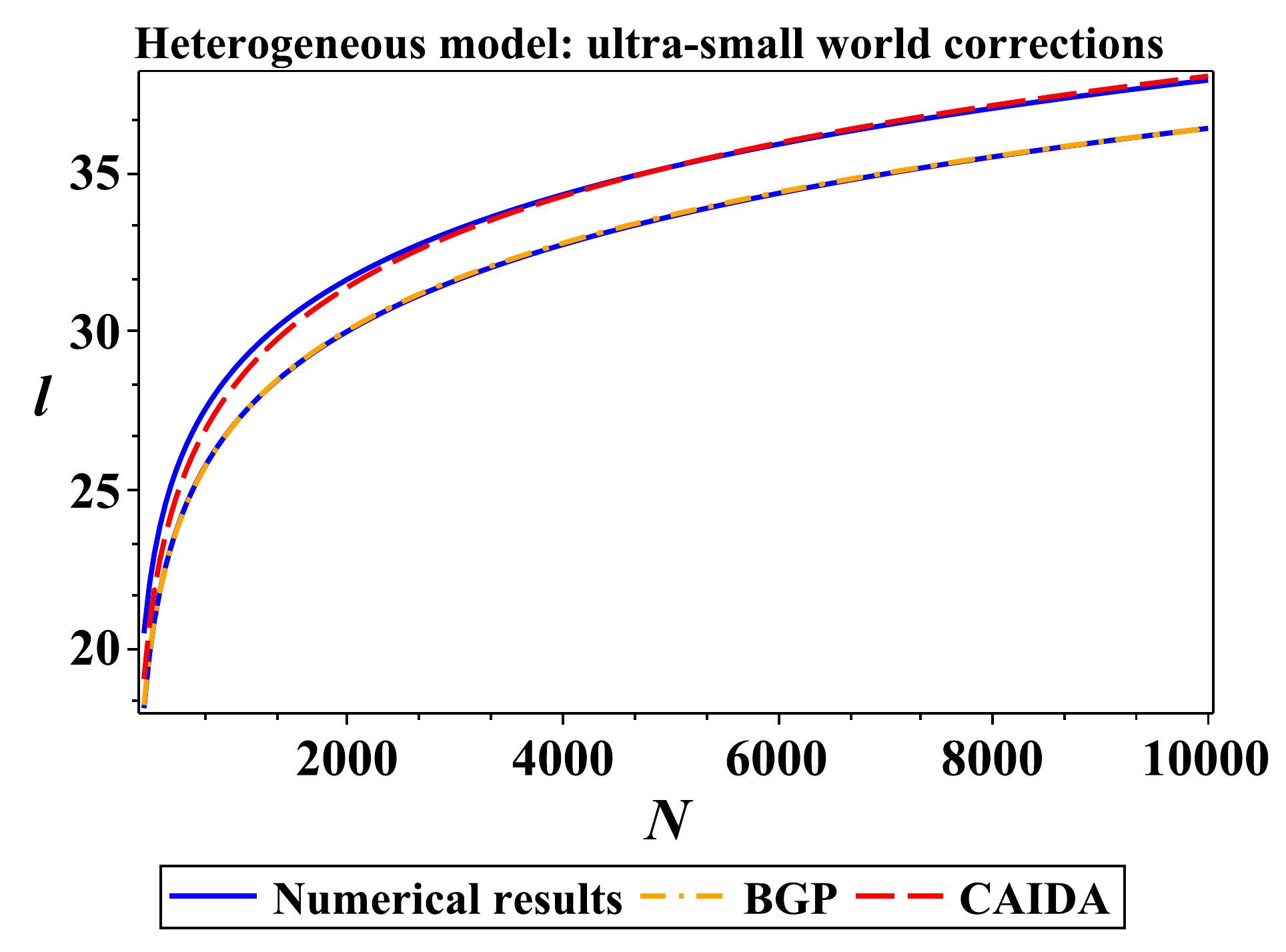} 
	\caption{Internet small-world properties: (a) Homogeneous model:  $l = 4.12\ln N $ (CAIDA), 
	 $l = 3.94\ln N $ (BGP).  (b) Heterogeneous model:  $l = 4.12\ln N $ (CAIDA), $l = 
	3.94\ln N $ (BGP). (c) ) Ultra-small world corrections:  $l = 4.12\ln N  + 0.045 \ln \ln N$ 
	(CAIDA), $l = 3.94\ln N  + 0.05 \ln\ln N $(BGP).}
	\label{fig4b}
\end{figure}

To compare our findings with real networks we need {to find the lowest number of 
steps required to pass from one node to other among pairs of nodes separated by the geodesic 
distance $l$}. {If we let} $l_0$ {denote} the mean geodesic distance per step, then the number 
of steps required to travel along the shortest path between {a pair} of nodes is given by $\ell = 
l/l_0$. 
The main {difficulty in the} computation of $\ell_0$ is {the} lack of analytical results for 
free-scale 
networks. To overcome this obstacle,  we note that for a high temperature the scale-free 
network becomes 
highly {randomized, which} allows us to employ the formula for the shortest distance in random 
networks given by {\cite{ARB}:}
\begin{align}
\ell_r = \frac{\ln N}{\ln \bar k}.
\label{L1g}
\end{align}
In the limit of $T \rightarrow \infty$, every pair of nodes is connected with probability 
$p=1/2$, and the computation of {the average node degree} yields $\bar k \rightarrow N/2$. In 
the same limit we obtain 
$\ell_r \rightarrow \ell_0 = \ln N / \ln(N/2)$ and
\begin{align}
l \rightarrow l_\infty = 2 R_\infty -  \frac{2}{\kappa(\gamma -1)},
\label{L2}
\end{align}
where
\begin{align}
  	R_\infty= \frac{2}{\kappa} \ln \bigg (2\Big (\frac{\gamma -1}{\gamma -2} 
  	\Big)\bigg ).
  	\label{R}
  \end{align}
Then  the geodesic distance per step can be written as  $l_0 = l_\infty/\ell_0$. Assuming that 
$l_0$ is a minimal geodesic distance per step ({a} ``quant'' of length), we obtain a qualitative 
formula for estimation of the {path length},
\begin{align}
\ell  =\frac{l}{l_0} = \frac{l}{l_\infty} \ell_0 .
\label{L3}
\end{align}
In Table 2 we compare our theoretical predictions with data available in the literature. 
We consider only the networks of large size, $N >10^3$. As one can see, the predictions of  Eq. 
\eqref{L3} are in a reasonable qualitative agreement with the average path lengths of real 
networks.

\vspace{0.25cm}
   \begin{center}
   \begin{tabular}{|l|c|c|c|c|c|c|c|c|c|}
    \hline 
    \rule[-1ex]{0pt}{2.5ex}{Network }& {Size $(N)$}& Average node degree $ (\bar k 
    )$&\textbf{$\boldsymbol 
    \gamma$} & $\ell_{real}$& $\ell_{rand}$ &$\ell_{pow}$  & $\ell$&Temperature (T) 
    &References\\ 
    \hline \hline
    \rule[-1ex]{0pt}{2.5ex} BGP & 17,446 &  4.68 & 2.16 & 3.69& 6.33 & -- & 4.06 &1.036 
    & 
    \cite{MKF}\\ 
      \hline 
    \rule[-1ex]{0pt}{2.5ex} CAIDA &  23,752  &  4.92 & 2.1 & -- &6.32 & --& 3.64& 
    1.067  
    & 
    \cite{Boguna2010}\\ 
      \hline 
    \rule[-1ex]{0pt}{2.5ex} Internet, router  & 150,000 & 2.66 & 2.4& 11&12.18& 7.47& 7.36 & 
    1.003& 
    \cite{ARB} \\ 
    \hline
  \rule[-1ex]{0pt}{2.5ex} Movie actors  & 212,250& 28.78 &2.3 & 4.54&3.65 &4.01 & 5.37 
  &1.007 &
  \cite{ARB}\\ 
      \hline
  \rule[-1ex]{0pt}{2.5ex} Co-authors, neuro &209,293 & 11.54 &2.1 & 6 & 5&3.86 & 3.92 & 
  1.0008&
  \cite{ARB}\\ 
          \hline
   \rule[-1ex]{0pt}{2.5ex} Co-authors, math & 70,975 & 3.9 &2.5 & 9.5 & 8.2 &6.53& 7.28 
   &1.0008 & 
   \cite{ARB}\\ 
                  \hline
    \end{tabular} 
    \end{center}
  \vspace{0.25 cm}
  {Table 2: The characteristics of some
  real networks: number of nodes ($N$),
average degree $ (\bar k )$, average path length $\ell_{real}$ and temperature ($T$). The 
columns $\ell_{rand}$, 
$\ell_{pow}$ and $\ell$ show values of the average path lengths for
 random network \eqref{L1},  power-law degree distribution and our model \eqref{L3}, 
 respectively.}

\end{widetext}


\begin{thebibliography}{36}%
\makeatletter
\providecommand \@ifxundefined [1]{%
 \@ifx{#1\undefined}
}%
\providecommand \@ifnum [1]{%
 \ifnum #1\expandafter \@firstoftwo
 \else \expandafter \@secondoftwo
 \fi
}%
\providecommand \@ifx [1]{%
 \ifx #1\expandafter \@firstoftwo
 \else \expandafter \@secondoftwo
 \fi
}%
\providecommand \natexlab [1]{#1}%
\providecommand \enquote  [1]{``#1''}%
\providecommand \bibnamefont  [1]{#1}%
\providecommand \bibfnamefont [1]{#1}%
\providecommand \citenamefont [1]{#1}%
\providecommand \href@noop [0]{\@secondoftwo}%
\providecommand \href [0]{\begingroup \@sanitize@url \@href}%
\providecommand \@href[1]{\@@startlink{#1}\@@href}%
\providecommand \@@href[1]{\endgroup#1\@@endlink}%
\providecommand \@sanitize@url [0]{\catcode `\\12\catcode `\$12\catcode
  `\&12\catcode `\#12\catcode `\^12\catcode `\_12\catcode `\%12\relax}%
\providecommand \@@startlink[1]{}%
\providecommand \@@endlink[0]{}%
\providecommand \url  [0]{\begingroup\@sanitize@url \@url }%
\providecommand \@url [1]{\endgroup\@href {#1}{\urlprefix }}%
\providecommand \urlprefix  [0]{URL }%
\providecommand \Eprint [0]{\href }%
\providecommand \doibase [0]{http://dx.doi.org/}%
\providecommand \selectlanguage [0]{\@gobble}%
\providecommand \bibinfo  [0]{\@secondoftwo}%
\providecommand \bibfield  [0]{\@secondoftwo}%
\providecommand \translation [1]{[#1]}%
\providecommand \BibitemOpen [0]{}%
\providecommand \bibitemStop [0]{}%
\providecommand \bibitemNoStop [0]{.\EOS\space}%
\providecommand \EOS [0]{\spacefactor3000\relax}%
\providecommand \BibitemShut  [1]{\csname bibitem#1\endcsname}%
\let\auto@bib@innerbib\@empty

\bibitem [{\citenamefont {Watts}\ and\ \citenamefont {Strogatz}(1998)}]{WDSS}%
    \BibitemOpen
    \bibfield  {author} {\bibinfo {author} {\bibfnamefont {D.~J.}\
    \bibnamefont {Watts}}\ and\ \bibinfo {author} {\bibfnamefont {S.~H.}\
    \bibnamefont {Strogatz}},\ }\bibfield  {title} {\enquote {\bibinfo {title}
    {Collective dynamics of `small-world' networks},}\ }\href@noop {} {\bibfield
    {journal} {\bibinfo  {journal} {Nature}\ }\textbf {\bibinfo {volume} {393}},\
    \bibinfo {pages} {440 -- 442} (\bibinfo {year} {1998})}\BibitemShut {NoStop}%
\bibitem [{\citenamefont {Boccaletti}\ \emph {et~al.}(2006)\citenamefont
  {Boccaletti}, \citenamefont {Latora}, \citenamefont {Moreno}, \citenamefont
  {Chavez},\ and\ \citenamefont {Hwang}}]{BSLV}%
  \BibitemOpen
  \bibfield  {author} {\bibinfo {author} {\bibfnamefont {S.}~\bibnamefont
  {Boccaletti}}, \bibinfo {author} {\bibfnamefont {V.}~\bibnamefont {Latora}},
  \bibinfo {author} {\bibfnamefont {Y.}~\bibnamefont {Moreno}}, \bibinfo
  {author} {\bibfnamefont {M.}~\bibnamefont {Chavez}}, \ and\ \bibinfo {author}
  {\bibfnamefont {D.-U.}\ \bibnamefont {Hwang}},\ }\bibfield  {title} {\enquote
  {\bibinfo {title} {Complex networks: Structure and dynamics},}\ }\href
  {\doibase https://doi.org/10.1016/j.physrep.2005.10.009} {\bibfield
  {journal} {\bibinfo  {journal} {Physics Reports}\ }\textbf {\bibinfo {volume}
  {424}},\ \bibinfo {pages} {175 -- 308} (\bibinfo {year} {2006})}\BibitemShut
  {NoStop}%
\bibitem [{\citenamefont {Newman}(2010)}]{NewmanIntroBook}%
  \BibitemOpen
  \bibfield  {author} {\bibinfo {author} {\bibfnamefont {M.}\ \bibnamefont
  {Newman}},\ }\href@noop {} {\emph {\bibinfo {title} {Networks: An
  Introduction}}}\ (\bibinfo  {publisher} {Oxford University Press, Inc.},\
  \bibinfo {address} {New York, NY, USA},\ \bibinfo {year} {2010})\BibitemShut
  {NoStop}%
\bibitem [{\citenamefont {Newman}\ \emph {et~al.}(2001)\citenamefont {Newman},
  \citenamefont {Strogatz},\ and\ \citenamefont {Watts}}]{NMSW}%
  \BibitemOpen
  \bibfield  {author} {\bibinfo {author} {\bibfnamefont {M.~E.~J.}\
  \bibnamefont {Newman}}, \bibinfo {author} {\bibfnamefont {S.~H.}\
  \bibnamefont {Strogatz}}, \ and\ \bibinfo {author} {\bibfnamefont {D.~J.}\
  \bibnamefont {Watts}},\ }\bibfield  {title} {\enquote {\bibinfo {title}
  {Random graphs with arbitrary degree distributions and their applications},}\
  }\href@noop {} {\bibfield  {journal} {\bibinfo  {journal} {Phys. Rev. E}\
  }\textbf {\bibinfo {volume} {64}},\ \bibinfo {pages} {026118} (\bibinfo
  {year} {2001})}\BibitemShut {NoStop}%
\bibitem [{\citenamefont {Newman}(2003)}]{MEJN1}%
  \BibitemOpen
  \bibfield  {author} {\bibinfo {author} {\bibfnamefont {M.~E.~J.}\
  \bibnamefont {Newman}},\ }\bibfield  {title} {\enquote {\bibinfo {title} {The
  structure and function of complex networks},}\ }\href@noop {} {\bibfield
  {journal} {\bibinfo  {journal} {SIAM Review}\ }\textbf {\bibinfo {volume}
  {45}},\ \bibinfo {pages} {167--256} (\bibinfo {year} {2003})}\BibitemShut
  {NoStop}%
\bibitem [{\citenamefont {Albert}\ and\ \citenamefont
  {Barab\'asi}(2002)}]{ARB}%
  \BibitemOpen
  \bibfield  {author} {\bibinfo {author} {\bibfnamefont {R.}\ \bibnamefont
  {Albert}}\ and\ \bibinfo {author} {\bibfnamefont {A.-L.}\
  \bibnamefont {Barab\'asi}},\ }\bibfield  {title} {\enquote {\bibinfo {title}
  {Statistical mechanics of complex networks},}\ }\href@noop {} {\bibfield
  {journal} {\bibinfo  {journal} {Rev. Mod. Phys.}\ }\textbf {\bibinfo {volume}
  {74}},\ \bibinfo {pages} {47--97} (\bibinfo {year} {2002})}\BibitemShut
  {NoStop}%
\bibitem [{\citenamefont {Park}\ and\ \citenamefont {Newman}(2004)}]{PJNM}%
  \BibitemOpen
  \bibfield  {author} {\bibinfo {author} {\bibfnamefont {J.}\ \bibnamefont
  {Park}}\ and\ \bibinfo {author} {\bibfnamefont {M.~E.~J.}\ \bibnamefont
  {Newman}},\ }\bibfield  {title} {\enquote {\bibinfo {title} {Statistical
  mechanics of networks},}\ }\href@noop {} {\bibfield  {journal} {\bibinfo
  {journal} {Phys. Rev. E}\ }\textbf {\bibinfo {volume} {70}},\ \bibinfo
  {pages} {066117} (\bibinfo {year} {2004})}\BibitemShut {NoStop}%
\bibitem [{\citenamefont {Bianconi}(2015)}]{BG1}%
  \BibitemOpen
  \bibfield  {author} {\bibinfo {author} {\bibfnamefont {G.}\
  \bibnamefont {Bianconi}},\ }\bibfield  {title} {\enquote {\bibinfo {title}
  {Interdisciplinary and physics challenges of network theory},}\ }\href
  {http://stacks.iop.org/0295-5075/111/i=5/a=56001} {\bibfield  {journal}
  {\bibinfo  {journal} {EPL}\ }\textbf {\bibinfo {volume}
  {111}},\ \bibinfo {pages} {56001} (\bibinfo {year} {2015})}\BibitemShut
  {NoStop}%
\bibitem [{\citenamefont {Krioukov}\ \emph {et~al.}(2009)\citenamefont
  {Krioukov}, \citenamefont {Papadopoulos}, \citenamefont {Vahdat},\ and\
  \citenamefont {Bogu\~n\'a}}]{KDPF1}%
  \BibitemOpen
  \bibfield  {author} {\bibinfo {author} {\bibfnamefont {D.}\ \bibnamefont
  {Krioukov}}, \bibinfo {author} {\bibfnamefont {F.}\ \bibnamefont
  {Papadopoulos}}, \bibinfo {author} {\bibfnamefont {A.}\ \bibnamefont
  {Vahdat}}, \ and\ \bibinfo {author} {\bibfnamefont {M.}\ \bibnamefont
  {Bogu\~n\'a}},\ }\bibfield  {title} {\enquote {\bibinfo {title} {Curvature
  and temperature of complex networks},}\ }\href@noop {} {\bibfield  {journal}
  {\bibinfo  {journal} {Phys. Rev. E}\ }\textbf {\bibinfo {volume} {80}},\
  \bibinfo {pages} {035101(R)} (\bibinfo {year} {2009})}\BibitemShut {NoStop}%
\bibitem [{\citenamefont {Krioukov}\ \emph {et~al.}(2010)\citenamefont
  {Krioukov}, \citenamefont {Papadopoulos}, \citenamefont {Kitsak},
  \citenamefont {Vahdat},\ and\ \citenamefont {Bogu\~n\'a}}]{KDPF2}%
  \BibitemOpen
  \bibfield  {author} {\bibinfo {author} {\bibfnamefont {D.}\ \bibnamefont
  {Krioukov}}, \bibinfo {author} {\bibfnamefont {F.}\ \bibnamefont
  {Papadopoulos}}, \bibinfo {author} {\bibfnamefont {M.}\ \bibnamefont
  {Kitsak}}, \bibinfo {author} {\bibfnamefont {A.}\ \bibnamefont {Vahdat}}, \
  and\ \bibinfo {author} {\bibfnamefont {M.}\ \bibnamefont
  {Bogu\~n\'a}},\ }\bibfield  {title} {\enquote {\bibinfo {title} {Hyperbolic
  geometry of complex networks},}\ }\href@noop {} {\bibfield  {journal}
  {\bibinfo  {journal} {Phys. Rev. E}\ }\textbf {\bibinfo {volume} {82}},\
  \bibinfo {pages} {036106} (\bibinfo {year} {2010})}\BibitemShut {NoStop}%
\bibitem [{\citenamefont {Bogu\~{n}\'{a}}\ \emph {et~al.}(2010)\citenamefont
  {Bogu\~{n}\'{a}}, \citenamefont {Papadopoulos},\ and\ \citenamefont
  {Krioukov}}]{Boguna2010}%
  \BibitemOpen
  \bibfield  {author} {\bibinfo {author} {\bibfnamefont {M.}\
  \bibnamefont {Bogu\~{n}\'{a}}}, \bibinfo {author} {\bibfnamefont
  {F.}\ \bibnamefont {Papadopoulos}}, \ and\ \bibinfo {author}
  {\bibfnamefont {D.}\ \bibnamefont {Krioukov}},\ }\bibfield  {title}
  {\enquote {\bibinfo {title} {Sustaining the Internet with hyperbolic
  mapping},}\ }\href {http://dx.doi.org/10.1038/ncomms1063} {\bibfield
  {journal} {\bibinfo  {journal} {Nature Communications}\ }\textbf {\bibinfo
  {volume} {1}},\ \bibinfo {pages} {62 EP --} (\bibinfo {year}
  {2010})}\BibitemShut {NoStop}%
\bibitem [{\citenamefont {Garlaschelli}\ and\ \citenamefont
  {Loffredo}(2009)}]{GDLM}%
  \BibitemOpen
  \bibfield  {author} {\bibinfo {author} {\bibfnamefont {D.}\ \bibnamefont
  {Garlaschelli}}\ and\ \bibinfo {author} {\bibfnamefont {M.~I.}\
  \bibnamefont {Loffredo}},\ }\bibfield  {title} {\enquote {\bibinfo {title}
  {{Generalized Bose-Fermi Statistics and Structural Correlations in Weighted
  Networks}},}\ }\href@noop {} {\bibfield  {journal} {\bibinfo  {journal}
  {Phys. Rev. Lett.}\ }\textbf {\bibinfo {volume} {102}},\ \bibinfo {pages}
  {038701} (\bibinfo {year} {2009})}\BibitemShut {NoStop}%
\bibitem [{\citenamefont {Serrano}\ \emph {et~al.}(2008)\citenamefont
  {Serrano}, \citenamefont {Krioukov},\ and\ \citenamefont
  {Bogu\~n\'a}}]{SMKD1}%
  \BibitemOpen
  \bibfield  {author} {\bibinfo {author} {\bibfnamefont {M. A.}\
  \bibnamefont {Serrano}}, \bibinfo {author} {\bibfnamefont {D.}\
  \bibnamefont {Krioukov}}, \ and\ \bibinfo {author} {\bibfnamefont {M.}\
  \bibnamefont {Bogu\~n\'a}},\ }\bibfield  {title} {\enquote {\bibinfo {title}
  {Self-similarity of complex networks and hidden metric spaces},}\ }\href
  {\doibase 10.1103/PhysRevLett.100.078701} {\bibfield  {journal} {\bibinfo
  {journal} {Phys. Rev. Lett.}\ }\textbf {\bibinfo {volume} {100}},\ \bibinfo
  {pages} {078701} (\bibinfo {year} {2008})}\BibitemShut {NoStop}%
\bibitem [{\citenamefont {Papadopoulos}\ \emph {et~al.}(2012)\citenamefont
  {Papadopoulos}, \citenamefont {Kitsak}, \citenamefont {Serrano},
  \citenamefont {Bogu{\~n}{\'a}},\ and\ \citenamefont {Krioukov}}]{PFKD}%
  \BibitemOpen
  \bibfield  {author} {\bibinfo {author} {\bibfnamefont {F.}\
  \bibnamefont {Papadopoulos}}, \bibinfo {author} {\bibfnamefont {M.}\
  \bibnamefont {Kitsak}}, \bibinfo {author} {\bibfnamefont {M.~{\'A}ngeles}\
  \bibnamefont {Serrano}}, \bibinfo {author} {\bibfnamefont {M.}\
  \bibnamefont {Bogu{\~n}{\'a}}}, \ and\ \bibinfo {author} {\bibfnamefont
  {D.}\ \bibnamefont {Krioukov}},\ }\bibfield  {title} {\enquote {\bibinfo
  {title} {Popularity versus similarity in growing networks},}\ }\href
  {https://doi.org/10.1038/nature11459} {\bibfield  {journal} {\bibinfo
  {journal} {Nature}\ }\textbf {\bibinfo {volume} {489}},\ \bibinfo {pages}
  {537 EP --} (\bibinfo {year} {2012})}\BibitemShut {NoStop}%
\bibitem [{\citenamefont {A.}\ \emph {et~al.}(2008)\citenamefont {A.},
  \citenamefont {M.},\ and\ \citenamefont {A.}}]{BABM}%
  \BibitemOpen
  \bibfield  {author} {\bibinfo {author} {\bibfnamefont {A.} \ \bibnamefont
  {Barrat}}, \bibinfo {author} {\bibfnamefont {M.} \ \bibnamefont {Barthelemy}}, \
  and\ \bibinfo {author} {\bibfnamefont {A.}\ \bibnamefont {Vespignani}},\
  }\href@noop {} {\emph {\bibinfo {title} {{Dynamical Processes on Complex
  Networks}}}}\ (\bibinfo  {publisher} {{Cambridge University Press}},\
  \bibinfo {year} {2008})\BibitemShut {NoStop}%
\bibitem [{\citenamefont {Verbeek}\ and\ \citenamefont {Suri}(2016)}]{VKSS}%
  \BibitemOpen
  \bibfield  {author} {\bibinfo {author} {\bibfnamefont {K.}\ \bibnamefont
  {Verbeek}}\ and\ \bibinfo {author} {\bibfnamefont {S.}\ \bibnamefont
  {Suri}},\ }\bibfield  {title} {\enquote {\bibinfo {title} {{Metric embedding,
  hyperbolic space, and social networks}},}\ }\href {\doibase
  https://doi.org/10.1016/j.comgeo.2016.08.003} {\bibfield  {journal} {\bibinfo
   {journal} {Computational Geometry}\ }\textbf {\bibinfo {volume} {59}},\
  \bibinfo {pages} {1 -- 12} (\bibinfo {year} {2016})}\BibitemShut {NoStop}%
\bibitem [{\citenamefont {Narayan}\ and\ \citenamefont {Saniee}(2011)}]{NOSI}%
  \BibitemOpen
  \bibfield  {author} {\bibinfo {author} {\bibfnamefont {O.}\ \bibnamefont
  {Narayan}}\ and\ \bibinfo {author} {\bibfnamefont {I.}\ \bibnamefont
  {Saniee}},\ }\bibfield  {title} {\enquote {\bibinfo {title} {Large-scale
  curvature of networks},}\ }\href {\doibase 10.1103/PhysRevE.84.066108}
  {\bibfield  {journal} {\bibinfo  {journal} {Phys. Rev. E}\ }\textbf {\bibinfo
  {volume} {84}},\ \bibinfo {pages} {066108} (\bibinfo {year}
  {2011})}\BibitemShut {NoStop}%
\bibitem [{\citenamefont {Bianconi}\ and\ \citenamefont
  {Rahmede}(2017)}]{BCRC}%
  \BibitemOpen
  \bibfield  {author} {\bibinfo {author} {\bibfnamefont {G.}\
  \bibnamefont {Bianconi}}\ and\ \bibinfo {author} {\bibfnamefont {C.}\
  \bibnamefont {Rahmede}},\ }\bibfield  {title} {\enquote {\bibinfo {title}
  {{Emergent Hyperbolic Network Geometry}},}\ }\href@noop {} {\bibfield
  {journal} {\bibinfo  {journal} {Scientific Reports}\ }\textbf {\bibinfo
  {volume} {7}},\ \bibinfo {pages} {41974 EP --} (\bibinfo {year}
  {2017})}\BibitemShut {NoStop}%
\bibitem [{\citenamefont {Bogu\~n\'a}\ and\ \citenamefont
  {Pastor-Satorras}(2003)}]{BMPSR}%
  \BibitemOpen
  \bibfield  {author} {\bibinfo {author} {\bibfnamefont {M.}\
  \bibnamefont {Bogu\~n\'a}}\ and\ \bibinfo {author} {\bibfnamefont {R.}\
  \bibnamefont {Pastor-Satorras}},\ }\bibfield  {title} {\enquote {\bibinfo
  {title} {Class of correlated random networks with hidden variables},}\ }\href
  {\doibase 10.1103/PhysRevE.68.036112} {\bibfield  {journal} {\bibinfo
  {journal} {Phys. Rev. E}\ }\textbf {\bibinfo {volume} {68}},\ \bibinfo
  {pages} {036112} (\bibinfo {year} {2003})}\BibitemShut {NoStop}%
\bibitem [{\citenamefont {Bogacz}\ \emph {et~al.}(2006)\citenamefont {Bogacz},
  \citenamefont {Burda},\ and\ \citenamefont {Waclaw}}]{BOGACZ2006587}%
  \BibitemOpen
  \bibfield  {author} {\bibinfo {author} {\bibfnamefont {L.}\ \bibnamefont
  {Bogacz}}, \bibinfo {author} {\bibfnamefont {Z.}\ \bibnamefont
  {Burda}}, \ and\ \bibinfo {author} {\bibfnamefont {B.}\ \bibnamefont
  {Waclaw}},\ }\bibfield  {title} {\enquote {\bibinfo {title} {Homogeneous
  complex networks},}\ }\href {\doibase
  https://doi.org/10.1016/j.physa.2005.10.024} {\bibfield  {journal} {\bibinfo
  {journal} {Physica A}\ }\textbf
  {\bibinfo {volume} {366}},\ \bibinfo {pages} {587 -- 607} (\bibinfo {year}
  {2006})}\BibitemShut {NoStop}%
\bibitem [{\citenamefont {Wang}\ \emph {et~al.}(2016)\citenamefont {Wang},
  \citenamefont {Li}, \citenamefont {Jin}, \citenamefont {Xiong},\ and\
  \citenamefont {Wu}}]{WANG}%
  \BibitemOpen
  \bibfield  {author} {\bibinfo {author} {\bibfnamefont {Z.}\ \bibnamefont
  {Wang}}, \bibinfo {author} {\bibfnamefont {Q.}\ \bibnamefont {Li}},
  \bibinfo {author} {\bibfnamefont {F.}\ \bibnamefont {Jin}}, \bibinfo
  {author} {\bibfnamefont {W.}\ \bibnamefont {Xiong}}, \ and\ \bibinfo
  {author} {\bibfnamefont {Y.}\ \bibnamefont {Wu}},\ }\bibfield  {title}
  {\enquote {\bibinfo {title} {Hyperbolic mapping of complex networks based on
  community information},}\ }\href {\doibase
  https://doi.org/10.1016/j.physa.2016.02.015} {\bibfield  {journal} {\bibinfo
  {journal} {Physica A}\ }\textbf
  {\bibinfo {volume} {455}},\ \bibinfo {pages} {104 -- 119} (\bibinfo {year}
  {2016})}\BibitemShut {NoStop}%
\bibitem [{\citenamefont {Ollivier}(2013)}]{OY1}%
  \BibitemOpen
  \bibfield  {author} {\bibinfo {author} {\bibfnamefont {Y.}\ \bibnamefont
  {Ollivier}},\ }\enquote {\bibinfo {title} {{A visual introduction to
  Riemannian curvatures and some discrete generalizations}},}\ in\ \href@noop
  {} {\emph {\bibinfo {booktitle} {Analysis and Geometry of Metric Measure
  Spaces: Lecture Notes of the 50th S\'eminaire de Math\'ematiques
  Sup\'erieures (SMS), Montr\'eal, 2011}}},\ \bibinfo {series} {{Crm
  Proceedings \& Lecture Notes}}, Vol.~\bibinfo {volume} {56},\ \bibinfo
  {editor} {edited by\ \bibinfo {editor} {\bibfnamefont {Galia}\ \bibnamefont
  {Dafni}}, \bibinfo {editor} {\bibfnamefont {Robert~John}\ \bibnamefont
  {McCann}}, \ and\ \bibinfo {editor} {\bibfnamefont {Alina}\ \bibnamefont
  {Stancu}}}\ (\bibinfo  {publisher} {AMS},\ \bibinfo
  {year} {2013})\BibitemShut {NoStop}%
\bibitem [{\citenamefont {Sreejith}\ \emph {et~al.}(2016)\citenamefont
  {Sreejith}, \citenamefont {Mohanraj}, \citenamefont {Jost}, \citenamefont
  {Saucan},\ and\ \citenamefont {Samal}}]{SRKM}%
  \BibitemOpen
  \bibfield  {author} {\bibinfo {author} {\bibfnamefont {R.~P.}\ \bibnamefont
  {Sreejith}}, \bibinfo {author} {\bibfnamefont {K.}\ \bibnamefont
  {Mohanraj}}, \bibinfo {author} {\bibfnamefont {J.}\ \bibnamefont
  {Jost}}, \bibinfo {author} {\bibfnamefont {E.}\ \bibnamefont {Saucan}}, \
  and\ \bibinfo {author} {\bibfnamefont {A.}\ \bibnamefont {Samal}},\
  }\bibfield  {title} {\enquote {\bibinfo {title} {Forman curvature for complex
  networks},}\ }\href {http://stacks.iop.org/1742-5468/2016/i=6/a=063206}
  {\bibfield  {journal} {\bibinfo  {journal} {Journal of Statistical Mechanics}\ }\textbf {\bibinfo {volume} {2016}},\ \bibinfo
  {pages} {063206} (\bibinfo {year} {2016})}\BibitemShut {NoStop}%
\bibitem [{\citenamefont {Saucan}\ \emph {et~al.}(2019)\citenamefont {Saucan},
  \citenamefont {Sreejith}, \citenamefont {Vivek-Ananth}, \citenamefont
  {Jost},\ and\ \citenamefont {Samal}}]{SESR}%
  \BibitemOpen
  \bibfield  {author} {\bibinfo {author} {\bibfnamefont {E.}\ \bibnamefont
  {Saucan}}, \bibinfo {author} {\bibfnamefont {R. P.}\ \bibnamefont {Sreejith}},
  \bibinfo {author} {\bibfnamefont {R. P.}\ \bibnamefont {Vivek-Ananth}},
  \bibinfo {author} {\bibfnamefont {J.}\ \bibnamefont {Jost}}, \ and\
  \bibinfo {author} {\bibfnamefont {A.}\ \bibnamefont {Samal}},\
  }\bibfield  {title} {\enquote {\bibinfo {title} {{Discrete Ricci curvatures
  for directed networks}},}\ }\href@noop {} {\bibfield  {journal} {\bibinfo
  {journal} {Chaos, Solitons \& Fractals}\ }\textbf {\bibinfo {volume} {118}},\
  \bibinfo {pages} {347 -- 360} (\bibinfo {year} {2019})}\BibitemShut {NoStop}%
\bibitem [{\citenamefont {Keller}(2011)}]{Keller2011}%
  \BibitemOpen
  \bibfield  {author} {\bibinfo {author} {\bibfnamefont {M.}\
  \bibnamefont {Keller}},\ }\bibfield  {title} {\enquote {\bibinfo {title}
  {Curvature, geometry and spectral properties of planar graphs},}\ }\href
  {\doibase 10.1007/s00454-011-9333-0} {\bibfield  {journal} {\bibinfo
  {journal} {Discrete {\&} Computational Geometry}\ }\textbf {\bibinfo {volume}
  {46}},\ \bibinfo {pages} {500--525} (\bibinfo {year} {2011})}\BibitemShut
  {NoStop}%
  \bibitem [{\citenamefont {Estrada}(2012)}]{EE2012}%
  \BibitemOpen
  \bibfield  {author} {\bibinfo {author} {\bibfnamefont {E.}\ \bibnamefont
  {Estrada}},\ }\bibfield  {title} {\enquote {\bibinfo {title} {{Complex
  networks in the Euclidean space of communicability distances}},}\ }\href
  {\doibase 10.1103/PhysRevE.85.066122} {\bibfield  {journal} {\bibinfo
  {journal} {Phys. Rev. E}\ }\textbf {\bibinfo {volume} {85}},\ \bibinfo
  {pages} {066122} (\bibinfo {year} {2012})}\BibitemShut {NoStop}%
  \bibitem [{\citenamefont {Estrada}\ \emph {et~al.}(2014)\citenamefont
  {Estrada}, \citenamefont {S\'anchez-Lirola},\ and\ \citenamefont {de~la
  Pe$\rm\tilde{n}$a}}]{EE2014}%
  \BibitemOpen
  \bibfield  {author} {\bibinfo {author} {\bibfnamefont {E.}\ \bibnamefont
  {Estrada}}, \bibinfo {author} {\bibfnamefont {M. G.}\ \bibnamefont
  {S\'anchez-Lirola}}, \ and\ \bibinfo {author} {\bibfnamefont
  {J.~A.}\ \bibnamefont {de~la Pe$\rm\tilde{n}$a}},\ }\bibfield
  {title} {\enquote {\bibinfo {title} {Hyperspherical embedding of graphs and
  networks in communicability spaces},}\ }\href {\doibase
  https://doi.org/10.1016/j.dam.2013.05.032} {\bibfield  {journal} {\bibinfo
  {journal} {Discrete Applied Mathematics}\ }\textbf {\bibinfo {volume}
  {176}},\ \bibinfo {pages} {53 -- 77} (\bibinfo {year} {2014})}\BibitemShut
  {NoStop}%
\bibitem [{\citenamefont {Horne}(2009)}]{poincare}%
  \BibitemOpen
  \bibfield  {author} {\bibinfo {author} {\bibfnamefont {Bill}\ \bibnamefont
  {Horne}},\ }\href@noop {} {\enquote {\bibinfo {title} {{Poincar\'{e}}},}\
  }\bibinfo {howpublished} {\url{http:/poincare.sourceforge.net/}} (\bibinfo
  {year} {2009})\BibitemShut {NoStop}%
\bibitem [{\citenamefont {Sabinin}(1999)}]{Sab1}%
  \BibitemOpen
  \bibfield  {author} {\bibinfo {author} {\bibfnamefont {L.~V.}\ \bibnamefont
  {Sabinin}},\ }\href@noop {} {\emph {\bibinfo {title} {{Smooth quasigroups and
  loops}}}}\ (\bibinfo  {publisher} {Kluwer Academic Publishers},\ \bibinfo
  {address} {Dordrecht},\ \bibinfo {year} {1999})\BibitemShut {NoStop}%
\bibitem [{\citenamefont {Nesterov}\ and\ \citenamefont
  {Sabinin}(2000{\natexlab{a}})}]{NS3}%
  \BibitemOpen
  \bibfield  {author} {\bibinfo {author} {\bibfnamefont {A.~I.}\ \bibnamefont
  {Nesterov}}\ and\ \bibinfo {author} {\bibfnamefont {L.~V.}\ \bibnamefont
  {Sabinin}},\ }\bibfield  {title} {\enquote {\bibinfo {title} {{Nonassociative
  geometry: towards discrete structure of spacetime}},}\ }\href@noop {}
  {\bibfield  {journal} {\bibinfo  {journal} {Phys. Rev. D}\ }\textbf {\bibinfo
  {volume} {62}},\ \bibinfo {pages} {081501(R)} (\bibinfo {year}
  {2000}{\natexlab{a}})}\BibitemShut {NoStop}%
\bibitem [{\citenamefont {Nesterov}\ and\ \citenamefont
  {Sabinin}(2000{\natexlab{b}})}]{NS3a}%
  \BibitemOpen
  \bibfield  {author} {\bibinfo {author} {\bibfnamefont {A.~I.}\ \bibnamefont
  {Nesterov}}\ and\ \bibinfo {author} {\bibfnamefont {L.~V.}\ \bibnamefont
  {Sabinin}},\ }\bibfield  {title} {\enquote {\bibinfo {title}
  {{Non-associative geometry and discrete structure of spacetime}},}\
  }\href@noop {} {\bibfield  {journal} {\bibinfo  {journal} {Comment. Math.
  Univ. Carolin.}\ }\textbf {\bibinfo {volume} {41,2}},\ \bibinfo {pages} {347
  -- 358} (\bibinfo {year} {2000}{\natexlab{b}})}\BibitemShut {NoStop}%
\bibitem [{\citenamefont {Sabinin}(1988)}]{S5}%
  \BibitemOpen
  \bibfield  {author} {\bibinfo {author} {\bibfnamefont {L.~V.}\ \bibnamefont
  {Sabinin}},\ }\bibfield  {title} {\enquote {\bibinfo {title} {Differential
  equations of smooth loops},}\ }\href@noop {} {\bibfield  {journal} {\bibinfo
  {journal} {Proc. of Sem. on Vector and Tensor Analysis}\ }\textbf {\bibinfo
  {volume} {23}},\ \bibinfo {pages} {133} (\bibinfo {year} {1988})}\BibitemShut
  {NoStop}%
\bibitem [{\citenamefont {Sabinin}(1989)}]{S6}%
  \BibitemOpen
  \bibfield  {author} {\bibinfo {author} {\bibfnamefont {L.~V.}\ \bibnamefont
  {Sabinin}},\ }\bibfield  {title} {\enquote {\bibinfo {title} {Differential
  geometry and quasigroups},}\ }\href@noop {} {\bibfield  {journal} {\bibinfo
  {journal} {Proc. Inst. Math. Siberian Branch of Ac. Sci. USSR}\ }\textbf
  {\bibinfo {volume} {14}},\ \bibinfo {pages} {208} (\bibinfo {year}
  {1989})}\BibitemShut {NoStop}%
\bibitem [{\citenamefont {Sabinin}(1994)}]{S7}%
  \BibitemOpen
  \bibfield  {author} {\bibinfo {author} {\bibfnamefont {L.~V.}\ \bibnamefont
  {Sabinin}},\ }\bibfield  {title} {\enquote {\bibinfo {title} {On differential
  equations of smooth loops},}\ }\href
  {http://stacks.iop.org/0036-0279/49/i=2/a=L18} {\bibfield  {journal}
  {\bibinfo  {journal} {Russian Mathematical Surveys}\ }\textbf {\bibinfo
  {volume} {49}},\ \bibinfo {pages} {172} (\bibinfo {year} {1994})}\BibitemShut
  {NoStop}%
  \bibitem [{\citenamefont {Nesterov}\ and\ \citenamefont
    {Mata}(2019{\natexlab{b}})}]{NHM}%
    \BibitemOpen
    \bibfield  {author} {\bibinfo {author} {\bibfnamefont {A.~I.}\ \bibnamefont
    {Nesterov}}\ and\ \bibinfo {author} {\bibfnamefont {H.}\ \bibnamefont
    {Mata }},\ }\bibfield  {title} {\enquote {\bibinfo {title}
    {How {nonassociative geometry describes a discrete structure of spacetime}},}\
    }\href@noop {} {\bibfield  {journal} {\bibinfo  {journal} {Frontiers in Math. Phys.}\ 
    }\textbf {\bibinfo {volume} {7}},\ \bibinfo {pages} {1
    -- 37} (\bibinfo {year} {2019}{\natexlab{b}})}\BibitemShut {NoStop}%
\bibitem [{\citenamefont {Nesterov}(2000)}]{N1}%
  \BibitemOpen
  \bibfield  {author} {\bibinfo {author} {\bibfnamefont {A.~I.}\ \bibnamefont
  {Nesterov}},\ }\bibfield  {title} {\enquote {\bibinfo {title} {{Principal
  $Q$-bundles}},}\ }in\ \href@noop {} {\emph {\bibinfo {booktitle} {Non
  {A}ssociative {A}lgebra and {I}ts {A}pplications}}},\ \bibinfo {editor}
  {edited by\ \bibinfo {editor} {\bibfnamefont {R.}~\bibnamefont {Costa}},
  \bibinfo {editor} {\bibfnamefont {H.}~\bibnamefont {{C}uzzo Jr.}}, \bibinfo
  {editor} {\bibfnamefont {A.}~\bibnamefont {{G}rishkov}}, \ and\ \bibinfo
  {editor} {\bibfnamefont {L.~A.}\ \bibnamefont {{P}eresi}}}\ (\bibinfo
  {publisher} {Marcel Dekker},\ \bibinfo {address} {New York},\ \bibinfo {year}
  {2000})\BibitemShut {NoStop}%
\bibitem [{\citenamefont {Nesterov}(2001)}]{N2}%
  \BibitemOpen
  \bibfield  {author} {\bibinfo {author} {\bibfnamefont {A.~I.}\ \bibnamefont
  {Nesterov}},\ }\bibfield  {title} {\enquote {\bibinfo {title} {Principal loop
  bundles: Toward nonassociative gauge theories},}\ }\href@noop {} {\bibfield
  {journal} {\bibinfo  {journal} {Int. J. Theor. Phys.}\ }\textbf {\bibinfo
  {volume} {40}},\ \bibinfo {pages} {339 -- 350} (\bibinfo {year}
  {2001})}\BibitemShut {NoStop}%
   \bibitem[{\citenamefont{Garlaschelli et~al.}(2013)\citenamefont{Garlaschelli,
            Ahnert}}]{CDLM}
          \bibinfo{author}{\bibfnamefont{D.}~\bibnamefont{Garlaschelli}},
            \bibinfo{author}{\bibfnamefont{M. I.} \bibnamefont{Loffredo}},
            ``Multispecies grand-canonical models for networks 
              with reciprocity.'',
            \bibinfo{journal}{Phys. Rev. E } \textbf{\bibinfo{volume}{73}},
            \bibinfo{pages}{015101(R)} (\bibinfo{year}{2006})\BibitemShut {NoStop}%
 \bibitem [{\citenamefont {Garlaschelli}\ \emph {et~al.}(2013)\citenamefont
        {Garlaschelli}, \citenamefont {Ahnert}, \citenamefont {Fink},\ and\
        \citenamefont {Caldarelli}}]{CDAS}%
        \BibitemOpen
        \bibfield  {author} {\bibinfo {author} {\bibfnamefont {D.}\ \bibnamefont
        {Garlaschelli}}, \bibinfo {author} {\bibfnamefont {S.~E.}\
        \bibnamefont {Ahnert}}, \bibinfo {author} {\bibfnamefont {T. M.~A.}\
        \bibnamefont {Fink}}, \ and\ \bibinfo {author} {\bibfnamefont {G.}\
        \bibnamefont {Caldarelli}},\ }\bibfield  {title} {\enquote {\bibinfo {title}
        {Low-temperature behaviour of social and economic networks},}\ }\href
        {\doibase 10.3390/e15083238} {\bibfield  {journal} {\bibinfo  {journal}
        {Entropy}\ }\textbf {\bibinfo {volume} {15}},\ \bibinfo {pages} {3148--3169}
        (\bibinfo {year} {2013})}\BibitemShut {NoStop}%
         \bibitem[{\citenamefont{Girvan and Newman}(2002)}]{GN}
                            \bibinfo{author}{\bibfnamefont{M.}~\bibnamefont{Girvan}} \bibnamefont{and}
                              \bibinfo{author}{\bibfnamefont{M. E. J.}~\bibnamefont{ Newman}},
                              \bibfield  {title} 
                                                    {\enquote {\bibinfo {title}
                                           {Community structure in social and biological networks},}\ 
                                                    }\href@noop {}
                              \bibinfo{journal}{PNAS}
                              \textbf{\bibinfo{volume}{99}}, \bibinfo{pages}{7821--7826} 
                              (\bibinfo{year}{2002}).   
          \bibitem [{\citenamefont {Erd\'eley}(1953)}]{AEWM}%
      \BibitemOpen
      \bibfield  {author} {\bibinfo {author} {\bibfnamefont {A.}\ \bibnamefont
      {Erd\'eley}}, \bibinfo {author} {\bibfnamefont {W.}\ \bibnamefont
        {Magnus}},\ and\ \bibinfo {author} {\bibfnamefont {F.}\ \bibnamefont
        {Oberhettinger}},\ }\href@noop {} {\emph {\bibinfo {title} { Higher Transcendental 
      Functions, Vol. I.}}}\ (\bibinfo  {publisher} {McGraw-Hill},\
      \bibinfo {address} {New York, NY, USA},\ \bibinfo {year} {1953})\BibitemShut
      {NoStop}%
  \bibitem [{\citenamefont {Krioukov}\ \emph {et~al.}(2006)\citenamefont
  {Krioukov}, \citenamefont {Papadopoulos}, \citenamefont {Vahdat},\ and\
  \citenamefont {Bogu\~n\'a}}]{MKF}%
  \BibitemOpen
  \bibfield  {author} {\bibinfo {author} {\bibfnamefont {P.}\ \bibnamefont
  {Mahadevan}}, \bibinfo {author} {\bibfnamefont {D.}\ \bibnamefont
  {Krioukov}}, \bibinfo {author} {\bibfnamefont {M.}\ \bibnamefont
  {Fomenkov}}, \bibinfo {author} {\bibfnamefont {B.}\ \bibnamefont
  {Huffaker}}, \bibinfo {author} {\bibfnamefont {X.}\ \bibnamefont
  { Dimitropoulos}}, \bibinfo {author} {\bibfnamefont {K.}\ \bibnamefont
  {Claffy}}, \and\ \bibinfo {author} {\bibfnamefont {A.}\ \bibnamefont
  {Vahdat}},  }\bibfield  {title} {\enquote {\bibinfo {title} {The Internet AS-Level Topology: Three Data Sources and One Definitive Metric},}\ }\href@noop {} {\bibfield  {journal}
  {\bibinfo  {journal} {ACM SIGCOMM Computer Communication Review (CCR)}\ }\textbf {\bibinfo {volume} {36}},\
  \bibinfo {pages} {17-26} (\bibinfo {year} {2006})}\BibitemShut {NoStop}%
            \bibitem[{\citenamefont{Cohen and Havlin}(2003)}]{CRHS}
            \bibinfo{author}{\bibfnamefont{R.}~\bibnamefont{Cohen}} \bibnamefont{and}
              \bibinfo{author}{\bibfnamefont{S.}~\bibnamefont{Havlin}},
              \bibfield  {title} 
                        {\enquote {\bibinfo {title}
               {Scale-Free Networks Are Ultrasmall},}\ 
                        }\href@noop {}
              \bibinfo{journal}{Phys. Rev. Lett.} \textbf{\bibinfo{volume}{90}},
              \bibinfo{pages}{058701} (\bibinfo{year}{2003}).
  \bibitem[{\citenamefont{Chung and Lu}(2002)}]{CFL}
              \bibinfo{author}{\bibfnamefont{F.}~\bibnamefont{Chung}} \bibnamefont{and}
                \bibinfo{author}{\bibfnamefont{L.}~\bibnamefont{Lu}},
                \bibfield  {title} 
                                      {\enquote {\bibinfo {title}
                             {The average distances in random graphs with given expected degrees},}\ 
                                      }\href@noop {}
                \bibinfo{journal}{PNAS}
                \textbf{\bibinfo{volume}{99}}, \bibinfo{pages}{15879} (\bibinfo{year}{2002}).   
  \end{thebibliography}


%

\end{document}